\documentclass[aps,prx,twocolumn,amsmath,amssymb,floatfix,superscriptaddress]{revtex4-1}
\usepackage{bbold,rotating}
\usepackage{comment}
\usepackage{physics}
\usepackage[dvipsnames]{xcolor}
\usepackage{amsmath,amsthm,amssymb,amsfonts,stmaryrd,wasysym,graphicx,multirow,color,textcomp}
\usepackage{hyperref}\usepackage{url}

\usepackage[utf8x]{inputenc}

\begin{document}

\title{Topological orders with classical Lie group symmetries from coupling electron wires}
\author{Pak Kau Lim} 
\author{Michael Mulligan}
\affiliation{Department of Physics and Astronomy, University of California, Riverside, California 92521, USA.}
\author{Jeffrey C.Y. Teo}
\affiliation{Department of Physics, University of Virginia, Charlottesville, Virginia 22904, USA.}
\affiliation{Theoretical Sciences Visiting Program, Okinawa Institute of Science and Technology Graduate University, Onna, 904-0495, Japan}

\begin{abstract}
We study the topological order that arises from chiral states with ${\rm SU}(N)$ or ${\rm SO}(N)$ 
edge-state symmetry.
This extends our previous study of topological orders that descend from the bosonic $E_8$ quantum Hall state.
We use exactly solvable models of coupled electron wires to construct states with ${\rm SU}(m)_n$, ${\rm SO}(m)_n$, or ${\rm Sp}(m)_n$ topological order for various levels $n$.
We use our constructions to write down string operators for various non-Abelian anyons.
We thereby provide a systematic, model derivation of quantum Hall states, topological superconductors, and spin liquids with emergent non-Abelian quasiparticle excitations, including those of Ising, metaplectic, and Fibonacci type. 
\end{abstract}

\maketitle
\tableofcontents
\section{Introduction}\label{sec:intro}

The $E_8$ quantum Hall state is the simplest short-range entangled 2d topological state of bosons, in which all edge modes move in the same direction.
We recently showed \cite{LimMulliganTeo2022A} how this state can be viewed as the ``parent" state for a variety of different Abelian and non-Abelian 2d topological orders.
This relation between the $E_8$ state and its ``children" is analogous to
 the relation between the $\nu=1$ integer quantum Hall (IQH) state of electrons and the various fractional quantum Hall (FQH) states that occur when the lowest Landau level is partially filled $\nu < 1$.
(Here, $\nu$ is the electrical filling fraction: the number of 2d electrons divided by the number of magnetic flux quanta.)
The present paper extends the previously-used methodology in \cite{LimMulliganTeo2022A} to large families of different parent states which, in contrast to the $E_8$ work, may be bosonic or fermionic and preserve or not preserve U(1) electromagnetic symmetry.
The broader picture is to write down the exactly solvable models for topological phases of electronic systems.
The presence of spin-1 local fields on the boundary motivates the study of WZW topological phases.
In a chiral phase of electrons, gapless spin-1 local fields on the boundary edge are even bosonic combinations of electrons and are the simplest bosonic local fields. Chiral CFTs generated by spin-1 local fields are classified by WZW algebras. 
We will show how such exactly solvable models allow for a rather systematic understanding for the emergence of quantum Hall states, topological superconductors, and spin liquids with Abelian and non-Abelian quasiparticle excitations of the Majorana, metaplectic, and Fibonacci type.

Our work in \cite{LimMulliganTeo2022A} used the coupled-wire construction~\cite{PhysRevLett.88.036401,PhysRevB.89.085101}: This is now a well established approach in which a target topological order is constructed from an exactly solvable model consisting of an array of coupled 1d electron wires.
The specific topological order that appears depends on the choice of inter and intra wire couplings and the nature of the interacting 1d electron liquid (e.g., the number of channels and the values of the short-ranged density-density interactions). 
Ref.~\cite{LimMulliganTeo2022A} used the construction in \cite{PhysRevB.100.085116} of the $E_8$ quantum Hall state at filling fraction $\nu=16$, given in terms of 11-channel electron wires, supplemented by appropriate inter/intra wire interactions.
This microscopic interacting electron construction is not unique; rather, it provides an explicit model that allows for detailed study.
Within this model \cite{PhysRevB.100.085116} of the $E_8$ state, we showed how the $E_8$ Kac-Moody (KM) Wess-Zumino-Witten (WZW) \cite{WessZumino71,WittenWZW,witten1984} edge-state symmetry can be used to engineer alternative inter/intra wire interactions that produce long-range entangled topological orders, different from the short-range entangled $E_8$ one.
(These ``alternative inter/intra wire interactions" become dominant at lower filling fractions $\nu < 16$.)

To understand how this works, it is helpful to recall that each coupled-wire model, for a given topological phase, such as the $E_8$ state -- initially understood to arise from interacting electron wires -- has an equivalent low-energy description in terms of coupled narrow strips of $E_8$ liquid (or whatever the target topological phase may be).
By a narrow strip of $E_8$ liquid, we mean an 8-channel non-chiral Luttinger liquid with $E_8$ KM symmetry.
Such an effective description is independent of non-universal microscopic details that define a specific interacting electron construction.
For example, the effective description is independent of the number of electron wires, however, it does depend on the filling fraction, which can manifest itself through the electrical Hall conductivity.

There are various ways to ``glue together" the narrow $E_8$ strips.
The most straightforward way produces $E_8$ topological order.
The Hamiltonian for this ``gluing" process takes the following schematic form:
\begin{align}
\begin{split}
    \mathcal{H}[\mathcal{G}]_{\rm edge} &= \mathcal{H}_0+
    \mathcal{H}^\mathcal{G}_{\rm inter},\\
    \mathcal{H}^\mathcal{G}_{\rm inter} &= u_{\rm inter}\sum_y
    {\bf J}^R_{y,\mathcal{G}}\cdot{\bf J}^L_{y+1,\mathcal{G}},
\end{split}
\end{align}
where, for the construction of the $E_8$ state, ${\cal G} = E_8$.
Above, $\mathcal{H}_0$ is the Hamiltonian for an array of un-coupled 8-channel non-chiral Luttinger liquids with $E_8$ KM symmetry.
The wires (i.e., the ``strips") in this array are taken to lie along the $x$ axis and to be regularly spaced along the $y$ axis.
$\mathcal{H}^\mathcal{G}_{\rm inter}$ describes the coupling between the degrees of freedom in these wires: Here,
${\bf J}^{R/L}_{y, {\cal G}}$ is a 248-dimensional vector of $E_8$ currents for right/left moving excitations on wire (or ``strip") $y$, and $u_{\rm inter}$ parameterizes the inter wire current-current interactions.
The dot product between currents runs over all 248 symmetry generators in the $E_8$ Lie algebra.

The alternative inter/intra wire interactions we studied in \cite{LimMulliganTeo2022A} were obtained by applying the conformal field theory technique known as conformal embedding to the (non-chiral) edge-state theory of each strip.
Conformal embedding allows the factorization of the $E_8$ edge-state degrees of freedom into representations of a product subgroup ${\cal G}_A \times {\cal G}_B \subset E_8$.
Example factorizations ${\cal G}_A \times {\cal G}_B$ include $SU(3) \times E_6$ and $G_2 \times F_4$ \cite{Bais:1986zs,PhysRevD.34.3092}.
Here and throughout, $\{SO(M) \}$, $\{SU(N)\}$, $\{ Sp(N) \}$, and $\{E_8, G_2, F_4 \}$ denote orthogonal, unitary, sympletic, and exceptional Lie algebras, with $N$ a positive integer.
This factorization allows for the construction of ${\cal G}_{A,B}$ symmetry currents ${\bf J}^{R/L}_{y, {\cal G}_{A,B}}$ out of the $E_8$ degrees of freedom.
Suppressing the wire, right/left, and ${\cal G}^{A/B}$ indices, these currents have an operator product expansion (OPE) \cite{francesco2012conformal}:
\begin{align}
    J_{a}(z)J_{b}(w)= \frac{k \delta_{ab}}{(z-w)^2}
    +\sum_cif^{abc}\frac{J_{c}(w)}{(z-w)}+\dots,
\end{align}
where $a,b$ denote Lie algebra indices, $z$ and $w$ are $(1+1)d$ spacetime coordinates along the edge of a given strip, $f^{abc}$ are Lie algebra structure constants, and  $k$ denotes the so-called level of the KM algebra.
The currents in the OPE above carry the same wire, right/left, and ${\cal G}^{A/B}$ labels.
Other OPEs, say, between a ${\cal G}^{A}$ current and a ${\cal G}^{B}$ current are nonsingular, i.e., the terms on the right-hand side of the OPEs vanish as $z \rightarrow w$.
(This is what it means for ${\cal G}_A$ and ${\cal G}_B$ to embed into ${\cal G} = E_8$ as the  product ${\cal G}_A \times {\cal G}_B$.)

The Sugawara construction is the final ingredient to be used with the conformal embedding technique.
In general, for a CFT with ${\cal G}$ KM symmetry currents ${\bf J}$ (again suppressing any wire, right/left, and symmetry labels), the stress-tenesor (and, in particular ${\cal H}_0$) admits the Sugawara construction,
\begin{align}
    T(z)=\frac{1}{2(k+g)}\sum_a :J_aJ_a:(z),
    \label{eq:SugawaraConstruction}
\end{align}
where $g$ is the dual Coxeter number of the Lie algebra of ${\cal G}$ and $: \cdot :$ denotes normal ordering.
The conformal embedding ${\cal G}_A \times {\cal G}_B \subset {\cal G}$ allows the stress-tensor to be rewritten as $T_{\mathcal{G}}(z)=T_{\mathcal{G}^A}(z)+T_{\mathcal{G}^B}(z)$, where each stress-tensor has the form in Eq.~\eqref{eq:SugawaraConstruction}, with appropriate currents summed over.
This decomposition and the central charge formula,
\begin{align}
c_\mathcal{G}=\frac{k{\rm dim}\mathcal{G}}{k+g},
\label{eq:CentralChargeForumla}
\end{align}
implies that the central charge of the ${\cal G} = E_8$ CFT is the sum of the central charges of the ${\cal G}_A$ and ${\cal G}_B$ CFTs, where ${\rm dim }\mathcal{G}$ is the dimension of the Lie algebra ${\cal G}$.

\begin{figure}[htbp]
\includegraphics[width=0.4\textwidth]{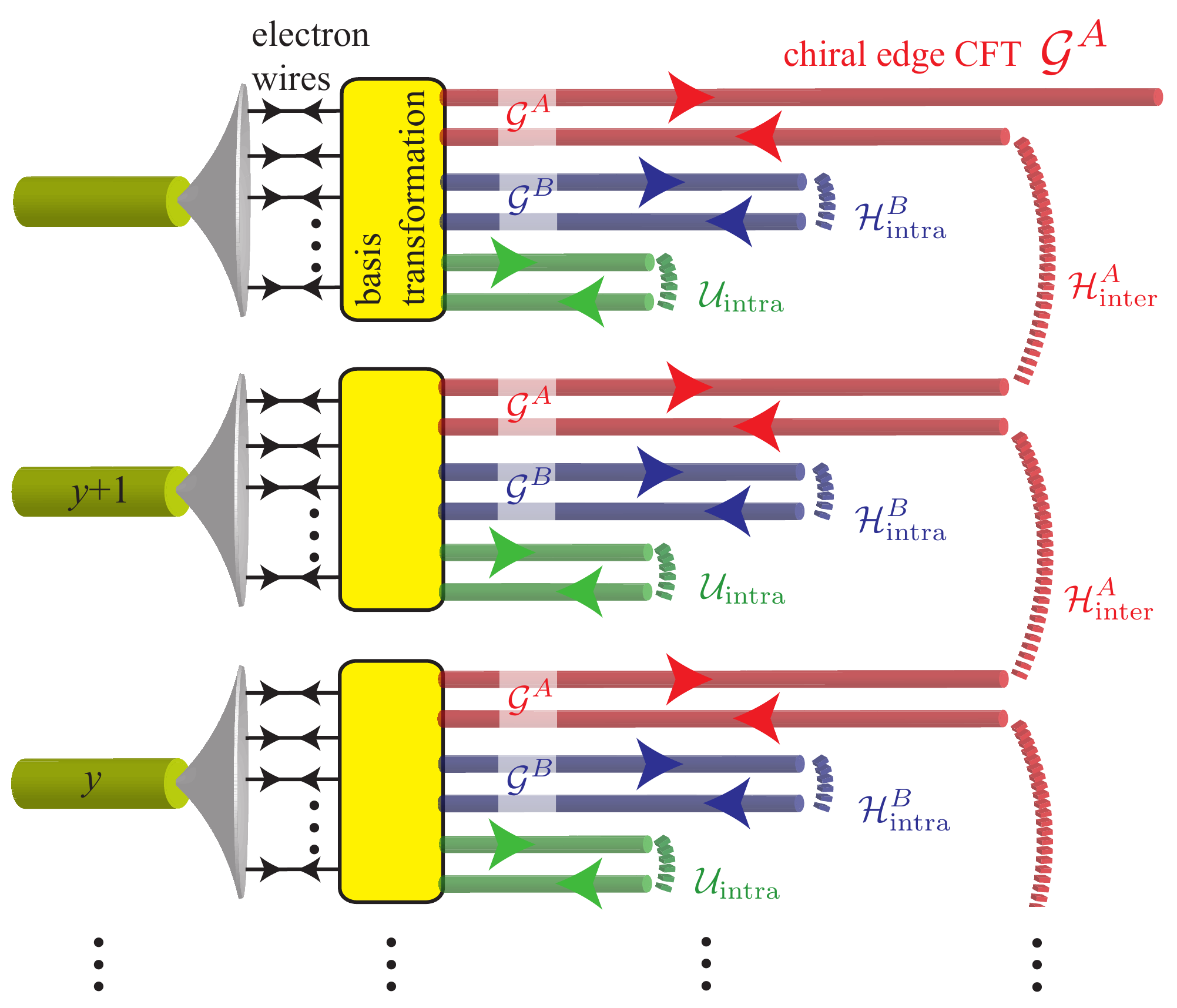}
\caption{Schematics of the coupled-wire construction in a 2D array of bundles of electron wires. The non-chiral $\mathcal{G}=\mathcal{G}^A\times\mathcal{G}^B$ CFT is realized in each bundle by a many-electron sine-Gordon interaction $\mathcal{U}_{\mathrm{intra}}$. The $A$ and $B$ sectors are gapped in the bulk via backscattering associating chiral KM currents in opposite directions by $\mathcal{H}^A_{\mathrm{inter}}$ and $\mathcal{H}^B_{\mathrm{intra}}$. The chiral $\mathcal{G}^A$ CFT is left un-gapped on the edges.}\label{fig:coupledWireConstruction}
\end{figure}

Equipped with the symmetry currents of the conformal embedding ${\cal G}_A \times {\cal G}_B \subset {\cal G}$ and using the Sugawara construction, we then showed how certain inter/intra wire interactions produce a quantum Hall state with either ${\cal G}_A$ or ${\cal G}_B$ topological order.
For instance, we showed that the combination of intra wire $E_6$ current-current interactions + inter wire $SU(3)$ current-current interactions produce the $SU(3)_1$ quantum Hall state.
(The ``1" subscript on $SU(3)_1$ denotes the the level of the algebra.)
The coupled-wire Hamiltonian for the ${\cal G}^A$ phase, depicted in figure~\ref{fig:coupledWireConstruction}, has the form: 
\begin{align}
    \mathcal{H}[\mathcal{G}^A] & = \mathcal{H}_0+ \mathcal{H}^{\mathcal{G}^A}_{\rm inter}+
    \mathcal{H}^{\mathcal{G}^B}_{\rm intra},
    \label{eq:chiralHamiltonian} 
\end{align}
with inter/intra wire interactions,
\begin{align}
\begin{split}
    \mathcal{H}^{\mathcal{G}^A}_{\rm inter}&= u_{\rm inter}\sum_y \mathbf{J}^R_{y,\mathcal{G}^A}\cdot
    \mathbf{J}^L_{y+1,\mathcal{G}^A},\\
    \mathcal{H}^{\mathcal{G}^B}_{\rm intra}&= u_{\rm intra}\sum_y \mathbf{J}^R_{y,\mathcal{G}^B}\cdot
    \mathbf{J}^L_{y,\mathcal{G}^B}.
\end{split}
\label{eq:generalinterintraHamilton}
\end{align}
The Hamiltonian for the ${\cal G}^B$ phase is obtained by exchanging ${\cal G}^A$ with ${\cal G}^B$.  
Importantly, all the interactions we considered are local, i.e., expressible in terms of products of the elementary boson/fermion creation and annihilation operators.
(These interactions are products of boson operators in the effective description consisting of $E_8$ strips and fermion operators in the microscopic electron description.)

The purpose of this paper is to apply this program to other topological phases with classical Lie group edge-state symmetries, different from the $E_8$ quantum Hall state.
Specifically, we will consider ``parent" $\mathcal{G}$-states with $\mathcal{G}=SU(mn)_1$, $SO(mn)_1$, or $SO(4mn)_1$ edge-state symmetries and the symmetry decompositions: 
\begin{align}
\begin{cases}
    &  {\rm SU}(m)_n\times SU(n)_m \subset {\rm SU}(mn)_1, \\
    & {\rm SO}(m)_n\times SO(n)_m \subset {\rm SO}(mn)_1,\\
    & {\rm Sp}(2m)_{n}\times {\rm Sp}(2n)_{m} \subset {\rm SO}(4mn)_1.
    \end{cases}
    \label{eq:ABCDDuality}
\end{align}
Here, $m$ and $n$ are positive integers.
Note that states with $SO(N)$ symmetry group do not have electrical $U(1)$ symmetry.
Following the procedure outlined above for the $E_8$ quantum Hall state, we will use these symmetry decompositions to engineer quantum Hall liquids, topological superconductors, or spin liquids with, respectively, $SU(N)_k$, $SO(N)_k$, or $Sp(N)_k$  edge-state symmetry, for some positive integer $k$. 
We thereby provide explicit constructions of various states with Abelian or non-Abelian topological orders.

The reminder of the paper is organized as follows.
In Sec.~\ref{Sec:AseriesPartonWireConstruction}, we construct the deconfined parton phase ${\rm U}(mn)_1/\mathbb{Z}_{mn}={\rm U}(1)_{mn}\times {\rm SU}(mn)_1$ of the Laughlin quantum Hall states as our parent phase. We then present the decomposition of the KM current algebra using the conformal embedding ${\rm SU}(m)_n\times {\rm SU}(n)_m \subset {\rm SU}(mn)_1$. 
The fractional quantum Hall states with chiral WZW CFTs given by 
(i) ${\rm U}(1)_{mn}\times {\rm SU}(mn)_1$, and (ii) ${\rm U}(1)_{mn}\times {\rm SU}(m)_n$ are formulated using the solvable coupled-wire model
(see also \eqref{eq:CaseIIHamiltonian}, and \eqref{eq:CaseIIIHamiltonian}). Special cases for $m=2$ are discussed and examples of topological orders are presented accordingly.
In Sec.~\ref{Sec.BDseriesAndTopOrder}, we review the origins of bosonic wires based on superconducting and spin liquids \cite{Teo_2023}. Such wires have ${\rm SO}(mn)_1$ symmetry and give rise to emergent Majorana fermions. We then construct coupled-wire models whose edge theories support ${\rm SO}(mn)_1$, ${\rm SO}(m)_n$ or ${\rm SO}(n)_m$ WZW CFTs. These constructions follow from the decomposition ${\rm SO}(m)_n\times{\rm SO}(n)_m \subset {\rm SO}(mn)_1$. 
In Sec.~\ref{Sec:CseriesTheSymplecticFermions}, we consider the emergence of symplectic fermions and their associated topological orders. These constructions use the conformal embedding ${\rm Sp}(2n)_m\times{\rm Sp}(2m)_n \subset {\rm SO}(4mn)_1$ and a description of the symplectic series in terms of the quaternion algebra.
We summarize and conclude in Sec.~\ref{Sec:DiscussionAndConclusion}.

\section{\texorpdfstring{$A_r$}{A} Series and Partons}
\label{Sec:AseriesPartonWireConstruction}

In this section, we show how fractional quantum Hall states with $U(1)_{mn}\times SU(m)_n$ topological order can arise from  ``partially occupying" the deconfined parton FQH states \cite{PhysRevB.99.245117} with $U(mn)_1/\mathbb{Z}_{mn}=U(1)_{mn}\times SU(mn)_1$ topological order. 
Fractional quantum Hall states with ${\rm SU}(m)_n$ topological order has previously been studied in  \cite{PhysRevB.95.125130,PhysRevB.100.115111,PhysRevLett.66.802}. In this paper, we construct using the coupled-wire model fractional quantum Hall states with similar topological order that partially filled deconfined parton Landau level.

\subsection{Generalized Parton Construction} 
\label{partonsection}

We begin with a description of the parton FQH states.
Prior work \cite{PhysRevB.99.245117} constructed such states with $U(3)_1/\mathbb{Z}_3$ topological order; we will generalize this to $U(mn)_1/\mathbb{Z}_{mn}$, for arbitrary positive integers $m,n$.
The parton phase can be motivated by imagining the fractionalization of a fundamental or elementary fermion (boson) $\Psi_{el}$ into fermionic partons $d^a$, for $a = 1, \ldots, mn$~\cite{PhysRevLett.66.802,PhysRevB.40.8079,doi:10.1142/S0217979292000840}:
\begin{align}
\Psi_{el} =d^1d^2\dots d^{mn}.
\label{eq:PartonAnsatz}
\end{align}
A FQH phase of the fundamental fermion (boson) is then realized as a particular topological ordering of the partons. 
(In the simplest of instances, a FQH phase can form when the partons enter an IQH phase.) 
$\Psi_{el}$ is ``fundamental" in the sense that it is composed of an integral combination of electron operators.
We will study cases where $\Psi_{el}$ carries electric charge $e$ ($2e$), when $\Psi_{el}$ is fermionic (bosonic). 
By ``electron operators," we are referring to spin-polarized electron creation/annihilation operators for the electrons in the underlying coupled-wire array that produce the parton FQH phase.
The factor of $2$ for even $mn$ reflects the fact that $\Psi_{el}$ is composed of an even number of electron operators.

One route to the parton phase (used in \cite{PhysRevB.99.245117}) begins with the parton decomposition \eqref{eq:PartonAnsatz}.
This decomposition has a $\mathbb{Z}_{mn}$ gauge redundancy by which $d^a\rightarrow e^{2\pi i/mn} d^a$.
Describing the parton phase by its edge-state theory, we may systematically obtain the correct edge-state Lagrangian by first treating $\mathbb{Z}_{mn}$ as a global symmetry to obtain a $U(mn)_1$ edge-state theory (i.e., an edge-state theory with $U(mn)_1$ KM symmetry) and subsequently gauging the $\mathbb{Z}_{mn}$ symmetry to obtain the edge-state theory for the parton FQH phase with $U(mn)_1/\mathbb{Z}_{mn}$ KM symmetry.
The edge-state Lagrangian for the parton FQH phase is
\begin{align}
\mathcal{L}_{U(mn)_1/\mathbb{Z}_{mn}} = \frac{1}{4\pi}K_{IJ}\partial_t\Tilde{\phi}^I\partial_\mathsf{x}\Tilde{\phi}^J  - \frac{1}{4\pi} \tilde V_{IJ} \partial_{\mathsf{x}}{\tilde \phi}^I\partial_{\mathsf{x}}{\tilde \phi}^J.
\label{eq:partionLagrangian}
\end{align}
Here, $\tilde \phi^I$ are chiral (left-moving) bosons to be defined later in \eqref{bigphibigtildephi}, $K=\left(K_{IJ}\right)$ is the $mn$-dimensional Cartan matrix of $U(mn)_1/\mathbb{Z}_{mn} = U(1)_{mn} \times SU(mn)_1$,
\begin{gather}
K=K_{U(1)_{mn}}\oplus K_{SU(mn)_1} = \begin{pmatrix}K_{U(1)_{mn}}&0\\0&K_{SU(mn)_1}\end{pmatrix}\label{eq:KUNlevel1},\\
K_{U(1)_{mn}}=mn,\quad K_{SU(mn)} = \left(\begin{smallmatrix}
2&-1 & &&&\\
-1& 2& -1 &&&\\
& -1& 2& &&\\
&&&\ddots&&\\
&&&&2&-1\\
&&&&-1&2
\end{smallmatrix}\right),\nonumber
\end{gather} where $I,J \in \{0, \ldots, mn-1 \}$.
Elements of $K$ above, not indicated explicitly, e.g., elements $K_{1p}$ for $p = 2, \ldots, mn-1$, are zero; we will use this notation throughout whenever convenient.
The Einstein summation convention is assumed, unless otherwise specified. The positive symmetric velocity matrix $\tilde V_{IJ}$ can be set to be proportional to $K_{IJ}$ by appropriate short-ranged density-density interactions, so that the free theory \eqref{eq:partionLagrangian} has the $U(mn)$ KM symmetry.
The above Lagrangian implies the bosons satisfy the equal-time commutation algebra:
\begin{align}    [\partial_\mathsf{x}\Tilde{\phi}^I(\mathsf{x}),\Tilde{\phi}^J(\mathsf{x}')] = 2\pi i(K^{-1})^{IJ} \delta(\mathsf{x}-\mathsf{x}').
\label{eq:etcrChevalley}
\end{align}
The bosons $\tilde \phi^I$ are $2\pi$-periodic: $\tilde \phi^I \sim \tilde \phi^I + 2 \pi N^I$, where $N^I$ is an arbitrary integer vector.
This periodicity implies that general quasiparticle creation/annihilation operators along the edge are given by vertex operators of the form $e^{i m_I \tilde \phi^I}$, for arbitrary integer vectors $m_I$; local fermion/boson operators are of the form $e^{i n^I K_{IJ} \tilde \phi^J}$, for arbitrary integer vectors $n^I$.
Local operators have trivial monodromy with any quasiparticle operator and may be understood to be a product of an integral number of elementary fermion/boson operators.

The parton phase has a conserved electric $U(1)$ symmetry and a corresponding characteristic electrical Hall response.
This response is determined by a so-called charge vector $t_I$.
In the coupled-wire construction of the parton FQH phase that we will describe later, the charge vector is
\begin{align}
t_I =\begin{cases}
 (1, 0, \ldots,0), \quad  (mn {\text{ odd}})\\
(2,0,\ldots,0), \quad  (mn {\text{ even}}).
\end{cases}
\end{align}
This charge vector implies that $\tilde \phi^0$ of the $U(1)_{mn}$ sector carries all of the electric charge along an edge; the $SU(mn)_1$ sector is electrically neutral. 
The fundamental fermion (boson) operator in the Laughlin charged sector is $\Psi_{el} = e^{i m n \tilde \phi^0}$.
The electrical Hall conductivity is 
\begin{align}
\sigma_{xy} = \nu\frac{e^2}{h},
\end{align}
where $\nu = t_I (K^{-1})^{IJ} t_J$.
In the parton phase, 
\begin{align}
\nu =
\begin{cases}
    \frac{1}{mn}, \quad (mn \text{ odd}),\\
    \frac{4}{mn}, \quad (mn \text{ even}).
\end{cases}
\end{align}
Here, $\nu$ is also the filling fraction that counts the ratio between the number of electric charge and number of magnetic flux quanta in this FQH phase.
In addition to the electrical response, the parton FQH phase has a thermal response characterized by the thermal Hall conductivity~\cite{PhysRevB.55.15832,2002NuPhB.636..568C},
\begin{align}
\kappa_{xy} = c\frac{\pi^2 k_B^2}{3 h}T,
\end{align}
where $c$ is the chiral central charge of the state and $T$ is the applied temperature gradient.
In the parton phase, $c = mn$, which is the total number of edge modes.

The $U(mn)$ symmetry becomes apparent under the following invertible transformation from the ``Chevalley" basis $\tilde\phi$ to the ``Cartan-Weyl" basis $\phi$.
\begin{align}
{\phi}^r=R_I^r \Tilde{\phi}^{I},
\label{cartanweylchevrelation}
\end{align}
where 
\begin{align}
R=\begin{pmatrix}
1 & 1 & & & &\\
1 & -1 & 1 & &  &\\
1 &  &-1 & 1 & & \\
\vdots & & &\ddots &  &\\
1 & &  &  &-1&
\end{pmatrix}.
\label{eq:Rmatrix}
\end{align}
Here $r=1,\ldots,mn$ and $I=0,\ldots,mn-1$ labels the rows and columns of the matrix.
The columns of $R$ are the simple roots of $U(1)_{mn} \times SU(mn)_1$, such that the $K$-matrix in \eqref{eq:KUNlevel1} decomposes into $K_{IJ} = \sum_r R^r_I R^r_J$.
Note that $R$ is not unimodular and has determinant $mn$.
In the Cartan-Weyl basis, the Lagrangian \eqref{eq:partionLagrangian} becomes 
\begin{align}
\mathcal{L}_{{\rm U}(mn)_1/\mathbb{Z}_{mn}} = \frac{1}{4\pi}\delta_{rs}\partial_t \phi^r\partial_\mathsf{x} \phi^s  - \frac{1}{4\pi} V_{rs} \partial_{\mathsf{x}} \phi^r\partial_{\mathsf{x}} \phi^s,
\label{eq:partionLagrangiancartanweyl}
\end{align}
where $V_{rs} = (R^{-1})_r^I \tilde V_{IJ} (R^{-1})_s^J=v\delta_{rs}$ is diagonal so that the theory is $U(mn)$ symmetric.
The Cartan-Weyl bosons obey the equal-time commutation relation
\begin{align}
[\partial_\mathsf{x}{\phi}^r(\mathsf{x}),{\phi}^s(\mathsf{x}')] =2\pi i\delta^{rs}\delta(\mathsf{x}-\mathsf{x}').
\end{align}
The Cartan-Weyl bosons have more complicated compactification conditions that mix various species $\tilde \phi^I$ according to the $R$ matrix ($\phi^r \sim \phi^r + 2 \pi R^r_I N^I$). The parton fermions are the vertex fields $d^r \sim e^{i \phi^r}$. (The $\sim$ symbol here indicates that we are suppressing a non-universal constant of engineering 
dimension $1/2$.) The fundamental fermion (boson) operator now is $\Psi_{el} = e^{i \phi^1} \cdots e^{i \phi^{mn}}$, which recovers the parton decomposition \eqref{eq:PartonAnsatz}. The partons are non-local fractional fields. They transform according to an internal $\mathbb{Z}_{mn}$ discrete gauge symmetry that sends $\phi^r\to \phi^r+2\pi/(mn)$. Any local vertex operator that is an integral combination of electrons must be $\mathbb{Z}_{mn}$ neutral and therefore invariant under the $\mathbb{Z}_{mn}$ transformation.

We close this subsection by presenting the ${\rm U}(mn)_1/\mathbb{Z}_{mn} = U(1)_{mn} \times SU(mn)_1$ edge-state symmetry currents of the parton FQH phase.
These currents (and the notation we use to present them) are used in the coupled-wire construction of the parton FQH phase; analogous ideas will be used in constructing the $SU(m)_n$ phases (and the other phases considered in later sections).

Using standard Lie algebra terminology, there are two types of currents: Cartan and root currents.
The Cartan current of $U(1)_{mn}$ is
\begin{align}
\mathtt{H}_{{\rm U}(1)_{mn}}
&=\frac{1}{\sqrt{mn}}\sum_{r=1}^{mn}\partial_{\mathsf{x}}{\phi}^r\sim (d^r)^{\dagger}X^{U(1)_{mn}}_{rs}d^s,\nonumber\\
X^{U(1)_{mn}}_{rs}&=\delta_{rs}/\sqrt{mn}
\label{eq:U1cartan}
\end{align}
where $r,s=1,\ldots, mn$.
The Cartan currents of $SU(mn)_1$ are 
\begin{gather} \begin{split}
\left[\mathtt{H}_{SU(mn)_1}\right]_p & =
\frac{1}{\sqrt{p(p+1)}}\left(\sum_{r=1}^p \partial_{\mathsf{x}}{\phi}^r-p\partial_{\mathsf{x}}{\phi}^{p+1}\right) 
\\&\sim (d^\dagger)^r X^p_{rs} d^s,\end{split}\label{eq:SUmnCartan}\end{gather}
\[
X^p_{rs}=\left\{\begin{array}{*{20}l}1/\sqrt{p(p+1)},&\mbox{for }1\leq r=s\leq p\\-p/\sqrt{p(p+1)},&\mbox{for }r=s=p+1\\0,&\mbox{otherwise}\end{array}\right.,
\]
where $p=1,\ldots, mn-1$. 
These Cartan currents form a maximal set of commuting operators in the affine Lie algebra. They span the operator space generated by diagonal parton densities $(d^r)^\dagger d^r$.

The root currents of $SU(mn)_1$ are the off-diagonal bilinear parton products
\begin{align}\begin{split}
[\mathtt{E}_{{\rm SU}(mn)_1}]_{\boldsymbol{\alpha}}&= 
e^{i(\phi^r-\phi^s)}\\&\sim d^r(d^s)^\dagger=d^{r'}X^{rs}_{r's'}(d^{s'})^\dagger\\
X^{rs}_{r's'}&=\delta^r_{r'}\delta^s_{s'}\end{split}\label{eq:SUmnRoots}.
\end{align}
Here, ${\boldsymbol{\alpha}=}
{\bf e}_r-{\bf e}_s$ are the root vectors of $SU(mn)$, where $r,s$ are distinct integers between $1$ and $mn$. The set of such root vectors $\boldsymbol{\alpha}$ is the root lattice $\Delta_{SU(mn)}$. ${\bf e}_{k}$ is the unit $mn$-vector whose $k^\mathrm{th}$ element equals 1 and is zero elsewhere.

\subsection{Coupled-Wire Construction of Parton FQH States}\label{Sec:CWPartonFQH}

In the last subsection, we defined the $U(mn)_1/\mathbb{Z}_{mn}=U(1)_{mn}\times SU(mn)_1$ parton CFT. In this subsection, we present a theoretical model based on electrons where the parton CFT emerges. In particular, we present the coupled-wire construction of the parton FQH phase with $U(1)_N \times SU(N)_1$ edge-state theory, for integer $N = mn > 1$.
(Recall by ``$U(1)_N \times SU(N)_1$ edge-state theory," we mean an edge-state theory with $U(1)_N \times SU(N)_1$ KM symmetry.)

Following \cite{PhysRevB.99.245117}, we introduce a 2D array of electron wires, aligned parallel to the $\mathsf{x}$ axis and in the presence of a transverse magnetic field ${\bf B} = B \hat {\bf z}$.
We choose the gauge ${\bf A} = - B \mathsf{y} \hat {\bf x}$.
The wires are grouped into bundles, each containing $N+2$ wires $j = 0, 1, \ldots, N+1$ and regularly spaced along the $\mathsf{y}$ axis. The vertical positions of the wires are $\mathsf{y} = y d+\delta_j$, where $y \in \mathbb{Z}$ labels the bundle, $d$ is the spacing between neighboring bundles, and $\delta_j$ is the relative vertical position of the $j^{\mathrm{th}}$ wire in a given bundle. 
Each wire $(y,j)$ carries a spin-polarized electron.
At low energies, each electron decomposes into a single Dirac electron, with one left ($L$) and one right ($R$) moving component. 
We denote 
$c^\sigma_{y j}(\mathsf{x})$ as the chiral component ($\sigma=R/L= \pm 1$) of 
the Dirac electron
in wire $j$ of bundle $y$
at the Fermi momentum point,
\begin{align}
k^\sigma_{yj}=\frac{eB}{\hbar c} \left(y d+\delta_j\right) +\sigma k_{F,j}.
\label{eq:generalShiftFermiMomenta}
\end{align}
Here $k_{F,j}$ is the ``bare" Fermi momentum that sets the electron number density (per unit length) $n^0_j=2k_{F,j}/(2\pi)$ on wire $j$.

We write the Dirac electrons in terms of chiral bosons $\Phi^\sigma_{yj}$ as
\begin{align}
c^\sigma_{yj}(\mathsf{x})\sim \exp\left[
i \big( \Phi^\sigma_{yj}(\mathsf{x}) + k^\sigma_{yj} \mathsf{x} \big)\right].
\label{eq:N2electronicchannels}
\end{align}
Before introducing interactions that couple the wires together, the free Lagrangian is
\begin{align}
\begin{split}
\mathcal{L}= \frac{1}{4 \pi} \sum_y\sum_{\sigma=\pm}\sum_{j=0}^{N+1} 
\sigma\partial_t\Phi^\sigma_{yj}
\partial_\mathsf{x}\Phi^\sigma_{yj} -\mathcal{H}_0,
\label{eq:LuttingerLagrangian}
\end{split}
\end{align}
where $\mathcal{H}_0$ is the kinetic Hamiltonian including density-density interactions between electrons within a bundle of wires,
\begin{align}
\mathcal{H}_0=\sum_y\sum_{\sigma,\sigma'=\pm}\sum_{j,j'= 0}^{N+1}v^{jj'}_{\sigma\sigma'}\partial_\mathsf{x}{\Phi}^\sigma_{yj}\partial_\mathsf{x}{\Phi}^{\sigma'}_{yj'},
\label{eq:LuttingerHamiltonian}
\end{align}
and $v^{jj'}_{\sigma\sigma'}$ is a positive definite symmetric matrix to be specified shortly.

The parton FQH phase is achieved by coupling together nearby bundles of wires in a particular way.
To motivate the necessary inter-bundle interaction, we first perform a basis transformation,
\begin{align}\begin{split}
    &\tilde \Phi^\sigma_{y \mu} = \sum_{\sigma' j} U^{\sigma \sigma'}_{\mu j} \Phi^{\sigma'}_{y j},\\
    &\tilde\Phi^\sigma_{y\rho}=N\tilde\phi^{\sigma,0}_y,\quad\tilde\Phi^\sigma_{yI}=(K_{SU(N)_1})_{IJ}\tilde\phi^{\sigma,J}_y,\\
    &\tilde\Phi^\sigma_{yc_1}=K_{c_1}\tilde\phi^\sigma_{yc_1},
    \quad\tilde\Phi^\sigma_{yc_2}=N\tilde\phi^\sigma_{yc_2},
\end{split}\label{bigphibigtildephi}\end{align}
where $\mu = \rho, c_1, c_2, 1, 2, \ldots, N-1$ and $I,J=1\ldots,N-1$. The $K$-matrix is the Cartan matrix of $U(N)$ presented in \eqref{eq:KUNlevel1}. $K_{c_1}=N$ if $N$ is odd, and $K_{c_1}=1$ if $N$ is even. $U^{\sigma \sigma'}=\left(U^{\sigma \sigma'}_{\mu j}\right)$ are $(N+2) \times (N+2)$ matrices with integral entries. Since $\tilde \Phi^\sigma_{y \mu}$ is an integral linear combination of the $\Phi^{\sigma'}_{y j}$, vertex operators of the form $\exp(i n^\mu \tilde \Phi^\sigma_{y \mu})$, for arbitrary integer-vector $n_\mu$, are local. 

The $U$ transformation allows the embedding of the parton Lagrangian $\mathcal{L}_{U(N)_1/\mathbb{Z}_N}$ \eqref{eq:partionLagrangian} inside a bundle of $N+2$ electron wires. After the basis transformation, the free Lagrangian \eqref{eq:LuttingerLagrangian} becomes 
\begin{align}
\begin{split}
\mathcal{L}&=
\frac{1 }{4 \pi} \sum_y\sum_{\sigma=\pm}
\sum_{I,J=0}^{N-1} \sigma K_{IJ}\partial_t\tilde{\phi}^{\sigma,I}_y\partial_\mathsf{x}\tilde{\phi}^{\sigma,J}_y\\
&\;\;\;\;+\frac{1}{4 \pi} \sum_y\sum_{\sigma=\pm}\sigma\left(K_{c_1}\partial_t\tilde{\phi}^{\sigma}_{yc_1}\partial_\mathsf{x}\tilde{\phi}^{\sigma}_{yc_1}+N\partial_t\tilde{\phi}^{\sigma}_{yc_2}\partial_\mathsf{x}\tilde{\phi}^{\sigma}_{yc_2}\right)\\
&\;\;\;\;-\mathcal{H}_0,
\end{split}
\label{afterUlagrangian}
\end{align} 
where $K=K_{U(1)_N}\oplus K_{SU(N)}$ is the $N$-dimensional matrix in \eqref{eq:KUNlevel1}. The velocity tensor $v^{jj'}_{\sigma\sigma'}$ in \eqref{eq:LuttingerHamiltonian} is tuned so that the free Hamiltonian becomes \begin{align}\begin{split}
\mathcal{H}_0&=\frac{1}{4 \pi} \sum_y\sum_{\sigma=\pm}\sum_{I,J=0}^{N-1} \tilde{V}_{IJ}\partial_\mathsf{x}\tilde{\phi}^{\sigma,I}_y\partial_\mathsf{x}\tilde{\phi}^{\sigma,J}_y
\\&\;\;\;\;+\sum_y\sum_{\sigma=\pm}\sum_{l,l'=1,2}\left(V_C\right)_{\sigma\sigma'}^{ll'}\partial_\mathsf{x}\tilde{\Phi}^{\sigma}_{y,c_l}\partial_\mathsf{x}\tilde{\Phi}^{\sigma'}_{y,c_{l'}},
\end{split}\end{align} where $\tilde{V}_{IJ}$ is identical to the $U(N)$ symmetric velocity tensor in \eqref{eq:partionLagrangiancartanweyl}, and $\left(V_C\right)_{\sigma\sigma'}^{ll'}$ is chosen so that an intra-bundle gapping interactions defined below in \eqref{cgapping} are relevant in the renormalization group sense.

We here present one particular integral $U$ matrix that transforms \eqref{eq:LuttingerLagrangian} into \eqref{afterUlagrangian}. For odd $N$, 
\begin{align}
\begin{split}
U^{++} &= 
\left(\begin{smallmatrix}
\frac{1+N}{2}{\bf e}_{N+2}\\
\frac{1+N}{2}{\bf e}_1\\
\sum_{a=1}^N (2a-N){\bf e}_{a+1}\\
2{\bf e}_{2}\\
\vdots\\
2{\bf e}_{N}\\
\end{smallmatrix}\right),
\\
U^{+-} &=
\left(\begin{smallmatrix}
\frac{1-N}{2}{\bf e}_{N+2}\\
\frac{1-N}{2}{\bf e}_1\\
\sum_{a=1}^{N} (N-2a+1){\bf e}_{a+1}\\
-{\bf e}_{2}-{\bf e}_{3}\\
\vdots\\
-{\bf e}_{N}-{\bf e}_{N+1}
\end{smallmatrix}\right),
\end{split}\label{eq:genSUNblocks}
\end{align}
$U^{--} = U^{++} \Lambda$, and $U^{-+} = U^{+-} \Lambda$, 
where $\Lambda$ is the $(N + 2) \times (N + 2)$ anti-diagonal matrix \[\Lambda=\begin{pmatrix}&&&1\\&&{\reflectbox{$\ddots$}}&\\&1&&\\1&&&\end{pmatrix}.\]
${\bf e}_a$ is the unit (row) vector in $\mathbb{R}^{N+2}$ whose $a^{\mathrm{th}}$ entry is 1 (where $a=j-1$, for $j=0,1,\ldots,N+1$). For even $N$, 
\begin{align}
\begin{split}
U^{++} &= 
\left(\begin{smallmatrix}
\frac{N}{2}{\bf e}_{1}+{\bf e}_{N+2}\\
-{\bf e}_{1}+{\bf e}_{N+2}\\
\sum_{a=1}^N (2a-N){\bf e}_{a+1}\\
2{\bf e}_{2}\\
\vdots\\
2{\bf e}_{N}
\end{smallmatrix}\right),
\\
U^{+-} &=
\left(\begin{smallmatrix}
\frac{2-N}{2}{\bf e}_{1}\\
{\bf e}_1\\
\sum_{a=1}^{N} (N-2a+1){\bf e}_{a+1}\\
-{\bf e}_{2}-{\bf e}_{3}\\
\vdots\\
-{\bf e}_{N}-{\bf e}_{N+1}
\end{smallmatrix}\right), \\
U^{--} &= \left(\begin{smallmatrix}
\frac{2+N}{2}{\bf e}_{1}\\
{\bf e}_{N+2}\\
\sum_{a=1}^N (N-2a+2){\bf e}_{a+1}\\
2{\bf e}_{N+1}\\
\vdots\\
2{\bf e}_{3}
\end{smallmatrix}\right), \\
U^{-+} &= 
\left(\begin{smallmatrix}
-\frac{N}{2}{\bf e}_{1}+{\bf e}_{N+2}\\
{\bf 0}\\
\sum_{a=1}^{N} (2a-N-1){\bf e}_{a+1}\\
-{\bf e}_{N}-{\bf e}_{N+1}\\
\vdots\\
-{\bf e}_{2}-{\bf e}_{3}
\end{smallmatrix}\right).
\end{split}\label{eq:genSUNevenblockSet1}
\end{align}

The ``integrated" modes $\tilde \Phi^\sigma_{c_1}$ and $\tilde \Phi^\sigma_{c_2}$ within each bundle are gapped using the following charge and momentum preserving sine-Gordon interactions (see Appendix \ref{sec:AseriesMomentumConserv} for momentum conservation).
\begin{align}
\mathcal{H}^{(c_1,c_2)}_{\rm intra} &=\sum_y\left(u_I\cos\Theta_{y}^{(I)}+u_{II}\cos\Theta_{y}^{(II)}\right),
\label{cgapping}
\end{align}
where
\begin{align}
\Theta_{y}^{(I)} = \Tilde{\Phi}^R_{c_1}-\Tilde{\Phi}^L_{c_1},\quad \Theta_{y}^{(II)} = \Tilde{\Phi}^R_{c_2}-\Tilde{\Phi}^L_{c_2}.
\label{eq:c12sineGordonsSUNeven}
\end{align}
The remaining boson modes $\tilde\phi^{\sigma,I}_y$, for $I = 0, \ldots, N-1$, generate the left and right propagating $U(N)_1/\mathbb{Z}_N=U(1)_N \times SU(N)_1$ parton CFTs on each bundle.

The Dirac partons $d^r$, $r=1,\ldots,N$, and Laughlin quasiparticle $\lambda$ on each bundle $y$ and each chiral sector $\sigma=R,L$ can be expressed as the vertex operators \begin{align}d^{\sigma,r}_y=e^{i\phi^{\sigma,r}_y},\quad\lambda^\sigma_y=e^{i\sum_{r=1}^N\phi^{\sigma,r}_y/N},\end{align} where $\phi^{\sigma,r}_y=R^r_I\tilde\phi^{\sigma,I}_y$ (see \eqref{cartanweylchevrelation} and \eqref{eq:Rmatrix}). These operators carry fractional electric charge $1/N$. They are non-local operators that can only appears in conjugate pairs of mutiplets. On any given bundle, the following chiral field combinations (with a fixed $\sigma$) are integral combinations of electrons. 
\begin{align}
\begin{split}
(\lambda^\sigma_y)^N&\sim d^{\sigma,1}_y\ldots d^{\sigma,N}_y\sim e^{i\sum_{r=1}^N\left(\phi^{\sigma,r}_y(\mathsf{x})+k^{\sigma,r}_y\mathsf{x}\right)} \\&= e^{i\left(\Tilde{\Phi}^\sigma_{y\rho}(\mathsf{x})+\Tilde{k}^\sigma_{y\rho}\mathsf{x}\right)},\\
d^{\sigma,r}_y{d^{\sigma,r+1}_y}^\dagger&\sim
e^{i\left(\phi^{\sigma,r}_y(\mathsf{x})-\phi^{\sigma,r+1}_y(\mathsf{x})\right)+i\left(k^{\sigma,r}_y-k^{\sigma,r+1}_y\right)\mathsf{x}} \\&= e^{i\left(\Tilde{\Phi}^\sigma_{y,I=r}(\mathsf{x})+\Tilde{k}^\sigma_{y,r}\mathsf{x}\right)}.
\end{split}
\label{eq:rootsasintegralofelectrons}
\end{align}
This is because the Chevalley bosons $\Tilde{\Phi}$ are integral combinations of the electron bosonized variables $\Phi$ from the $U$ transformation in \eqref{bigphibigtildephi}. The locality of the second identity in \eqref{eq:rootsasintegralofelectrons} implies all parton conjugate pairs $d^r{d^{r'}}^\dagger$ on any given wire and chiral sector are integral. In addition, the non-chiral conjugate pairs of Laughlin quasiparticles $\lambda^R_y{\lambda^L_y}^\dagger$ and Dirac partons $d^{R,r}_y{d^{L,r'}_y}^\dagger$, for any $r,r'=1,\ldots,N$, are effectively local at energies lower than that of the intra-bundle potential. Each of these conjugate pairs operators is a vertex operator $e^{i\sum_{\sigma,r}a_{\sigma r}\phi^{\sigma r}_y}$, whose exponent $\sum_{\sigma,r}a_{\sigma r}\phi^{\sigma r}_y$ equals an integral linear combination of the electron bosonized variable $\Phi^\sigma_{yj}$ plus a (fractional) combination $b_I\Theta^{(I)}_y+b_{II}\Theta^{(II)}_y$ of the sine-Gordon angle variables in \eqref{eq:c12sineGordonsSUNeven}. Under the intra-bundle potential \eqref{cgapping}, the sine-Gordon angle variables are pinned to their ground state expectation values $\langle\Theta^{(I)}_y\rangle$ and $\langle\Theta^{(II)}_y\rangle$ at energies $\ll u_I,u_{II}$. Therefore, the non-chiral conjugate pairs $\lambda^R_y{\lambda^L_y}^\dagger$ and $d^{R,r}_y{d^{L,r'}_y}^\dagger$ are effectively integral electronic (up to the phase $e^{i(b_I\langle\Theta^{(I)}_y\rangle+b_{II}\langle\Theta^{(II)}_y\rangle)}$). 

By further backscattering the left and right moving $U(1)_N\times SU(N)_1$ currents in opposite directions between neighboring wires, we obtain the parton FQH state with a fully gapped bulk. The charge and momentum preserving inter-bundle current backscattering interactions are
\begin{align}
\begin{split}
&\mathcal{H}[{\rm U}(1)_N\times {\rm SU}(N)_1]\\
&=\mathcal{H}_0
+\mathcal{H}^{(c_1,c_2)}_{\rm intra}+
\mathcal{H}^{{\rm U}(1)_N}_{\rm inter}+\mathcal{H}^{{\rm SU}(N)_1}_{\rm inter}
\end{split}
\label{eq:CaseIIHamiltonian}
\end{align}
where 
\begin{widetext}
\begin{align}
\begin{split}
    \mathcal{H}^{{\rm U}(1)_{N}}_{\rm inter}&= 
u_{\rm inter}\sum_y\left[\mathtt{H}_{{\rm U}(1)_{N}}\right]^{R\dagger}_y\left[\mathtt{H}_{{\rm U}(1)_{N}}\right]^{L}_{y+1}
+
u_{\rm inter}\sum_y \left(
\left[\mathtt{E}_{{\rm U}(1)_{N}}\right]^{R\dagger}_{y,\boldsymbol{\alpha}_{{\rm U}(1)_N}}\left[\mathtt{E}_{{\rm U}(1)_{N}}\right]^{L}_{y+1,\boldsymbol{\alpha}_{{\rm U}(1)_N}}
 + h.c. \right)
\\
&=
u_{\rm inter}\sum_y 
\left(\frac{1}{mn}\sum_{r,s=1}^N\partial_{\mathsf{x}}\phi^{R}_{y,r}\partial_{\mathsf{x}}\phi^L_{y+1,s}
-2\cos\left(\boldsymbol{\alpha}_{{\rm U}(1)_N}\cdot\boldsymbol{\theta}_{y+1/2}\right)\right),
\end{split}
\label{eq:U1Ninter}
\end{align}
and
\begin{align}
\begin{split}
\mathcal{H}^{{\rm SU}(N)_1}_{\rm inter}&= 
u_{\rm inter}\sum_y \left(\sum_{p=1}^{N-1}\left[\mathtt{H}_{{\rm SU}(N)_1}\right]^{R\dagger}_{y,p}\left[\mathtt{H}_{{\rm SU}(N)_1}\right]^{L}_{y+1,p}
+
\sum_{\boldsymbol{\alpha}\in \Delta_{SU(N)}}
\left[\mathtt{E}_{{\rm SU}(N)_1}\right]^{R\dagger}_{y,\boldsymbol{\alpha}}\left[\mathtt{E}_{{\rm SU}(N)_1}\right]^{L}_{y+1,\boldsymbol{\alpha}}
\right) \\
&=u_{\rm inter}\sum_y
\left(\sum_{p=1}^{N-1}\left[\mathtt{H}_{{\rm SU}(N)_1}\right]^{R\dagger}_{y,p}\left[\mathtt{H}_{{\rm SU}(N)_1}\right]^{L}_{y+1,p}
-\sum_{\boldsymbol{\alpha}\in\Delta_{{\rm SU}(N)}}\cos(\boldsymbol{\alpha}\cdot\boldsymbol{\theta}_{y+1/2})\right)\label{eq:SUNinterlevel1}.
\end{split}
\end{align}
\end{widetext}
In the above, $\left[\mathtt{H}_{{\rm U}(1)_{N}}\right]^{L, R}_{y}$, $\left[\mathtt{H}_{{\rm SU}(N)_{1}}\right]^{L, R}_{y}$, and $\left[\mathtt{E}_{{\rm SU}(N)_{1}}\right]^{L, R}_{y}$ are $L,R$ KM currents on bundle $y$ of the form given in Eqs.~\eqref{eq:U1cartan}, \eqref{eq:SUmnCartan}, and \eqref{eq:SUmnRoots}; $\left[\mathtt{E}_{{\rm U}(1)_{N}}\right]^{L, R}_{y}$ is a vertex operator of the form in \eqref{eq:SUmnRoots} with root vector $\boldsymbol{\alpha}_{{\rm U}(1)_N} = \sum_{j=1}^N{\bf e}_j$.
The arguments of the cosine potentials are $\boldsymbol{\theta}_{y+1/2}=(\theta_{y+1/2,1},\ldots,\theta_{y+1/2,N})$ where $\theta_{y+1/2,a} \equiv \phi^R_{y,a}-\phi^L_{y+1,a}$.
The arguments of the cosine potentials mutually commute and may therefore be pinned.
This pinning gaps out the bulk degrees of freedom, while the modes living on the top ($y_{\rm max}$) and bottom ($y_{\rm min}$) bundles of wires remain gapless with chiral edge-state Lagrangian of the form \eqref{eq:partionLagrangiancartanweyl}.

We notice in passing that the electrically neutral $SU(N)_1$ currents carry vanishing $\mathsf{x}$-momentum and therefore can also be backscattered within the same bundle by an intra-bundle interaction $\mathcal{H}^{SU(N)_1}_{\mathrm{intra}}\sim[{\bf J}_{SU(N)_1}]^L_y\cdot[{\bf J}_{SU(N)_1}]^R_y$. In this case, the resulting FQH state $\mathcal{H}[U(1)_N]=\mathcal{H}^{U(1)_N}_{\mathrm{inter}}+\mathcal{H}^{SU(N)_1}_{\mathrm{intra}}$ will only carry chiral edge mode for $U(1)_N$, and not for $SU(N)_1$. It will belong in the same FQH phase as the Laughlin $\nu=1/N$ state, where the partons are confined. On the other hand, our current model \eqref{eq:CaseIIHamiltonian} describes a distinct FQH state where the $1/N$-charged partons can emerge as deconfined gapped excitations in the bulk. The $\mathcal{H}^{SU(N)_1}_{\mathrm{intra}}$ and $\mathcal{H}^{SU(N)_1}_{\mathrm{inter}}$ interactions in general can be present simultaneously and compete. The complete model $\mathcal{H}^{U(1)_N}_{\mathrm{inter}}+\mathcal{H}^{SU(N)_1}_{\mathrm{intra}}+\mathcal{H}^{SU(N)_1}_{\mathrm{inter}}$ may provide a platform to study the critical transition of parton decomfinement.

\subsection{Conformal Embedding and Descendant Topological Order}

In this subsection, we will use the conformal embedding (also referred to as level-rank duality) $SU(m)_n\times SU(n)_m\subseteq SU(mn)_1$ to demonstrate how FQH states with ${\rm U}(1)_{mn}\times {\rm SU}(m)_n$ or ${\rm U}(1)_{mn}\times {\rm SU}(n)_m$ non-Abelian topological order can arise from partially filling the parton FQH state $U(1)_{mn}\times SU(mn)_1$.

\subsubsection{Symmetry embedding}\label{sec:LRDAseries}

We first consider the embeddings of matrix tensor products ${\rm GL}(m,\mathbb{C})\times {\rm GL}(n,\mathbb{C})\subset {\rm GL}(mn,\mathbb{C})$. Let $\mathbb{1}_p$ be the $p\times p$ identity matrix.
Then, for any $m\times m$ matrix $A_{ab}$ in ${\rm GL}(m,\mathbb{C})$ and $n\times n$ matrix $B_{cd}$ in ${\rm GL}(n,\mathbb{C})$, they can be embedded into matrices of the tensor space $\mathbb{C}^{mn}=\mathbb{C}^m\otimes\mathbb{C}^n$ by the mappings $A\to(A\otimes\mathbb{1}_n)$ and $B\to(\mathbb{1}_m\otimes B)$. Moreover, the two embeddings commute because $A\otimes B = (A\otimes \mathbb{1}_n)(\mathbb{1}_m\otimes B)= (\mathbb{1}_m\otimes B)(A\otimes \mathbb{1}_n)$.
We adopt the following notation for tensor products: $(A\otimes B)_{rs}= A_{ab}B_{cd}\delta^{(a-1)n+c}_r\delta^{(b-1)n+d}_s$, 
where $r,s=1,\ldots, mn$. 
The counting of rows in $(A\otimes B)$, indexed by $(a,c)$, is determined by $r=(a-1)n+c$; the counting of columns, indexed by $(b,d)$, is determined by $s=(b-1)n+d$. 
Thus, the matrix embeddings are \begin{align}\begin{split} 
\big(A\otimes \mathbb{1}_n
\big)_{rs}
&= 
A_{ab}\delta_{cd}\delta^{(a-1)n+c}_r\delta^{(b-1)n+d}_s,\\
\left(\mathbb{1}_m\otimes B\right)_{rs} &=
\delta_{ab}B_{cd}\delta^{(a-1)n+c}_r\delta^{(b-1)n+d}_s.\label{tensorembedding}\end{split}\end{align}

Let us apply this to ${\rm SU}(m)_n \times {\rm SU}(n)_m \subset {\rm SU}(mn)_1$. The ${\rm SU}(m)$ Lie algebra is spanned by traceless Hermitian $m$-dimensional matrices $X$, and they can be embedded inside the ${\rm SU}(mn)$ Lie algebra, $X\to X^A=(X\otimes\mathbb{1}_n)$, using the first equation in \eqref{tensorembedding}. In particular, the Cartan generators $X^p$ and root matrices $X^{ab}$ defined in \eqref{eq:SUmnCartan} and \eqref{eq:SUmnRoots} now lives inside ${\rm SU}(mn)$ using 
\begin{align}
\begin{split}
(X^A)^p_{rs}
&=\frac{1}{\sqrt{p(p+1)}}
\sum_{c=1}^n\Big(\sum_{q=1}^p\delta^{(q-1)n+c}_r
\delta^{(q-1)n+c}_s\\
&\;\;\;\;-p\delta^{pn+c}_r\delta^{pn+c}_s\Big).
\end{split}
\label{eq:SUmnCartanMetric}\\
\begin{split}
(X^A)^{pq}_{rs}&=
(X^{pq})_{ab}\delta_{cd}\delta^{(a-1)n+c}_r
\delta^{(b-1)n+c}_s\\
&=
\sum_{c=1}^n\delta^{(p-1)n+c}_r\delta^{(q-1)n+c}_s.
\end{split}
\end{align}

We can apply this matrix embedding technology to the parton edge-state theory described in Sec.~\ref{partonsection} to define the ${\rm SU}(m)_n \times {\rm SU}(n)_m$ currents to be parton bilinears.
We begin with ${\rm SU}(m)_n$.
In terms of the partons $d^a$, the Cartan operators are $\sum_{r,s}$
$(d^r)^\dagger(X^A)^p_{rs}d^s$; the ladder operators are $\sum_{r,s}$$d^r(X^A)^{pq}_{rs}(d^s)^\dagger$ and $\sum_{r,s}(d^r)^\dagger(X^A)^{pq}_{rs}d^s$. In terms of the Cartan-Weyl bosons $\phi_a$, these Cartan and root operators are:

\begin{align}
\begin{split}
[\mathtt{H}_{{\rm SU}(m)_n}]_p
&=\sum_{c=1}^n
\frac{1}{\sqrt{p(p+1)}}
\Bigg(
\sum_{a=1}^p\partial_\mathsf{x}{\phi}_{(a-1)n+c}\\
&\;\;\;\;-p\partial_\mathsf{x}{\phi}_{pn+c}\Bigg),\\
[\mathtt{E}_{{\rm SU}(m)_n}]_{\boldsymbol{\alpha}}
&=\sum_{c=1}^n e^{i({\alpha}^c_{ab})^j(\phi_j(\mathsf{x})+k_j\mathsf{x})}.
\label{eq:SUmcurrentCartan}
\end{split}
\end{align}
Here, $p=1,\ldots, m-1$, where $m-1$ is the rank of ${\rm SU}(m)$, and 
\begin{align}
\boldsymbol{\alpha}^c_{ab}=\pm\left(
{\bf e}_{(a-1)n+c}-{\bf e}_{(b-1)n+c}\right),
\label{eq:rootvectorsSUmn}
\end{align}
for $1\leq a<b\leq m$ and $c = 1, \ldots, n$, with $n$ the level of ${\rm SU}(m)$, and
${\bf e}_\ell$ are the unit vectors in $\mathbb{R}^{mn}$.

The same construction applies to ${\rm SU}(n)_m$.
The matrix representations of Cartan generators and root operators are 
$(X^B)^p_{rs}$=$\delta_{ab}(X^p)_{cd}\delta^{(a-1)n+c}_r\delta^{(b-1)n+d}_s$
and $(X^B)^{pq}_{rs}$=$\delta_{ab}(X^{pq})_{cd}\delta^{(a-1)n+c}_r\delta^{(b-1)n+d}_s$, respectively, which can be used to construct the corresponding currents in terms of partons, as above.
Written in terms of the Cartan-Weyl bosons, these operators are:
\begin{align}
\begin{split}
[\mathtt{H}_{{\rm SU}(n)_m}]_q&
=\sum_{a=1}^m
\frac{1}{\sqrt{q(q+1)}}
\Bigg(
\sum_{b=1}^q\partial_\mathsf{x}{\phi}_{(a-1)n+b}\\
&\;\;\;\;-q\partial_\mathsf{x}{\phi}_{(a-1)n+q+1}\Bigg),\\
[\mathtt{E}_{{\rm SU}(n)_m}]_{\boldsymbol{\alpha}}
&=\sum_{a=1}^m e^{i({\alpha}^a_{cd})^j(\phi_j(\mathsf{x})+k_j\mathsf{x})}.
\end{split}
\label{eq:SUncurrentCartan}
\end{align}
Here, $q=1,\ldots, n-1$, where $n-1$ is the rank of $SU(n)$, and 
\begin{align}
\boldsymbol{\alpha}^a_{cd}=\pm\left(
{\bf e}_{(a-1)n+c}-{\bf e}_{(a-1)n+d}\right),
\label{eq:rootvectorSUnm}
\end{align}
for $1\leq c< d\leq n$ and $a = 1, \ldots, m$.

\subsubsection{Descendant states}
\label{SUdescendants}

We now apply this technology to construct fractional quantum Hall states with ${\rm U}(1)_{mn}\times {\rm SU}(m)_n$ or ${\rm U}(1)_{mn}\times {\rm SU}(n)_m$ topological orders. 
The starting point is the free Lagrangian of the electron wires in \eqref{afterUlagrangian}. By introducing $\mathcal{H}^{(c_1,c_2)}_{\rm intra}$ in \eqref{cgapping}, the integrated fermion modes $\tilde \Phi^\sigma_{y c_1}$ and $\tilde \Phi^\sigma_{y c_1}$ are gapped.
On each bundle there remains a non-chiral $U(1)_{mn} \times SU(mn)_1$ parton theory into which we embed $U(1)_{mn} \times SU(m)_n \times SU(n)_m$.
Note that the embedded symmetry currents are local since they are combinations of the integral vertex operators appearing in \eqref{eq:rootsasintegralofelectrons}.

The full model Hamiltonians for the ${\rm U}(1)_{mn}\times {\rm SU}(m)_n$ or ${\rm U}(1)_{mn}\times {\rm SU}(n)_m$ states, using current backscattering potentials, is
\begin{widetext}
\begin{align}
\begin{split}
&\mathcal{H}\left[{\rm U}(1)_N\times {\rm SU}(m)_n\right]=\mathcal{H}_0
+\mathcal{H}^{(c_1,c_2)}_{\rm intra}+\mathcal{H}^{{\rm SU}(n)_m}_{\rm intra}
+\mathcal{H}^{{\rm U}(1)_N}_{\rm inter}+\mathcal{H}^{{\rm SU}(m)_n}_{\rm inter}.
\end{split}
\label{eq:CaseIIIHamiltonian}
\end{align}
The intra- and inter-wire potentials of the neutral sectors are defined by 
\begin{align}
\begin{split}
\mathcal{H}^{{\rm SU}(n)_m}_{\rm intra}&=
u_{\rm intra}\sum_y
\left(\sum_{q=1}^{n-1}
[\mathtt{H}_{{\rm SU}(n)_m}]^{R\dagger}_{y,q}
[\mathtt{H}_{{\rm SU}(n)_m}]^L_{y,q}
+
\sum_{\boldsymbol{\alpha}}
\left[\mathtt{E}_{{\rm SU}(n)_m}\right]^{R\dagger}_{y,\boldsymbol{\alpha}}\left[\mathtt{E}_{{\rm SU}(n)_m}\right]^{L}_{y,\boldsymbol{\alpha}}
\right),
\end{split}\label{eq:SUnmintra}
\end{align}
\begin{align}
\begin{split}
\mathcal{H}^{{\rm SU}(m)_n}_{\rm inter}&=
u_{\rm inter}\sum_y
\left(\sum_{p=1}^{m-1}
[\mathtt{H}_{{\rm SU}(m)_n}]^{R\dagger}_{y,p}
[\mathtt{H}_{{\rm SU}(m)_n}]^L_{y+1,p}
+
\sum_{\boldsymbol{\alpha}}
\left[\mathtt{E}_{{\rm SU}(m)_n}\right]^{R\dagger}_{y,\boldsymbol{\alpha}}\left[\mathtt{E}_{{\rm SU}(m)_n}\right]^{L}_{y+1,\boldsymbol{\alpha}}
\right),
\end{split}\label{eq:SUmninter}
\end{align}
\end{widetext}
where $\mathcal{H}^{{\rm U}(1)_N}_{\rm inter}$ is given in \eqref{eq:U1Ninter}.
The $U(1)_{mn}\times SU(n)_m$ state can be obtained by exchanging ${\rm SU}(m)_n $ with $ {\rm SU}(n)_m$ in the formulas above.
We conjecture that the $SU(m)_n$ and $SU(n)_m$ current backscattering potentials gap all neutral degrees of freedom in the bulk. If so, the model leaves behind an edge-state theory on each boundary with ${\rm U}(1)_{mn}\times {\rm SU}(m)_n$ KM symmetry.
The Hamiltonian above preserves momentum and charge conservation.
The latter follows from the fact that current operators $[\mathtt{E}_\mathcal{G}]^{R}_{\boldsymbol{\alpha}}$ and $[\mathtt{E}_\mathcal{G}]^{L}_{\boldsymbol{\alpha}}$ carry equal electric charge.
We refer the proof of momentum conservation to Appendix \ref{sec:AseriesMomentumConserv}.

\subsection{Topological Order Examples}

In this subsection, we will present the topological order carried by the parton FQH states as well as its non-Abelian descendants.

\subsubsection{Emergent \texorpdfstring{$\mathbb{Z}_N$}{ZN} gauge theory in the Abelian \texorpdfstring{$U(1)_N\times SU(N)_1$}{U(1)xSU(N)}}

The topological order of $U(1)_N\times SU(N)_1$ is identical to the $\mathbb{Z}_N$ orbifold of $U(N)_1$ because of the deconfined partons $d^a=e^{i\phi^a}$. Each parton field is rotated by a complex phase $e^{2\pi i/N}$ under the discrete $\mathbb{Z}_N$ gauge transformation, whereas the local electronic field operator $\Psi_{\rm el}=e^{i(\phi^1+\ldots+\phi^N)}$ is gauge neutral. 
The different partons all belong in the same anyon class $[d]=\mathrm{span}\{d^1,\ldots,d^N\}$. Because the $SU(N)_1$ current $d^a(d^b)^\dagger$ is an integral local field and belong to the vacuum anyon class. The parton anyon class $[d]$ can be regarded as the fundamental gauge charge. (It can be made bosonic by combining the fermionic parton with an electron.) It obeys the fusion rule $[d]^N=1$ because $\Psi_{\rm el}=d^1\ldots d^N$ belongs to the trivial vacuum class 1. 

The Laughlin quasiparticle $\lambda=e^{i(\phi^1+\ldots+\phi^N)/N}$, which is the primitive non-local field in $U(1)_N$, carries a gauge flux component. It has non-trivial mutual braiding statistics with the gauge charge $[d]$ \begin{align}
\langle\lambda(z)d^a(w)\rangle=
\frac{1}{(z-w)^{1/N}}
\label{eq:gaugechargeandfluxmutualstat}
\end{align}
with the monodromy phase $e^{2\pi i/N}$. It obeys the order $N$ fusion rule $\lambda^N=1$ and carries spin $h_\lambda=\frac{1}{2N}$. Combinations of the Laughlin quasiparticle $\lambda$ and the deconfined parton $[d]$ generate all anyon classes in the $U(1)_N\times SU(N)_1$ topological order. For example, the primitive primary field sector in $SU(N)_1$ consists of the vertex fields $e^{i\phi^a-i\sum_{b=1}^N\phi^b/N}$, for $a=1,\ldots,N$. This sector is the fusion of $\mathcal{E}^1=[d]\times\lambda^{-1}$.

\subsubsection{Emergent Ising, Fibonacci and metaplectic anyons in \texorpdfstring{$SU(2)_n\times SU(n)_2\subseteq SU(2n)_1$}{SU(2)nxSU(n)2}}\label{ss:sumnchain}
Here we consider the example of topological orders that arise from the embedding of $SU(2)_n\times SU(n)_2$ in $SU(2n)_1$.
This is motivated by:
(i) the $\mathbb{Z}_N$ parafermion decomposition of $SU(2)_n$ current operators, where the parafermions CFT is the coset $SU(2)_n/U(1)_{2n}$;
(ii) and the emergence of Fibonacci and metaplectic topological orders.

The root operators of ${\rm SU}(2)_n$ in \eqref{eq:SUmcurrentCartan} are the decomposable raising operators
\begin{align}
\begin{split}
\mathtt{E}^{+}_{{\rm SU}(2)_n}
=\sum_{c=1}^n e^{i(\phi_c-\phi_{n+c})}
=\xi\times \Psi,
\label{eq:SU2LevelnCurrents}
\end{split}
\end{align}
and the lowering operator $\mathtt{E}^- = (\mathtt{E}^+)^\dagger$.
The second identity above suggests the parafermion decomposition \cite{GEPNER198710,Ginsparg88}, with $\xi$ a primary field in a ${\rm U}(1)_{2n}$ sector, and $\Psi$ the $\mathbb{Z}_n$ parafermion primary field in the $SU(2)_n/U(1)_{2n}$ coset.
For any given level $n$, we have 
\begin{align}
\begin{split}
\xi &=\sqrt{n} e^{i\theta_\perp/n}, 
\quad 
\Psi = \frac{1}{\sqrt{n}}
\sum_{ c=1}^n
e^{i(\theta_{c}-\theta_\perp/n)},
\end{split}
\end{align} 
where $\theta_\perp = \theta_1+\dots+\theta_{n}$, and $\theta_c=\phi_c-\phi_{n+c}$. 
To obtain a $\mathbb{Z}_n$ primary field
\begin{align}
\begin{split}
\Psi_k = \frac{1}{\sqrt{C^n_k}} \sum_{1\leq {j_1} <\ldots < {j_k}\leq n}
e^{i(\theta_{j_1} +\ldots + \theta_{j_k}-k\theta_\perp/n)},
\end{split}
\end{align}
we fuse $\Psi$ $k$-time with itself, where $k=1,\ldots,n-1$.
$\sqrt{C^n_k}$ is the normalized constant scaled by
$C^n_k = n!/[k!(n-k)!]$. 
Readers can refer to Appendix \ref{Appendix:ParafermionDecomp} for explicit vertex operator representations of parafermions for cases $n=2,3$, and $4$.

Interactions of the ${\rm SU}(2)_n$ model can be now expressed by 
backscattering opposite chiral $\mathbb{Z}_n$ parafermion fields and coupling to a sine-Gordon potential. 
The operator $\mathcal{O}^{{\rm U}(1)_{2n}}_{y+\epsilon}$ is introduced through opposite chiral backscattering of $\xi$ from the ${\rm U}(1)$ sector.
We adopt the notation that when $\epsilon=0$, it refers to the intrawire backscattering; and when $\epsilon=1$ it refers to the interwire backscattering.
Hence, 
\begin{align}
\begin{split}
\mathcal{U}^{{\rm SU}(2)_n}_{\epsilon}
&=
u_{\epsilon}\sum_y \Bigg(
[\mathtt{H}_{{\rm SU}(2)_n}]^{R\dagger}_{y}
[\mathtt{H}_{{\rm SU}(2)_n}]^{L}_{y+\epsilon}\\
&\;\;\;\;+\left(
\left(J^+\right)^R_{y}\left(J^-\right)^L_{y+\epsilon}+h.c.\right)\Bigg)\\
&=
u_{\epsilon}
\sum_y
\Bigg([\mathtt{H}_{{\rm SU}(2)_n}]^{R\dagger}_{y}
[\mathtt{H}_{{\rm SU}(2)_n}]^{L}_{y+\epsilon}\\
&\;\;\;\;+
\sum_{s=+,-}
\left[\mathtt{E}^s_{{\rm SU}(2)_n}\right]^{R\dagger}_y
\left[\mathtt{E}^s_{{\rm SU}(2)_n}\right]^{L}_{y+\epsilon/2}\Bigg)\\
&\rightarrow 
u_{\epsilon}\sum_y
\Bigg([\mathtt{H}_{{\rm SU}(2)_n}]^{R\dagger}_{y}
[\mathtt{H}_{{\rm SU}(2)_n}]^{L}_{y+\epsilon}\\
&\;\;\;\;+
\left\langle \mathcal{O}^{{\rm U}(1)_{2n}}_{y+\epsilon}
\right\rangle
\Psi^{R\dagger}_y\Psi^L_{y+\epsilon}\Bigg).
\end{split}
\label{eq:Znparafermion}
\end{align}
The sine-Gordon potential is pinned at a finite ground state expectation value. Subsequently, \eqref{eq:Znparafermion} opens up a mass gap for the counter-propagating $\mathbb{Z}_n$ parafermions sector~\cite{Fateev91}.

In the B sector, based on \eqref{eq:SUncurrentCartan}, the ${\rm SU}(n)_2$ root operator is
\begin{align}
\begin{split}
\left[\mathtt{E}_{{\rm SU}(n)_2}\right]^\sigma_{y,cd}
&\sim d^\sigma_{y,c}d^{\sigma\dagger}_{y,d}
+d^\sigma_{y,n+c}d^{\sigma\dagger}_{y,n+d}.
\end{split}
\end{align}
Backscattering opposite chiral root operators gives the gapping potential
\begin{align}
\begin{split}
\mathcal{U}^{{\rm SU}(n)_2}_{\epsilon} &= 
u_\epsilon
\sum_y
\Bigg([\mathtt{H}_{{\rm SU}(n)_2}]^{R\dagger}_{y}
[\mathtt{H}_{{\rm SU}(n)_2}]^{L}_{y+\epsilon}\\
&\;\;\;\;+\Big(
\sum_{1\leq c<d\leq n}
\left[\mathtt{E}_{{\rm SU}(n)_2}\right]^{R\dagger}_{y,cd}
\left[\mathtt{E}_{{\rm SU}(n)_2}\right]^{L}_{y+\epsilon,cd}\\
&\;\;\;\;+h.c.\Big)\Bigg).
\end{split}
\label{eq:SUnL2Potential}
\end{align}
For the rest of the discussion on topological orders, we assume that \eqref{eq:SUnL2Potential} opens up a finite bulk excitation energy gap.

The topological order in the bulk corresponds to the conformal field theory on the edge.
First, the ${\rm SU}(N)_1$ topological phase for general $N$ can be constructed from \eqref{eq:CaseIIHamiltonian} using the neutral modes in \eqref{eq:rootsasintegralofelectrons}. The topological phase supports $N$ Abelian anyons superselection sector $\mathcal{E}^m$ each containing $C^N_m$ independent primary fields,
\begin{align}
\begin{split}
\mathcal{E}^m ={\rm span}\left\{
e^{i\left(\phi_{a_1}+\ldots+\phi_{a_m}-m\phi_\perp/N\right)}
\right\}_{1\leq a_1<\ldots<a_m\leq N},
\end{split}
\label{eq:SUNL1primaries}
\end{align}
with $\phi_\perp=\phi_1+\ldots+\phi_N$.
They carry spin $h(\mathcal{E}^m)=m(N-m)/2N$, and 
obey fusion rules $\mathcal{E}^k\times\mathcal{E}^\ell = \mathcal{E}^{k+\ell}$ and $\mathcal{E}^0=\mathcal{E}^N=1$.
Primary fields within a superselection sector are closed under the rotation of $SU(N)_1$ WZW KM algebra.
When $N=2n$, we here present the three examples of the decomposition $SU(2)_n\times SU(n)_2$ for $n=2,3,4$. They contain Ising, Fibonacci, metaplectic anyons respectively. Each one of them is a primary field in the respective WZW algebras:
\begin{itemize}
\item[(I)] $SU(2)_2=SO(3)_1=Sp(2)_2$,
\item[(II)] $SU(2)_3$ and $SU(3)_2$, 
\item[(III)] $SU(2)_4=SO(3)_2=SU(3)_1/\mathbb{Z}_2$ and $SU(4)_2=SO(6)_2=SU(6)_1/\mathbb{Z}_2$.
\end{itemize}

For case (I), when $n=2$, the parafermion $\Psi^\sigma_y$ has spin $1/2$ and is self-conjugate, i.e.~is its own anti-particle. $\Psi^\sigma_y=(\Psi^\sigma_y)^\dagger$ is a Majorana fermion. 
The ${\rm SU}(2)_2$ and ${\rm SO}(3)_1$ WZW algebras are identical. The $SU(2)_2$ current interaction \eqref{eq:Znparafermion} is the two-fermion backscattering potential $i(\xi^R)^\dagger\xi^L\Psi^R\Psi^L$, which under a mean-field approximation becomes the single-fermion potential $m_\xi(\xi^R)^\dagger\xi^L+im_\Psi\Psi^R\Psi^L$, where $m_\xi=\langle i\Psi^R\Psi^L\rangle$ and $m_\Psi=\langle(\xi^R)^\dagger\xi^L\rangle$. The resulting topological phase under \eqref{eq:CaseIIIHamiltonian} and \eqref{eq:Znparafermion} supports Ising topological order. The edge chiral $SU(2)_2$ CFT
carries primary fields superselection sectors $1, [f], [\sigma]$ corresponding to the vacuum, the emergent fermion, and the Ising sectors. 
Using the parafermion decomposition, 
the superselection sectors of fermion and Ising twist field in ${\rm SU}(2)_2$ (the A-sector, i.e.~the first $SU(2)$ in the $SU(2)_2\times SU(2)_2\subseteq SU(4)_1$ embedding) are 
\begin{align}
\begin{split}
[f]&=\mathrm{span}\bigg\{\xi,\xi^\dagger,\Psi\bigg\},\quad\xi=e^{i(\phi_1+\phi_2-\phi_3-\phi_4)/2},\\
[\sigma] &=\mathrm{span}\bigg\{
\sigma e^{\pm i(\phi_1+\phi_2-\phi_3-\phi_4)/4}
\bigg\}.
\end{split}
\label{eq:superselectionSU2L2A}
\end{align}
They form the $j=1$ and $j=1/2$ irreducible representations of $SU(2)_2$. $\Psi$ is the emergent Majorana/$\mathbb{Z}_2$ parafermion (see also \eqref{eq:parafdecompSU2L2}). $\sigma$ in the above equation is the Ising twist field of $\Psi$ carrying spin $h=1/16$. It has a $\pi$-monodromy with the Majorana fermion $\Psi$. The vertex fields $e^{\pm i(\phi_1+\phi_2-\phi_3-\phi_4)/4}$ are twist fields of the Dirac fermion $\xi$. Each one of them carries spin $h=1/8$, and has a $\pi$-monodromy with $\xi$. The two twist fields $\sigma$ and $e^{\pm i(\phi_1+\phi_2-\phi_3-\phi_4)/4}$ are bound to each other and cannot be individually present without the other. This is because each one has a $\pi$-braiding phase with the local boson $\xi\Psi$, which is the raising current operator of $SU(2)_2$. Combining them, the total spin of the primary fields in the superselection sector $[\sigma]$ is $h_{[\sigma]}=3/16$. The primary field presentation for the B-sector -- the second $SU(2)$ in the $SU(2)_2\times SU(2)_2\subseteq SU(4)_1$ embedding -- follows a similar structure. The primary fields satisfy fusion rules
\begin{align}
f\times f=1, \quad f\times \sigma=\sigma, \quad \sigma\times \sigma = 1+f.
\end{align}
The spins, quantum dimensions, and dimensions of the superselection sectors are summarized in Table \ref{tab:CFTdataSU2at2}. 
\begin{table}[htbp]
\setlength{\tabcolsep}{34.4pt} 
\renewcommand{\arraystretch}{1.} 
\centering
\caption{The spin $h$, quantum dimension $d$, and number of fields $\#$ of each non-trivial primary sector of ${\rm SU}(2)_2$.}
\label{tab:CFTdataSU2at2}
\begin{tabular}{lccl}
  \hline 
  \hline
       & $f$   & $\sigma$\\
  \hline  
  $h$  & $\frac{1}{2}$ & $\frac{3}{16}$\\
  $d$  & $1$   & $\sqrt{2}$\\
  $\#$ & $3$   & $2$\\
  \hline 
  \hline
\end{tabular}
\end{table}

Now we move on to case (II) -- the topological phases with $SU(2)_3$ and $SU(3)_2$ edge CFTs constructed from the $SU(2)_3\times SU(3)_2\subseteq SU(6)_1$ embedding. The ${\rm SU}(2)_3$ theory from the $A$ sector can be decomposed into a ${\rm U}(1)_6$ sector and a $\mathbb{Z}_3$ parafermion CFT, which is the coset $SU(2)_3/U(1)_6$.
The primary fields in the ${\rm U}(1)$ sector can be generated by $\Xi=e^{i(\phi_1-\phi_4+\phi_2-\phi_5+\phi_3-\phi_6)/6}$.
The $\mathbb{Z}_3$ parafermion CFT is connected to the chiral sector of three-state Potts model at criticality~\cite{DOTSENKO1984312,Zamolodchikov:1990bz} with central charge $c=4/5$. Operators of the three-state Potts model can be generated by the energy density $(\tau)$ and spin $(\sigma)$ operators of the $\mathbb{Z}_3$ model \cite{DOTSENKO1984312}. 
The spin operators can be obtained from the OPEs of the energy density operator and the $\mathbb{Z}_3$ parafermionic primaries, namely,
$\sigma^+=\tau\times \Psi$ and 
$\sigma^-=\tau\times \Psi^2$. $\tau$ has conformal dimension of $2/5$, and $\sigma^{\pm}$ are of $1/15$. The $\mathbb{Z}_3$ parafermionic primaries $\Psi, \Psi^2$ has conformal dimension of $2/3$ and can be identified with the charged fields $\Sigma^+,\Sigma^-$ in the three-state Potts. Alternatively, these aforementioned primaries are the scaling fields $\Phi_{r,s}$ of the minimal model $\mathcal{M}(6,5)$ at central charge $4/5$, where
$\Phi_{2,1}=\tau$, $\Phi_{3,3}=\sigma^{\pm}$, and $\Phi_{4,3}=\Sigma^{\pm}$. 
The locality of the $SU(2)_3$ current \eqref{eq:SU2LevelnCurrents} allows the following primary fields from the tensor product of $U(1)_6\times\mathbb{Z}_3$, and they are the non-trivial primary fields of $SU(2)_3$.
\begin{align}
\begin{split}
[j=1/2] &=\mathrm{span}\left\{\Xi^\dagger \sigma^+, \Xi\sigma^-\right\},\\
[j=1]&=\mathrm{span}\left\{\tau,\Xi^2\sigma^+,(\Xi^2)^\dagger\sigma^-\right\},\\
[j=3/2]&=\mathrm{span}\left\{
\Xi^3, \Xi\Psi^2,\Xi^\dagger\Psi,(\Xi^3)^\dagger
\right\}.
\end{split}\label{SU23anyons}
\end{align}
They obey the following fusion rules \cite{francesco2012conformal}: 
\begin{gather}
[3/2]\times[3/2]=1, \quad [3/2]\times[1]=[1/2],\nonumber\\
[1/2]\times[1]=[3/2], \quad
[1]\times[1]=1+[1], \\
[1]\times[1/2]=[3/2]+[1/2], \quad [1/2]\times[1/2]=1+[1],\nonumber
\end{gather} where $1=[j=0]$ is the vacuum class.
The primary fields $[j=\frac{1}{2}]$ and $[j=1]$ have the quantum dimension of golden ratio carried by the Fibonacci anyon $\tau$, which has spin $h_\tau=2/5$, quantum dimension $d_\tau=(1+\sqrt{5})/2$ and lives inside the $[1]$ class. Fields in $[3/2]$ are Abelian semions and all have free field vertex operator representations.
The ${\rm SU}(2)_3$ primaries are summarized in Sec.~\ref{tab:CFTdataSU2at3}.
(Interested readers can refer to \cite{PhysRevB.103.235118,mongg2,PhysRevB.91.235112,LimMulliganTeo2022A} for other constructions of Fibonacci topological order.)
\begin{table}[htbp]
\setlength{\tabcolsep}{22.5pt} 
\renewcommand{\arraystretch}{1.} 
\centering
\caption{The spin $h$, quantum dimension $d$, and number of fields $\#$ of each non-trivial primary sector of ${\rm SU}(2)_3$.}
\label{tab:CFTdataSU2at3}
\begin{tabular}{lcccl}
  \hline 
  \hline
   & $[\frac{1}{2}]$ & $[1]$ & $[\frac{3}{2}]$\\[1mm]
  \hline  
  $h$  & $\frac{3}{20}$ & $\frac{2}{5}$ & $\frac{3}{4}$\\
  $d$  & $\frac{1+\sqrt{5}}{2}$ & $\frac{1+\sqrt{5}}{2}$ & $1$\\
  $\#$ & $2$ & $3$ & $4$\\
  \hline 
  \hline
\end{tabular}
\end{table}

We now present the anyon classes of $SU(3)_2$. By using the coset identification $SU(3)_2=SU(6)_1/SU(2)_3$ (which comes from the conformal embedding $SU(2)_3\times SU(3)_2\subseteq SU(6)_1$), the topological order of $SU(3)_2$ is identical to the reduced tensor product $SU(6)_1\boxtimes\overline{SU(2)_3}$. Here $\overline{SU(2)_3}$ is the time-reversal conjugate of $SU(2)_3$. Its anyons $\overline{[j]}$ have the fusion rules as those in $SU(2)_3$, but they have conjugated spins $h_{\overline{[j]}}=-h_{[j]}$ (mod 1) and braiding phases. The tensor product $\boxtimes$ is relative to the anyon condensation~\cite{PhysRevB.79.045316} of the local bosonic pair of semions $b=\mathcal{E}^3\times\overline{[j=3/2]}$ in $SU(6)_1\times\overline{SU(2)_3}$, where $\mathcal{E}^3$ is the semion class in $SU(6)_1$ (see \eqref{eq:SUNL1primaries} for the primary fields $\mathcal{E}^m$ of $SU(N)_1$). The anyon selection rule restricts the deconfined anyons of ${\rm SU}(3)_2$ to be the anyon tensor products $\mathcal{E}^m\times\overline{[j]}$ in $SU(6)_1\times\overline{SU(2)_3}$ that have trivial monodromy braiding with the local boson $b$. The non-trivial $SU(3)_2$ anyons are \begin{align}\begin{split}
[a]&=\mathcal{E}^2\times\overline{[0]}\equiv\mathcal{E}^5\times\overline{[3/2]},\\
[a^\ast]&=\mathcal{E}^4\times\overline{[0]}\equiv\mathcal{E}^1\times\overline{[3/2]},\\
[\bar{\tau}]&=\mathcal{E}^0\times\overline{[1]}\equiv\mathcal{E}^3\times\overline{[1/2]},\\
[E]&=\mathcal{E}^2\times\overline{[1]}\equiv\mathcal{E}^5\times\overline{[1/2]},\\
[E^\ast]&=\mathcal{E}^4\times\overline{[1]}\equiv\mathcal{E}^1\times\overline{[1/2]},
\end{split}\label{SU32anyons}\end{align} where anyon products are equivalent $\equiv$ modulo the local boson $b$.
The spins and fusion rules of the $SU(3)_2$ anyons can be determined by the product structure. The spins and quantum dimensions are summarized in table~\ref{tab:CFTdataSU3at2}. The fusion rules are \begin{align}
\begin{split}
[a]\times[a]&=[a^*], \quad [a^*]\times[a^*]=[a], \quad [a]\times[a^*]=1\\
[\bar\tau]\times[\bar\tau]&=1+[\bar\tau], \quad [a]\times[\bar{\tau}]=[{E}], \quad [a^*]\times[\bar{\tau}]=[E^*].
\end{split}
\end{align}

\begin{table}[htbp]
\setlength{\tabcolsep}{17.8pt} 
\renewcommand{\arraystretch}{1.} 
\centering
\caption{The spin $h$, quantum dimension $d$, and number of fields $\#$ of each non-trivial primary sector of ${\rm SU}(3)_2$.}
\label{tab:CFTdataSU3at2}
\begin{tabular}{lccc}
  \hline 
  \hline
       & $[a]$ or $[a^*]$ & $[\bar{\tau}]$ & $[E]$ or $[E^*]$\\[1mm]
  \hline  
  $h$  & $\frac{2}{3}$ & $\frac{3}{5}$ & $\frac{4}{15}$\\
  $d$  & $1$ & $\frac{1+\sqrt{5}}{2}$ & $\frac{1+\sqrt{5}}{2}$ \\
  $\#$ & $6$ & $8$ & $3$\\
  \hline 
  \hline
\end{tabular}
\end{table}

Each $SU(3)_2$ primary is a superselection sector of fields that rotate irreducibly under the $SU(3)_2$ WZW algebra. We begin with the Abelian $[a]$ and $[a^\ast]$ sectors. The primary fields in $[a]$ are linear combinations of the following six vertex fields
\begin{align}
\begin{split}
\mathcal{E}^2_{14}, \mathcal{E}^2_{25}, \mathcal{E}^2_{36}, 
\frac{\mathcal{E}^2_{15}+\mathcal{E}^2_{24}}{\sqrt{2}},
\frac{\mathcal{E}^2_{16}+\mathcal{E}^2_{34}}{\sqrt{2}},
\frac{\mathcal{E}^2_{26}+\mathcal{E}^2_{35}}{\sqrt{2}},
\end{split}\label{SU3level2a}
\end{align}
where $\mathcal{E}^2_{j_1j_2}=e^{i(\phi_{j_1}+\phi_{j_2}-2\phi_\perp/6)}$ are the vertex fields in the $\mathcal{E}^2$ class of ${\rm SU}(6)_1$ (see \eqref{eq:SUNL1primaries}). These six fields are the ones in $\mathcal{E}^2$ that are decoupled from the $SU(2)_3$ sub-algebra of $SU(6)_1$. They have trivial OPE with the $SU(2)_3$ currents and therefore solely represent the $SU(3)_2$ sub-algebra. The primary fields in $[a^\ast]$ are spanned by the Hermitian conjugate of the vertices in \eqref{SU3level2a}.

Next, we move on to the Fibonacci super-selection sector $[\bar{\tau}]$ in $SU(3)_2$. It is spanned by eight primary fields. To see this, we observe the field products in $[\tau]\times[\bar{\tau}]$, -- where $[\tau]=[j=1]$ is the Fibonacci primary field sector in $SU(2)_3$ -- are the $SU(6)_1$ WZW currents outside of those in $SU(2)_3\times SU(3)_2$. On each bundle in each chiral sector, the field product pairs in $[\tau]\times[\bar{\tau}]$ are local spin $h=2/5+3/5=1$ bosons that extend the $SU(2)_3\times SU(3)_2$ WZW algebra to the full $SU(6)_1$ WZW algebra. The difference in dimensions of $SU(6)$ and $SU(2)\times SU(3)$ is $35-(3+8)=24$. There are 3 linearly independent fields in $[\tau]=[j=1]$. Therefore, there must be $24/3=8$ linearly independent fields in $[\bar{\tau}]$ so that the product $[\tau]\times[\bar{\tau}]$ accounts for the 24 missing currents outside of $SU(2)_3\times SU(3)_2$ in $SU(6)_1$. The eight fields in $[\bar{\tau}]$ form the adjoint representation of $SU(3)$. The affine $SU(6)_1$ Lie algebra, as a vector space of the currents, decomposes into \begin{align}SU(6)_1=SU(2)_3\oplus SU(3)_2\oplus\left([\tau]\times[\bar\tau]\right).\label{SU6currentsplitting}\end{align}

We notice this field counting method holds for general $SU(m)_n\times SU(n)_m\subseteq SU(mn)_1$ embeddings. The conformal dimension of a primary field sector $[V]$ in ${\rm SU}(m)$ at level $n$ is given by \cite{Ginsparg88, francesco2012conformal}:
\begin{align}
    h_{V} = \frac{C_{V}/2}{n+g}
    \label{eq:spinCasimirSU}
\end{align}
where $C_{V}$ is the quadratic Casimir invariant of the irreducible representation $[V]$, and $g$ the dual Coxeter number of the algebra. $g=m$ for $SU(m)$. 
In particular, the adjoint representation $[A]$ of $SU(m)$ has dimension $\dim[A]=\dim(SU(m))=m^2-1$ and the quadratic Casimir invariant $C_{A}=2m$~\cite{francesco2012conformal}. It corresponds to a primary field sector $[A]$ in the $SU(m)_n$ WZW CFT if $n\geq2$. The scaling dimension according to \eqref{eq:spinCasimirSU} is $h_A=m/(m+n)$.  
Therefore, in the conformal embedding where ${\rm SU}(m)_n\times{\rm SU}(n)_m\subseteq {\rm SU}(mn)_1$, the scaling dimensions of the adjoint primary fields in the two sectors combine to 
\begin{align}
\begin{split}
h_{A}^{SU(m)_n} +h_{A}^{SU(n)_m}=\frac{m}{n+m}+\frac{n}{m+n}=1.
\end{split}
\label{eq:SpinHighestWeightAdjointRep}
\end{align}
There are $(m^2-1)(n^2-1)$ fields in the pair tensor product $[A]_{SU(m)_n}\times[A]_{SU(n)_m}$. This number equals the difference in the numbers of current operators, $\dim(SU(mn))-\dim(SU(m))-\dim(SU(n))$. The spin-1 bosons in the adjoint pair are exactly the currents in $SU(mn)_1$ outside of $SU(m)_n$ and $SU(n)_m$. This is summarized by the branching rule \begin{align}\begin{split}SU(mn)_1&=SU(m)_n\oplus SU(n)_m\\&\quad\quad\quad\oplus\left([A]_{SU(m)_n}\times[A]_{SU(n)_m}\right)\end{split}\label{eq:SU2nL1dimensionalityIdentity}\end{align} that generalizes \eqref{SU6currentsplitting}. Since all $SU(mn)_1$ currents are local bosons in the coupled-wire constructions, so are the field pair product in $[A]_{SU(m)_n}\times[A]_{SU(n)_m}$.

\begin{figure}[htbp]
\centering
\includegraphics[width=0.2\textwidth]{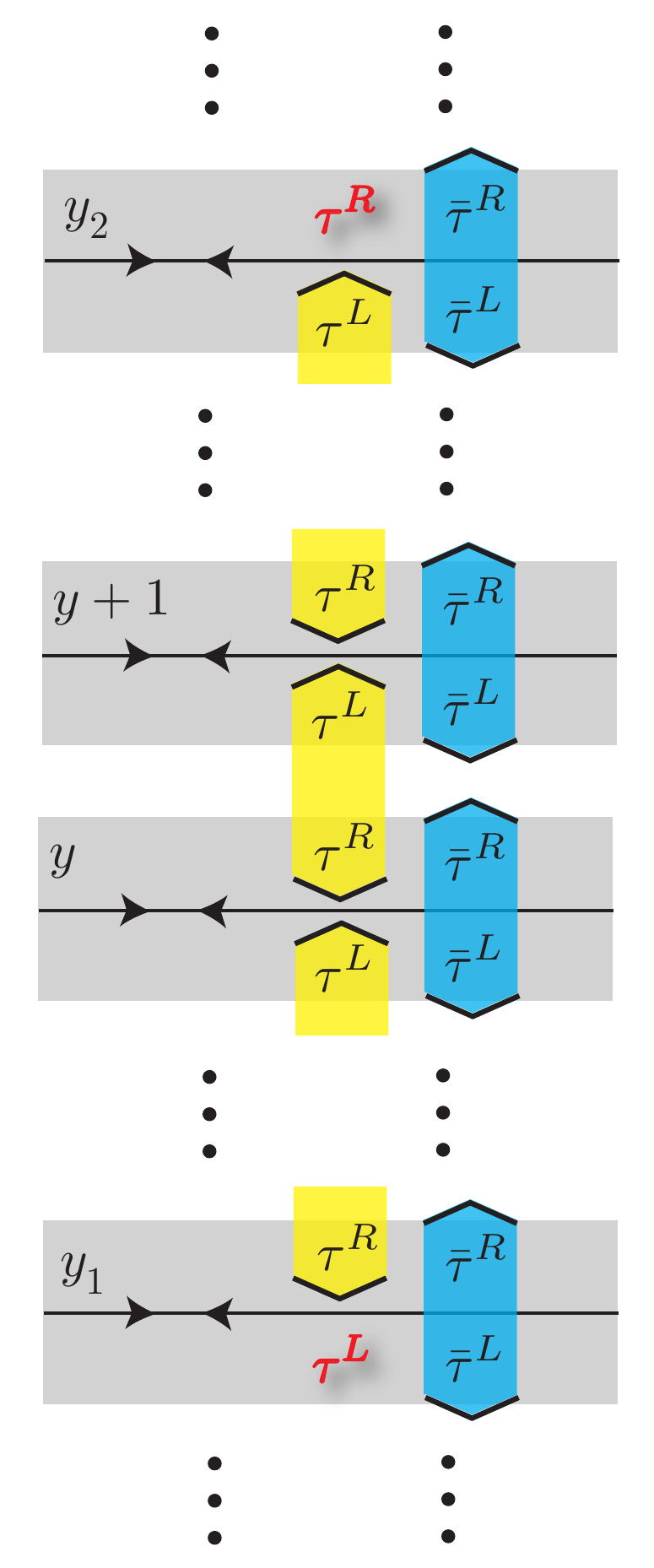}
\caption{Fibonacci excitation pair created by the string of local operators $\mathcal{S}$ in \eqref{FibonacciString}. Yellow (blue) brackets are ground state expectation values of Fibonacci pairs as a result of the inter-wire (intra-wire) potential. The unpaired Fibonacci fields (highlighted in red) $\tau^L_{y_1}$ and $\tau^R_{y_2}$ become anyon excitations at the two ends of the string.}\label{FibonacciWire}
\end{figure}

Going back to the case when $m=2$ and $n=3$, the identification of Fibonacci pairs $[\tau]\times[\bar\tau]$ in $SU(2)_3\times SU(3)_2$ as local $SU(6)_1$ currents, $J^\sigma_y\sim\tau^\sigma_y\Bar{\tau}^\sigma_y$ on each bundle $y$ and in each chiral sector $\sigma=R,L$, allows us to write down an explicit string of local boson operators that creates a Fibonacci anyon excitation at each end.
\begin{align}
\begin{split}
\mathcal{S}&=\prod_{y=y_1}^{y_2}J^L_y(\mathsf{x})J^R_y(\mathsf{x})^\dagger
\sim 
\prod_{y=y_1}^{y_2}\tau^L_y\tau^R_y
\prod_{y=y_1}^{y_2}\bar{\tau}^L_y\bar{\tau}^R_y\\
&\xrightarrow{\mbox{\tiny at low energy}}\tau^L_{y_1}\left(\prod_{y=y_1}^{y_2-1}
\langle\tau^R_y\tau^L_{y+1}\rangle
\right)\tau^R_{y_2}
\left(\prod_{y=y_1}^{y_2-1}
\langle\bar{\tau}^L_y\bar{\tau}^R_{y}\rangle
\right),
\end{split}\label{FibonacciString}
\end{align}
where $J^\sigma_y$ can be chosen to be an arbitrary current operator in $SU(6)_1$ that falls outside of $SU(2)_3\times SU(3)_2$ (see \eqref{SU6currentsplitting}), i.e.~the most singular terms in the OPE of $J^\sigma_y(z)$ and currents from $SU(2)_3$ or $SU(3)_2$ are proportional to $1/(z-w)$ and not $1/(z-w)^2$.  
Neighboring Fibonacci field pairs from opposite chiral sectors at low energy are pinned to the ground state expectation values by the $SU(2)_3$ and $SU(3)_2$ current backscattering potentials. In the $SU(2)_3$ phase where $SU(2)_3$ (or $SU(3)_2$) currents are back-scattered between (resp.~within) bundles, $\tau^L_y\tau^R_{y+1}\to\langle\tau^L_y\tau^R_{y+1}\rangle$ and $\bar\tau^L_y\bar\tau^R_y\to\langle\bar\tau^L_y\bar\tau^R_y\rangle$. This leaves behind two dangling Fibonacci modes $\tau^L_{y_2}$ and $\tau^R_{y_1}$ at the two ends of the string. (See figure~\ref{FibonacciWire}.) 

As for the remaining primary field super-selection sectors $[E]$ and $[E^\ast]$ in $SU(3)_2$, they correspond to the two fundamental three-dimensional irreducible representations of $SU(3)$. To see this, under the conformal embedding $SU(2)_3\times SU(3)_2\subseteq SU(6)_1$, the $SU(6)_1$ primary field sector $\mathcal{E}^2$ splits into two parts (c.f.~\eqref{SU32anyons}) \begin{align}\mathcal{E}^2=\left([j=0]\times[a]\right)\oplus\left([j=1]\times[E]\right).\label{SU6E2split1}\end{align}
For instance, the spin of $[j=1]\times[E]$ combines to $2/5+4/15=2/3$, which is identical to that of $\mathcal{E}^2$ and $[a]$. There are $C^6_2=15$ linearly independent primary fields in $\mathcal{E}^2$. The six of them in \eqref{SU3level2a} generate $[a]=[j=0]\times[a]$. The remaining nine generate $[j=1]\times[E]$. Since $[j=1]$ is the three-dimensional vector representation of $SU(2)_3$ (see \eqref{SU23anyons}), $[E]$ must be of dimension $9/3=3$. The same counting holds for $[E^\ast]$ because $\mathcal{E}^4$ splits into $[a^\ast]$ and $[j=1]\times[E^\ast]$. Apart from \eqref{SU6E2split1}, the other $SU(6)_1$ primary field super-selection sectors split into $SU(2)_3\times SU(3)_2$ components according to the branching rules \begin{align}\begin{split}&\mathcal{E}^1=\left[j=1/2\right]\times\left[E^\ast\right],\\&\mathcal{E}^3=\left(\left[j=3/2\right]\times1\right)\oplus\left(\left[j=1/2\right]\times\left[\bar\tau\right]\right),\end{split}\label{SU6E2split2}\end{align} and similarly for the conjugates $\mathcal{E}^4=(\mathcal{E}^2)^\dagger$ and $\mathcal{E}^5=(\mathcal{E}^1)^\dagger$.

The branching rules \eqref{SU6E2split1} and \eqref{SU6E2split2} allow us to write down strings of electron operators that create the anyon excitations on each end. These open strings are similar to those that created the Fibonacci anyons in \eqref{FibonacciString} and figure~\ref{FibonacciWire}. They are products of local operators over a range of consecutive bundles \begin{align}\mathcal{S}^m=\prod_{y=y_1}^{y_2}\mathcal{E}^{L,m}_y(\mathsf{x})\mathcal{E}^{R,m}_y(\mathsf{x})^\dagger,\label{Emstring}\end{align} where $m=1,\ldots,5$. On each wire in low-energy, the conjugate pair of $SU(6)_1$ primaries $\mathcal{E}^{L,m}_y{\mathcal{E}^{R,m}_y}^\dagger$ is effectively an integral product of electrons up to ground state expectation values $\langle\Theta^{(I)}_y\rangle$ and $\langle\Theta^{(II)}_y\rangle$ of the gapped $c_{1,2}$ modes. Any arbitrary anyon conjugate pair can be created by choosing an appropriate primary field in $\mathcal{E}^m$ and applying the operator string $\mathcal{S}^m$ on a ground state. We here demonstrate the pair creation of the Abelian anyon $[a]$ and the non-Abelian anyon $[E]$ in a $U(1)_6\times SU(3)_2$ FQH model. First, we apply the string $\mathcal{S}^2$ in \eqref{Emstring} and choose the $SU(6)_1$ primary $\mathcal{E}^2$ on each chiral sector and bundle to be one (or any normalized linear combination) of the primary fields in \eqref{SU3level2a}. Since these fields all belong in the $[a]$ primary field sector in $SU(3)_2$, the string in low-energy decomposes into \begin{align}\mathcal{S}^2&\xrightarrow{\mbox{\tiny at low energy}}a^L_{y_1}\left(\prod_{y=y_1}^{y_2-1}\left\langle {a^R_y}^\dagger a^L_{y+1}\right\rangle\right){a^R_{y_2}}^\dagger\label{E2E2string1}\end{align} where $a^\sigma_y$ is the chosen primary field $\mathcal{E}^2$ in \eqref{SU3level2a} at $(y,\sigma)$. By the inter-bundle backscattering $\mathcal{H}_{\mathrm{inter}}^{SU(3)_2}$, conjugate field pairs between neighboring bundles are pinned to their ground state expectation values $\langle{a^R_y}^\dagger a^L_{y+1}\rangle$, which is a $U(1)$ phase. This leaves behind the unpaired fields $a^L_{y_1}$ and ${a^R_{y_2}}^\dagger$ at the two ends and creates the separated conjugate pair of Abelian anyon excitations. 

Next, we again apply the string $\mathcal{S}^2$ in \eqref{Emstring} but instead choose a new $SU(6)_1$ primary $\mathcal{E}^2$ that is orthogonal to the $[a]$ sector (i.e.~the most singular term in its OPE with any of the fields in \eqref{SU3level2a} is at most proportional to $1/(z-w)^h$ but not $1/(z-w)^{2h}$, where $h=2/3$ is the spin of $\mathcal{E}^2$). According to the branching rule \eqref{SU6E2split1}, this field $\mathcal{E}^2$ decomposes into the tensor product $\tau\times E$, where $\tau$ belongs to the $[j=1]$ sector in $SU(2)_3$ and $E$ belongs to the $[E]$ sector in $SU(3)_2$. Therefore, in low-energy, \begin{align}\begin{split}\mathcal{S}^2&\sim\prod_{y=y_1}^{y_2}\tau^L_y\tau^R_y\prod_{y=y_1}^{y_2}E^L_y{E^R_y}^\dagger\\
&\xrightarrow{\mbox{\tiny at low energy}}\left(\prod_{y=y_1}^{y_2}\left\langle\tau^L_y\tau^R_y\right\rangle\right)\\&\quad\quad\quad\quad\times E^L_{y_1}\left(\prod_{y=y_1}^{y_2-1}\left\langle {E^R_y}^\dagger E^L_{y+1}\right\rangle\right){E^R_{y_2}}^\dagger,\end{split}\label{E2E2string2}\end{align} where the ground state expectation values are pinned by the intra/inter-bundle current backscattering potentials $\mathcal{H}_{\mathrm{intra}}^{SU(2)_3}$ and $\mathcal{H}_{\mathrm{inter}}^{SU(3)_2}$. This leaves behind the separated conjugate pair of non-Abelian excitations in anyon classes $[E]$ and $[E^\ast]$ at the two ends of the string. If the same string is applied to the $SU(2)_3$ state instead of $SU(3)_2$, it will create a pair of Fibonacci anyons $\tau$. All other anyon pairs can be created using different string operators $\mathcal{S}^m$ in \eqref{Emstring} with appropriate choices of primary fields $\mathcal{E}^m$ and branching rules \eqref{SU6E2split2}.

Lastly, we present case (III) -- the topological phases with $SU(2)_4$ and $SU(4)_2$ edge CFTs constructed from the $SU(2)_4\times SU(4)_2\subseteq SU(8)_1$ embedding. The ${\rm SU}(2)_4$ theory can be decomposed into a ${\rm U}(1)_8$ sector and a $\mathbb{Z}_4$ parafermion CFT, which is the coset $SU(2)_4/U(1)_8$. The $SU(2)_4$ current raising operator can be decomposed according to \eqref{eq:SU2LevelnCurrents} (also see \eqref{eq:SU2L4parafermiondecomposition1} and \eqref{eq:SU2L4parafermiondecomposition2}).
We now define $\Xi\equiv e^{i\sum_{j=1}^4\theta_j/8}$ where $\theta_j=\phi_j-\phi_{4+j}$ that generates ${\rm U}(1)_8$. They carry spin
$h_{\Xi^q}=q^2/16$ for $q=-3,\ldots,4$.
The $\mathbb{Z}_4$ parafermion is $\Psi$ from \eqref{eq:SU2L4parafermiondecomposition2}. For $\Psi_k$, the corresponding spin is $h_k=k(4-k)/4$ with $k=0,1,2,3$.
The primary fields of 
the $\mathbb{Z}_4$ parafermion CFT was described 
in \cite{cmp/1104178892, Teo_2023} by identifying the theory with the $U(1)_6/\mathbb{Z}_2$ orbifold CFT. They are the 
spin $1/3$ vertex operators
$\phi_3^{1,2}$, that can be identified with $\Psi_{1,3}$; 
spin $1$ operator $j$ that can be identified with $\Psi_2$; the rest are the operators $\phi_{1,2}$, twist fields 
$\sigma_{1,2}$ and $\tau_{1,2}$, which are non-Abelian and do not have free field expression.
Their conformal and quantum dimensions are included in Table~\ref{tab:Z4parafermionCFTdata}.
\begin{table}[htbp]
\setlength{\tabcolsep}{10.8pt} 
\renewcommand{\arraystretch}{1.} 
\centering
\caption{The spin $h$ and quantum dimension $d$ of $\mathbb{Z}_4$ parafermion CFT. Notation of primary fields are based on \cite{cmp/1104178892}.}
\label{tab:Z4parafermionCFTdata}
\begin{tabular}{lccccccl}
  \hline 
  \hline
     & $\phi^{1,2}_3$ & $j$ & $\phi_1$
     & $\phi_2$ & $\sigma_{1,2}$ & $\tau_{1,2}$
  \\[1mm]
  \hline  
  $h$  & $\frac{3}{4}$ & $1$ & $\frac{1}{12}$ & $\frac{1}{3}$ & $\frac{1}{16}$ & $\frac{9}{16}$\\
  $d$  & $1$ & $1$ & $2$ & $2$ & $\sqrt{3}$ & $\sqrt{3}$\\
  \hline 
  \hline
\end{tabular}
\end{table}

The ${\rm SU}(2)_4$ topological order can now be obtained from the reduced tensor product $U(1)_8\times\mathbb{Z}_4$ relative to the anyon condensation of the bosonic product $\Xi^2\times\phi_3^1$, which is the raising current operator $\mathtt{E}^+_{SU(2)_4}$ (along with its conjugate $\mathtt{E}^-_{SU(2)_4}={\mathtt{E}^+_{SU(2)_4}}^\dagger$ and its square $\Xi^4\times j$).
The primary field superselection sector of ${\rm SU}(2)_4$ is labeled by $[j]$, with 
$j=0,1/2,1,3/2,2$ with spin $h_j=j(j+1)/6$.
Under the parafermion decomposition, each of the $SU(2)_3$ super-sector is spanned by a set of primary field products in $U(1)_8\times\mathbb{Z}_4$ that rotate irreducibly under $\mathtt{E}^\pm_{SU(2)_4}$.
\begin{align}
\begin{split}
[j=1/2]&=\mathrm{span}\left\{\Xi\sigma_1,\Xi^{-1}\sigma_2\right\},\\
[j=1]&=\mathrm{span}\left\{\phi_2,\Xi^2\phi_1,\Xi^{-2}\phi_1\right\},\\
[j=3/2]&=\mathrm{span}\left\{\Xi\tau_1,\Xi^{-1}\tau_2,\Xi^3\sigma_2,\Xi^{-3}\sigma_1\right\},
\\
[j=2]&=\mathrm{span}\left\{\Xi^4, \Xi^2\phi_3^2, j, \Xi^{-2}\phi_3^1, \Xi^{-4}\right\}.
\end{split}\label{SU24=U1Z8}
\end{align}
Alternatively, modular data like quantum dimensions and fusion algebras of a WZW KM algebras are readily encoded in the modular $\mathcal{S}$-matrix. 
The modular $\mathcal{S}$-matrix can be obtained from the ribbon identity that relates the $2\pi$ braiding phase of anyon $j,j'$ over a fixed fusion channel $k$ \cite{kitaev2006anyons}.
Specifically,
\begin{align}
\mathcal{S}_{ij}=\frac{1}{\mathcal{D}}\sum_k d_k N_{ij}^k
\frac{\theta_k}{\theta_i\theta_j},
\end{align}
where $d_k$ is the quantum dimension of a superselection sector, and $\mathcal{D}=\sqrt{\sum_k d_k^2}$ is the total quantum dimension. $\theta_k = e^{2\pi i h_k}$ is the topological spin of anyone $k$, where $h_k$ is the conformal scaling dimension of its corresponding primary field.
The general formula for the modular $\mathcal{S}$-matrix of ${\rm SU}(2)_n$ is given by the formula~\cite{francesco2012conformal}
\begin{align}
\begin{split}
\mathcal{S}_{j,j'}(n)=\sqrt{\frac{2}{n+2}}
\sin{\frac{(2j+1)(2j'+1)\pi}{n+2}}.
\label{eq:generalSUnLkSmatrix}
\end{split}
\end{align}
For $n=4$, we have
\begin{align}
\begin{split}
\mathcal{S}_{j,j'} = \frac{2}{\sqrt{3}}\sin\frac{(2j+1)(2j'+1)\pi}{6} 
\end{split}
\end{align}
where $j,j',k = 0,1/2,\ldots, 2$. Using the Verlinde formula~\cite{Verlinde88}, one can obtain all fusion algebra, and quantum dimensions of anyons, by computing the fusion tensor
\begin{align}
\mathcal{N}_{jj'}^{k}=\sum_\sigma\frac{\mathcal{S}_{j\sigma}\mathcal{S}_{j'\sigma}\mathcal{S}^*_{k\sigma }}{\mathcal{S}_{0\sigma}}.
\label{eq:VerlindeFormula}
\end{align}
The topological data of ${\rm SU}(2)_4$ is summarized in Table.\ref{tab:CFTdataSU2L4}.
The $SU(2)_4$ anyons obey the fusion rules
\begin{gather}
[1/2]\times[1/2] = [0]+[1], \quad 
[1/2]\times[2]=[3/2],\nonumber\\
[2]\times[2]=[0],\nonumber\\
 [1/2]\times[1]=[1/2]+[3/2],\nonumber\\
[1]\times[1]=[0]+[1]+[2].
\end{gather}
${\rm SU}(2)_4$ supports metaplectic topologial order \cite{PhysRevB.87.165421,10.1063/1.4914941,BRUILLARD20162364}. It is identical to the $SO(3)_2$ WZW algebra and the $SU(3)_1/\mathbb{Z}_2$ orbifold CFT~\cite{Teo_2023}. It has two spin $1$ bosons, and two metaplectic anyons with quantum dimension of $\sqrt{3}$ and one with quantum dimension of $2$. The metaplectic anyon $[j=1]$, which has integral quantum dimension, is capable of computing the Kauffman polynomial that is \#P-hard \cite{PhysRevB.87.165421}.
\begin{table}[htbp]
\setlength{\tabcolsep}{15.5pt} 
\renewcommand{\arraystretch}{1.} 
\centering
\caption{The spin $h$, quantum dimension $d$, and number of fields $\#$ of each non-trivial primary sector of ${\rm SU}(2)_4$.}
\label{tab:CFTdataSU2L4}
\begin{tabular}{cccccc}
  \hline 
  \hline
   & $[\frac{1}{2}]$ & $[1]$ & $[\frac{3}{2}]$ & $[2]$ &\\[1mm]
  \hline  
  $h$  & $\frac{1}{8}$ & $\frac{1}{3}$ & $\frac{5}{8}$ & $1$\\
  $d$  & $\sqrt{3}$ & $2$ & $\sqrt{3}$ & $1$\\
  $\#$ & $2$ & $3$ & $4$ & $5$\\
  \hline 
  \hline
\end{tabular}
\end{table}

Now we move on to $SU(4)_2$. From the conformal embedding $SU(2)_4\times SU(4)_2\subseteq SU(8)_1$, the $SU(2)_4$ CFT is identical to the coset $SU(8)_1/SU(2)_4$.  Its topological order is the same as the reduced tensor product $SU(8)_1\boxtimes\overline{SU(2)_4}$ between $SU(8)_1$ and the time-reversal conjugate of $SU(2)_4$. The tensor product is relative to the anyon condensation of the boson pair $\mathcal{E}^4\times\overline{[j=2]}$ (see \eqref{eq:SUNL1primaries} for the $SU(N)_1$ primary fields $\mathcal{E}^m$). The ${\rm SU}(4)_2$ primary fields are in 1-1 correspondence with the deconfined anyons in $SU(8)_1\boxtimes\overline{SU(2)_4}$ that have trivial mutual statistics with the condensed boson $\mathcal{E}^4\times\overline{[j=2]}$. The Abelian ones are (up to the condensed boson) $[a]=\mathcal{E}^4\times\overline{[j=0]}$ that has spin $h=1$ as well as $[b]=\mathcal{E}^2\times\overline{[0]}$ and its anti-particle $[b^\ast]=\mathcal{E}^6\times\overline{[0]}$ that both have spin $h=3/4$. Primary fields in these Abelian superselection sectors have free field expressions.
The $SU(8)_1$ primary field sectors of $\mathcal{E}^2$ and $\mathcal{E}^4$ consist of vertex operators $\mathcal{E}^2_{j_1j_2}\equiv e^{i(\phi_{j_1}+\phi_{j_2}-2\phi_\perp/8)}$ for $1\leq j_1<j_2\leq 8$, and $\mathcal{E}^4_{j_1 j_2 j_3 j_4}\equiv e^{i(\sum_{k=1}^4\phi_{j_k}-4\phi_\perp/8)}$ for $1\leq j_1<j_2<j_3<j_4\leq 8$. 
The $20$ dimensional superselection sector $[a]$ is spanned by the vertex fields combinations
\begin{align}
\begin{split}
[a]&=
\left\{\begin{array}{@{} c @{}}
\mathcal{E}^4_{1256},\mathcal{E}^4_{1357},\mathcal{E}^4_{1458},\mathcal{E}^4_{2367}, \mathcal{E}^4_{2468}, \mathcal{E}^4_{3478},\\
\frac{\mathcal{E}^4_{1257}+\mathcal{E}^4_{1356}}{\sqrt{2}},
\frac{\mathcal{E}^4_{1258}+\mathcal{E}^4_{1456}}{\sqrt{2}},
\frac{\mathcal{E}^4_{1267}+\mathcal{E}^4_{2356}}{\sqrt{2}},
\frac{\mathcal{E}^4_{1268}+\mathcal{E}^4_{2456}}{\sqrt{2}},\\
\frac{\mathcal{E}^4_{1358}+\mathcal{E}^4_{1457}}{\sqrt{2}},
\frac{\mathcal{E}^4_{1367}+\mathcal{E}^4_{2357}}{\sqrt{2}},
\frac{\mathcal{E}^4_{1378}+\mathcal{E}^4_{3457}}{\sqrt{2}},
\frac{\mathcal{E}^4_{1468}+\mathcal{E}^4_{2458}}{\sqrt{2}},\\
\frac{\mathcal{E}^4_{1478}+\mathcal{E}^4_{3458}}{\sqrt{2}},
\frac{\mathcal{E}^4_{2368}+\mathcal{E}^4_{2467}}{\sqrt{2}},
\frac{\mathcal{E}^4_{2378}+\mathcal{E}^4_{3467}}{\sqrt{2}},
\frac{\mathcal{E}^4_{2478}+\mathcal{E}^4_{3468}}{\sqrt{2}},\\
\frac{\mathcal{E}^4_{1278}+\mathcal{E}^4_{3456}+
\mathcal{E}^4_{1368}+\mathcal{E}^4_{2457}}{2},
\frac{\mathcal{E}^4_{1368}+\mathcal{E}^4_{2457}+
\mathcal{E}^4_{1467}+\mathcal{E}^4_{2358}}{2}
\end{array}\right\}
\end{split}\label{eq:SU4L2supersectora}
\end{align}
in $\mathcal{E}^4$ that are decoupled from and have non-singular OPE with the $SU(2)_4$ WZW sub-algebra. 
Similarly, $[b]$ is the $10$ dimensional superselection sector spanned by the $SU(2)_4$-invariant vertex fields 
\begin{align}
\begin{split}
[b]&=
\left\{\begin{array}{@{} c @{}}
\mathcal{E}^2_{15}, \mathcal{E}^2_{26}, \mathcal{E}^2_{37}, \mathcal{E}^2_{48}, \frac{\mathcal{E}^2_{16}+\mathcal{E}^2_{25}}{\sqrt{2}},
\frac{\mathcal{E}^2_{17}+\mathcal{E}^2_{35}}{\sqrt{2}},\\
\frac{\mathcal{E}^2_{18}+\mathcal{E}^2_{45}}{\sqrt{2}},
\frac{\mathcal{E}^2_{27}+\mathcal{E}^2_{36}}{\sqrt{2}},
\frac{\mathcal{E}^2_{28}+\mathcal{E}^2_{46}}{\sqrt{2}},
\frac{\mathcal{E}^2_{38}+\mathcal{E}^2_{47}}{\sqrt{2}}
\end{array}\right\}.
\end{split}\label{eq:SU4L2supersectorb}
\end{align}
Fields within each sector can be rotated irreducibly into one another by the ${\rm SU}(4)_2$ KM currents.
The Abelian supersector $[{b}^\ast]$ is the Hermitian conjugate of $[b]$. 

For the non-Abelian primary field superselection sectors of ${\rm SU}(4)_2$, we label them by borrowing the notations from table~\ref{tab:Z4parafermionCFTdata}. 
First, we label the twist fields to be $[\bar\tau_1]$, $[\bar\tau_2]$, $[\bar\sigma_1]$, and $[\bar\sigma_2]$. They can be identified (up to the condensed boson) with the following pair product anyons in $SU(8)_1\times\overline{SU(2)_4}$. 
\begin{enumerate}
\item $[\bar{\sigma}_1]=\mathcal{E}^5\times\overline{[j=3/2]}$ and $[\bar{\sigma}_2]=\mathcal{E}^3\times\overline{[j=3/2]}$. Both carry spin $h=5/16$. 
\item $[\bar{\tau}_1]=\mathcal{E}^5\times\overline{[j=1/2]}$ and $[\bar{\tau}_2]=\mathcal{E}^5\times\overline{[j=1/2]}$. Both carry spin $h=13/16$. 
\end{enumerate}
Second, we label the remaining metaplectic~\cite{10.1063/1.4914941,BRUILLARD20162364} primary fields to be 
$[\bar{\phi}_2]=\mathcal{E}^4\times\overline{[j=1]}$, which carries spin $h=2/3$, and
$[\bar{\phi}_1]=\mathcal{E}^2\times\overline{[j=1]}$, which carries spin $h=5/12$.
The fusion rules of ${\rm SU}(4)_2$ are generated by
\begin{gather}
[a]\times [a]=1,\quad [a]\times[b]=[b^\ast],\quad [b]\times [b] = [a]\nonumber\\
[b]\times[\bar{\phi}_1]=[\bar{\phi}_2],\quad [b]\times[\bar{\phi}_2]=[\bar{\phi}_1],\nonumber\\
[b]\times[\bar\sigma_1]=[\bar\tau_2],\quad [b]\times[\bar\tau_2]=[\bar\tau_1],\nonumber\\
\quad [b]\times[\bar\tau_1]=[\bar\sigma_2],\quad [b]\times[\bar\sigma_2]=[\bar\sigma_1]\nonumber\\
[\bar{\sigma}_1]\times [\bar{\sigma}_2] = [\bar{\tau}_1]\times [\bar{\tau}_2] = 1+[\bar{\phi}_2],\nonumber\\
[\bar{\phi}_i]\times[\bar{\phi}_i] = 1+[a]+[\bar{\phi}_2]\nonumber\\ 
[\bar\phi_2]\times[\bar\sigma_j]=[\bar\phi_2]\times[\bar\tau_j]=[\bar\sigma_j]+[\bar\tau_j]
\label{eq:fusionrulesSU4L2}
\end{gather}
where $i,j=1,2$.

The corresponding modular $\mathcal{S}$-matrix is:
\begin{widetext}
\begin{align}
    \mathcal{S}_{j,j'}=\frac{1}{2\sqrt{6}}\left(\begin{smallmatrix}
   1& 1 &1 &1 &\sqrt{3} &\sqrt{3} &\sqrt{3} &\sqrt{3} &2 &2\\
   1& -1 &1 &-1 &-i\sqrt{3} &i\sqrt{3} &-i\sqrt{3} &i\sqrt{3} &2 &-2\\   
   1& 1&1 &1 &-\sqrt{3} &-\sqrt{3}  &-\sqrt{3}  &-\sqrt{3}  &2 &2\\
   1& -1 &1 &-1 &i\sqrt{3} &-i\sqrt{3} &i\sqrt{3} &-i\sqrt{3} &2 &-2 \\
   \sqrt{3}& -i\sqrt{3}&-\sqrt{3} &i\sqrt{3} &\sqrt{\frac{3}{2}}(1-i) &-\sqrt{\frac{3}{2}} (1+i) &-\sqrt{\frac{3}{2}}(1-i) &\sqrt{\frac{3}{2}}(1+i) &0 &0\\
   \sqrt{3}& i\sqrt{3}&-\sqrt{3} &-i\sqrt{3} &-\sqrt{\frac{3}{2}}(1+i) &\sqrt{\frac{3}{2}}(1-i) &\sqrt{\frac{3}{2}}(1+i) &-\sqrt{\frac{3}{2}}(1-i) &0 &0\\
   \sqrt{3}& -i\sqrt{3}&-\sqrt{3} &i\sqrt{3} &-\sqrt{\frac{3}{2}}(1-i) &\sqrt{\frac{3}{2}}(1+i) &\sqrt{\frac{3}{2}}(1-i) &-\sqrt{\frac{3}{2}}(1+i) &0 &0\\
  \sqrt{3}&  i\sqrt{3}&-\sqrt{3} &-i\sqrt{3} &\sqrt{\frac{3}{2}}(1+i) &-\sqrt{\frac{3}{2}}(1-i) &-\sqrt{\frac{3}{2}}(1+i) &\sqrt{\frac{3}{2}}(1-i) &0 &0\\
  2&2&2 &2 &0 &0 &0 &0 &-2 &-2\\
   2&-2 &2 &-2 &0 &0 &0 &0 &-2 &2
    \end{smallmatrix}\right),
\end{align}
with topological sectors label 
$j,j' =\{0,1,2,\ldots, 9\}=\left\{1, b, a, b^*,
\bar{\sigma}_1, \bar{\tau}_2,\bar{\tau}_1,\bar{\sigma}_2,\bar{\phi}_2,\bar{\phi}_1
\right\}$.
\end{widetext}

The primary fields' scaling dimensions and quantum dimensions of $SU(4)_2$ are listed in table~\ref{tab:CFTdataSU4L2}. The $SU(4)_2$ theory and the $\mathbb{Z}_4$ parafermion CFT (see table~\ref{tab:Z4parafermionCFTdata}) share identical primary field fusion rules. For example, the Abelian anyons $1,[b],[a],[b^\ast]$ in $SU(4)_2$ and $1,\phi_3^1,j,\phi_3^2$ both form a $\mathbb{Z}_4$ fusion group. 
The metaplectic super-sector $[\bar{\phi}_2]$ forms the adjoint representation of ${\rm SU}(4)_2$. Therefore, it carries 15 primary fields since $\dim(SU(4))=15$. In the $SU(2)_4\times SU(4)_2\subseteq SU(8)_1$ conformal embedding, field products in $[j=1]\times[\bar\phi_2]$ 
accounts for (a) the 45 local $SU(8)_1$ currents $J$ outside of $SU(2)_4\times SU(4)_2$ (see \eqref{eq:SU2nL1dimensionalityIdentity}), as well as (b) the 45 non-local spin-1 vertex fields in $\mathcal{E}^4$ in $SU(8)_1$ outside of the $[j=2]$ super-sector in \eqref{SU24=U1Z8} from $SU(2)_4$ and the $[a]$ super-sector in \eqref{eq:SU4L2supersectora} from $SU(4)_2$. As vector spaces, we have the decomposition \begin{align}\begin{split}&SU(8)_1=SU(2)_4\oplus SU(4)_2\oplus\left([j=1]\times[\bar\phi_2]\right),\\&\mathcal{E}^4=[j=2]\oplus[a]\oplus\left([j=1]\times[\bar\phi_2]\right).\end{split}\label{SU8splitting1}\end{align}
There are 45 fields in both cases because $45$ equals $\dim(SU(8))-\dim(SU(2))-\dim(SU(4))$ as well as $\#\mathcal{E}^4-\#[j=2]-\#[a]$. Since there are 3 fields in $[j=1]$, there must be $15=45/3$ fields in $[\bar\phi_2]$. 

According to the branching rules, \begin{align}\begin{split}&\mathcal{E}^2=[b]\oplus\left([j=1]\times[\bar\phi_1]\right)\\&\mathcal{E}^6=[b^\ast]\oplus\left([j=1]\times[\bar\phi_1]\right).\end{split}\label{SU8splitting2}\end{align}
field products in $[j=1]\times[\bar\phi_1]$ 
splits into components that belong to (a) the $SU(8)_1$ vertex fields in $\mathcal{E}^2$ outside of the $[b]$ sector in \eqref{eq:SU4L2supersectorb} from $SU(4)_2$, and (b) the $SU(8)_1$ fields in $\mathcal{E}^6$ outside of $[b^\ast]$.  
There are 18 fields in each case because $18=\#\mathcal{E}^2-\#[b]$. Since $\#[j=1]=3$, there are $6=18/3$ fields in $[\bar\phi_1]$. The $SU(8)_1$ primary field super-sectors $\mathcal{E}^{1,3,5,7}$ with odd powers splits into $SU(2)_4\times SU(4)_2$ components according to \begin{align}\begin{split}&\mathcal{E}^1=[j=1/2]\times[\bar\sigma_1],\\&\mathcal{E}^3=\left([j=1/2]\times[\bar\tau_2]\right)\oplus\left([j=3/2]\times[\bar\sigma_2]\right),\\&\mathcal{E}^5=\left([j=1/2]\times[\bar\tau_1]\right)\oplus\left([j=3/2]\times[\bar\sigma_1]\right),\\&\mathcal{E}^7=[j=1/2]\times[\bar\sigma_2].\end{split}\label{SU8splitting3}\end{align} Dimension counting from $\#\mathcal{E}^m=C^8_m$ and $\#[j]=2j+1$ forces $\#[\sigma_{1,2}]=4$ and $\#[\tau_{1,2}]=20$. The branching rules \eqref{SU8splitting1}, \eqref{SU8splitting2} and \eqref{SU8splitting3} allows us to construct strings of electron operators $\mathcal{S}^m=\prod_{y=y_1}^{y_2}\mathcal{E}^{L,m}_y(\mathsf{x})\mathcal{E}^{R,m}_y(\mathsf{x})^\dagger$ (c.f.~\eqref{Emstring}) so that when operating on a ground state, it creates a conjugate excitation pair in an arbitrary anyon class in the $SU(2)_4$ or $SU(4)_2$ topological phase given an appropriate choice of the $SU(8)_1$ primary field $\mathcal{E}^m$ (c.f.~\eqref{FibonacciString}, \eqref{E2E2string1} and \eqref{E2E2string2}).

\begin{table}[htbp]
\setlength{\tabcolsep}{9.4pt} 
\renewcommand{\arraystretch}{1.} 
\centering
\caption{The spin $h$, quantum dimension $d$, and number of fields $\#$ of each non-trivial primary sector of ${\rm SU}(4)_2$.}
\label{tab:CFTdataSU4L2}
\begin{tabular}{lccccccl}
  \hline 
  \hline
       & $[a]$ & $[b],[b^*]$ & $[\bar{\sigma}_{1,2}]$  & $[\bar{\tau}_{1,2}]$  & $[\bar{\phi}_2]$ & $[\bar{\phi}_1]$ \\[1mm] 
  \hline  
  $h$  & $1$ & $\frac{3}{4}$ & $\frac{5}{16}$ & $\frac{13}{16}$ & $\frac{2}{3}$ & $\frac{5}{12}$\\
  $d$  & $1$ & $1$ & $\sqrt{3}$ & $\sqrt{3}$ & $2$  & $2$\\
  $\#$ & $20$ & $10$ & $4$ & $20$ & $15$ & $6$\\
  \hline 
  \hline
\end{tabular}
\end{table}

We conclude this section by commenting on a few cases where the electrically neutral $SU(m)_n$ topological sector contains a discrete gauge symmetry and the $SU(m)_n$ WZW CFT on the edge can be identified with an orbifold CFT $\mathcal{G}_k/G$~\cite{TeoHughesFradkin2015,Teo_2023,LimMulliganTeo2022A}, for some gauge group $G$ and WZW algebra extension $\mathcal{G}_k$ of $SU(m)_n$. These examples are \begin{align}\begin{split}&SU(2)_4=SO(3)_2=\frac{SU(3)_1}{\mathbb{Z}_2},\\&SU(4)_2=SO(6)_2=\frac{SU(6)_1}{\mathbb{Z}_2},\\&SU(8)_1=\frac{(E_7)_1}{\mathbb{Z}_2},\\&SU(9)_1=\frac{(E_8)_1}{\mathbb{Z}_3},\quad SU(3)_3=\frac{SO(8)_1}{\mathbb{Z}_3}.\end{split}\label{SUmnorbifolds}\end{align} In these examples, the gauge groups are the Abelian cyclic groups $\mathbb{Z}_2$ or $\mathbb{Z}_3$. The WZW algebra extensions $\mathcal{G}_k$ are simply-laced and have level $k=1$. Therefore, the corresponding $\mathcal{G}_1$ topological order is Abelian and can be described by an Abelian Chern-Simons field theory whose $K$-matrix is the Cartan matrix of the $\mathcal{G}$ Lie algebra. The $SU(m)_n$ topological order in the bulk is referred to as a twist liquid~\cite{TeoHughesFradkin2015} where the global $G$-symmetry in the $\mathcal{G}_1$ Abelian topological phase is gauged~\cite{PhysRevB.100.115147}.

The $SU(m)_n$ WZW algebras in each example in \eqref{SUmnorbifolds} is the sub-algebra in $\mathcal{G}_1$ consisting with currents that are unchanged under the $G$ symmetry. Moreover, there are bosonic $SU(m)_n$ primary field(s) in super-selection sector(s) $[\mathcal{Z}]$ with spin $h_{\mathcal{Z}}=1$ that carry gauge charges. They irreducibly and non-trivially represent the gauge group $G$. Since $G$ here is cyclic, this means they transform according to $\mathcal{Z}\to-\mathcal{Z}$ if $G=\mathbb{Z}_2$, and $\mathcal{Z}\to e^{\pm2\pi i/3}\mathcal{Z}$ if $G=\mathbb{Z}_3$. These spin-1 bosons would extend the $SU(m)_n$ WZW algebra to $\mathcal{G}_1$, if they were local. This means $SU(m)_n\oplus[\mathcal{Z}]=\mathcal{G}_1$ for $G=\mathbb{Z}_2$, or $SU(m)_n\oplus[\mathcal{Z}]\oplus[\mathcal{Z}^\ast]=\mathcal{G}_1$ for $G=\mathbb{Z}_3$. For the examples listed in \eqref{SUmnorbifolds}, \begin{align}\begin{split}&SU(2)_4\oplus[j=2]=SU(3)_1,\\&SU(4)_2\oplus[a]=SU(6)_1,\\&SU(8)_1\oplus\mathcal{E}^4=\left(E_7\right)_1,\\&SU(9)_1\oplus\mathcal{E}^3\oplus\mathcal{E}^6=\left(E_8\right)_1,\\&SU(3)_3\oplus{\bf 10}\oplus\overline{\bf 10}=SO(8)_1,\end{split}\end{align} where $[j=2]$ and $[a]$ are bosonic primary fields in $SU(2)_4$ and $SU(4)_2$ (see table~\ref{tab:CFTdataSU2L4} and \ref{tab:CFTdataSU4L2}), $\mathcal{E}^{m=3,6}$ are the bosonic $SU(9)_1$ primary fields defined in \eqref{eq:SUNL1primaries}, and ${\bf 10}$ and $\overline{\bf 10}$ are the two Abelian bosonic $SU(3)_3$ primary field super-sectors~\cite{TeoHughesFradkin2015} that form a ten-dimensional irreducible representation of the $SU(3)$ Lie algebra. For instance, in each of these cases, the dimension of $SU(m)$ and the number of fields in the gauge charge(s) $[\mathcal{Z}]$ add up to the dimension of the extended Lie algebra $\mathcal{G}$.

However, unlike the $SU(m)_n$ currents, the primary fields in $[\mathcal{Z}]$ are {\em non-local} and are {\em not} integral combinations of electrons, despite being bosonic. Consequently, the $SU(m)_n$ WZW CFT on the edge of the electronic topological phase does {\em not} extend to $\mathcal{G}_1$. The anyon class corresponding to $[\mathcal{Z}]$ does not condense (in the anyon condensation sense~\cite{PhysRevB.79.045316}) and the bulk $SU(m)_n$ topological order remains. In particular, there are twist fields $[\Sigma]$ in the $SU(m)_n$ CFT that correspond to deconfined gauge fluxes in the bulk. For $SU(2)_4$, these gauge fluxes are $[j=\frac{1}{2},\frac{3}{2}]$ (see table~\ref{tab:CFTdataSU2L4}). For $SU(4)_2$, they are $[\bar\sigma_{1,2}]$ and $[\bar\tau_{1,2}]$ (see table~\ref{tab:CFTdataSU4L2}). For $SU(9)_1$, they are $\mathcal{E}^{m=1,2,4,5,7,8}$ (see~\eqref{eq:SUNL1primaries}), and for $SU(3)_3$, they are ${\bf 3},\overline{\bf 3},{\bf 6},\overline{\bf 6},{\bf 15},\overline{\bf 15}$ (see~\cite{TeoHughesFradkin2015}). These gauge fluxes $[\Sigma]$ and gauge charges $[\mathcal{Z}]$ have non-trivial mutual braiding monodromy phases $-1$ for $G=\mathbb{Z}_2$ or $e^{\pm2\pi i/3}$ for $G=\mathbb{Z}_3$.

In general, the $SU(m)_n$ WZW CFT on the edge of the topological phases constructed in this section always persists for any $m\geq2$ and $n\geq1$. Even if there are bosonic spin-1 primary fields $[\mathcal{Z}]$, they are always fractional. The non-locality is a result of the conformal embedding $SU(m)_n\times SU(n)_m\subseteq SU(mn)_1$ used in the coupled wire construction. If the primary fields in $[\mathcal{Z}]$ were local and were to extend the $SU(m)_n$ algebra, they would either (a) be currents living inside the $SU(mn)_1$, or (b) belong in a spin-1 primary field super-sector of $SU(mn)_1$ that extend the ${\rm SU}(m)_n$
algebra. First, case (a) does not apply because the only $SU(mn)_1$ currents that live outside of $SU(m)_n$ and $SU(n)_m$ are in the tensor product $[A]_{SU(m)_n}\times[A]_{SU(n)_m}$ between the adjoint representations of $SU(m)_n$ and $SU(n)_m$ (see \eqref{eq:SU2nL1dimensionalityIdentity}). None of them are primary fields solely of $SU(m)_n$ or $SU(n)_m$. Second, case (b) is impossible as well. This is because the parton bundle construction dictates that all the $SU(mn)_1$ primary fields $\mathcal{E}^j$ in \eqref{eq:SUNL1primaries} are non-local. Therefore, even if some of them are bosonic, such as the ones in $SU(8)_1$ and $SU(9)_1$, they will never extend ${\rm SU}(m)_n$
to a larger local WZW algebra.

On the other hand, there are examples beyond $SU(m)_n$ WZW algebras where the anyon condensation of gauge charges and the extension of WZW algebras occur. In the next section, we will see that the coupled-wire construction that uses backscattering of the KM currents in the orthogonal WZW algebras $SO(n)_2$ will result in a topological phase with a different topological order. We will show that the $\mathbb{Z}_2$ gauge charge $[S]$ in $SO(n)_2=SU(n)_1/\mathbb{Z}_2$ will in fact be local and will be an integral combination of electrons under the conformal embedding $SO(2)_n\times SO(n)_2\subseteq SO(2n)_1$. This is because the primary fields in $[S]$ are currents in $SO(2n)_1$. On the edge, they will extend the WZW algebra to $SO(n)_2\oplus[S]=SU(n)_1$. In the bulk, $[S]$ will anyon condense and reduce the topological order to the Abelian $SU(n)_1$, where the $\mathbb{Z}_2$ symmetry is global rather than an internal gauge symmetry.

\section{\texorpdfstring{$B_r$}{B} and \texorpdfstring{$D_r$}{D} Series, Emergent Dirac and Majorana Fermions}\label{Sec.BDseriesAndTopOrder}

In this section, we construct and study states with electrically neutral topological phases with a $SO(m)_n$ topological order. These are superconducting or spin liquid phases of electrons. The $SO(m)_n$ topological phase ``partially occupies" the parent bosonic $SO(mn)_1$ state where non-local Majorana fermions emerges. The construction relies on the ${\rm SO}(m)_n\times {\rm SO}(n)_m \subset {\rm SO}(mn)_1$ conformal embedding.

\subsection{Coupled-Wire Construction of Topological Superconductors and Spin Liquids}
\label{sec:BosoicWiresOfSOFamily}
We construct neutral topological phases with $SO(N)_1$ topological order and chiral edge WZW CFT, where $N$ is an arbitrary integer greater than 1. 
Following the strategy described previously, we consider a 2D array of bundles, each consisting of $\mathcal{N}$ wires.
Each wire now hosts a single {\it spinful} electron with Rashba spin-orbit interaction, which, at low energies, decomposes into a Dirac electron $c^\sigma_{y,sj}$ with chirality $\sigma = R/L = +/-$, spin $s=\uparrow(\downarrow)=+(-)$, and $j = 1, \ldots, {\cal N}$ labels a given wire in the bundle $y$.
The spin-orbit interaction, parameterized by momentum $k_{SO}$, splits the spin-degenerate Fermi points. 
 
The Dirac electrons can be represented by vertex operators of chiral bosons $\Phi^\sigma_{y,sj}$ as follows:
\begin{align}
c^\sigma_{y,sj}(\mathsf{x})\sim 
\exp\left[i \Phi^\sigma_{y,sj}(\mathsf{x})+i(\sigma k_f+sk_{SO})\mathsf{x}
    \right],
\label{eq:withSOFermiMomentum}
\end{align} 
where $k^\sigma_s=\sigma k_f+sk_{SO}$ is the Fermi momentum about which $c^\sigma_{y,sj}$ is defined.
The Luttinger liquid Lagrangian density is 
\begin{align}
\begin{split}
\mathcal{L}&=\frac{1}{4\pi}
\sum_y\sum_{s,\sigma=\pm}\sum_{j=1}^{\mathcal{N}}
\sigma\partial_t\Phi^\sigma_{y,sj}\partial_{\mathsf{x}}\Phi^\sigma_{y,sj}-\mathcal{H}_0,
\end{split}
\label{eq:LuttingerLagrangianDensity}
\end{align} 
where the free boson Hamiltonian density, including intra-bundle density-density interactions, is
\begin{align}
\mathcal{H}_0=    
\sum_yv^{jj',ss'}_{\sigma\sigma'}\partial_\mathsf{x}{\Phi}^\sigma_{y,sj}\partial_\mathsf{x}{\Phi}^{\sigma'}_{y,s'j'}.
\label{eq:LuttingerHamiltonianDensity}
\end{align}

Our first objective is to introduce an intra-bundle many-body interaction that gaps all local fermion excitations and leaves behind a non-chiral $SO(N)_1$ bosonic CFT on each bundle. This can be achieved by either an Umklapp back-scattering or superconducting pairing potential \cite{Teo_2023}. On all bundle $y$, we impose one of the interactions below:
\begin{align}
\begin{split}
\mathcal{U}_{\rm Umklapp} &= u\cos\theta_\rho\sim
u\prod_{j=1}^\mathcal{N}\prod_{s=\uparrow,\downarrow} c^{L\dagger}_{sj}c^R_{sj}+h.c.,\\
\mathcal{U}_{\rm SC} &= \Delta \cos\varphi_\rho\sim
\Delta\prod_{j=1}^\mathcal{N}\prod_{s=\uparrow,\downarrow}
c^L_{sj}c^R_{sj}+h.c.,
\end{split}
\label{eq:SOfermionicgaps}
\end{align}
where the sine-Gordon angle variables are
\begin{align}\begin{split}
\theta_\rho =\sum_{sj}(\Phi^R_{sj}-\Phi^L_{sj}), \quad
\varphi_\rho =\sum_{sj}(\Phi^L_{sj}+\Phi^R_{sj}).
\end{split}\end{align}
When one of these interactions is present and is relevant, the charge degrees of freedom of each bundle are gapped. (1) In the Umklapp case, the potential is a $2\mathcal{N}$-body scattering process that violates momentum conservation.
This interaction appears at low energies in the presence of a charge density wave with commensurate momentum. 
It can also arise in a 1D lattice with a commensurate filling so that the total momentum transfer $4\mathcal{N}k_f$ is an integer multiple of $2\pi/a$, where $a$ is the microscopic lattice constant. $\mathcal{U}_{\mathrm{Umklapp}}$ in \eqref{eq:SOfermionicgaps} leads to a spin liquid state where all excitations with energy below $u$ are electrically neutral. (2)  In the superconducting case, the pairing potential preserves momentum, but violates charge conservation. It can be induced via the proximity effect by Cooper pair tunneling from a bulk superconductor. The $U(1)$ symmetry is broken to $\mathbb{Z}_2$ by $\mathcal{U}_{\mathrm{SC}}$ in \eqref{eq:SOfermionicgaps} and a charge $Q$ is indistinguishable from $Q+2e$ because of the Cooper pair condensate.

Next, we perform a basis transformation of the bosons that isolates $\varphi_\rho$ or $\theta_\rho$ and decouples it from the rest of the boson in each bundle. The basis transformation depends on whether the gapping potential is given by the Umklapp process or the superconducting pairing.
In the Umklapp case, we define
\begin{align}
    2\phi^\sigma_\rho = \varphi_0+\sigma\theta_\rho, \quad 
    \phi^\sigma_\ell = \varphi_0-\Phi^\sigma_{s_\ell j_\ell},
    \label{eq:transformationUmklapp}
\end{align}
whereas in the superconducting case, we define
\begin{align}
    2\phi^\sigma_\rho = \varphi_\rho+\sigma\theta_0, \quad 
    \phi^\sigma_\ell = \theta_0-\sigma\Phi^\sigma_{s_\ell j_\ell}.
    \label{eq:transformationSuper}
\end{align} 
In either case, the non-chiral conjugate bosons, in addition to $\varphi_\rho$ and $\theta_\rho$, are
\begin{align}
    \varphi_0 = \left(\Phi^R_{\uparrow1}+\Phi^L_{\downarrow1}\right)/2, \quad
    \theta_0 = \left(\Phi^R_{\uparrow1}-\Phi^L_{\downarrow1}\right)/2.
\end{align}
The above spin and wire index assignment $(s_\ell,j_\ell)$ for a given $\ell$ are chosen to account for all the electron modes $\Phi^\sigma_{s_\ell j_\ell}$, except $\Phi^R_{\uparrow1}$ and $\Phi^L_{\downarrow1}$. For example, for a given chiral sector $\sigma=\pm$, they can be chosen to be \begin{align}(s_\ell,j_\ell)=\left\{\begin{array}{*{20}l}(\sigma,\ell+1),&\mbox{if $\ell=1,\ldots,\mathcal{N}-1$}\\(-\sigma,\ell-\mathcal{N}+1),&\mbox{if $\ell=\mathcal{N},\ldots,2\mathcal{N}-1$}\end{array}\right..\end{align} The charge boson $\phi^\sigma_\rho$ now decouples from the remaining $2\mathcal{N}-1$ non-local Dirac fermions $d^\sigma_\ell\sim e^{i\phi^\sigma_\ell}$, for $\ell=1,\ldots,2\mathcal{N}-1$. 
In this new basis, the free boson Lagrangian density is
\begin{align}
\begin{split}
\mathcal{L}
&=\frac{1}{4\pi}\sum_{\sigma=R,L}
\sigma\left[
2\partial_\mathsf{x}\phi^\sigma_\rho\partial_t
\phi^\sigma_\rho +
\sum_{\ell=1}^{2\mathcal{N}-1}
\partial_\mathsf{x}\phi^\sigma_\ell\partial_t
\phi^\sigma_\ell
\right]-\mathcal{H}_0,\\
\mathcal{H}_0&=\frac{v}{4\pi}
\sum_{\sigma}
\sum_{\ell=1}^{2\mathcal{N}-1}
\left(\partial_\mathsf{x}\phi^\sigma_\ell\right)^2+
v_\rho\left[
g_\rho\left(\partial_\mathsf{x}\varphi\right)^2+\frac{1}{g_\rho}\left(\partial_\mathsf{x}\theta\right)^2
\right],
\end{split}
\label{eq:diagonalHamiltonianDensitySO}
\end{align}
where $2\phi^\sigma_\rho=\varphi+\sigma\theta$, i.e.~$(\varphi,\theta)=(\varphi_0,\theta_\rho)$ in the Umklapp case, or $(\varphi,\theta)=(\varphi_\rho,\theta_0)$ in the superconducting case. The velocity tensor $v^{jj',ss'}_{\sigma\sigma'}$ in \eqref{eq:LuttingerHamiltonianDensity} is tuned so that the boson velocities in $\mathcal{H}_0$ are diagonal and isotropic among the Dirac fermions densities $\partial_\mathsf{x}\phi^\sigma_\ell$, which are decoupled from $\partial_\mathsf{x}\varphi$ and $\partial_\mathsf{x}\theta$. 
This fine-tuning can be relaxed once a bulk energy gap opens up from the current backscattering interactions introduced later in this section. The Umklapp potential $\mathcal{U}_{\mathrm{Umklapp}}$ in \eqref{eq:SOfermionicgaps} is relevant in the RG sense, when $g_\rho<4$. The pairing potential $\mathcal{U}_{\mathrm{SC}}$ in \eqref{eq:SOfermionicgaps} is relevant when $g_\rho>1/4$. The Luttinger parameter $g_\rho$ in either case can be tuned to the corresponding desired range by the presence of appropriate density interactions in the electron wires. In both cases, the potential $\mathcal{U}_{\mathrm{Umklapp}}$ or $\mathcal{U}_{\mathrm{SC}}$ gaps all local fermionic degrees of freedom and leaves behind the gapless $\phi_\ell^\sigma$ modes. Each bundle is now at a low-energy fixed point described by the bosonic $SO(4\mathcal{N}-2)_1$ WZW CFT whose gapless excitations are bosonic pairs of Dirac fermions, $d^\sigma_\ell d^{\sigma'}_{\ell'},{d^\sigma_\ell}^\dagger d^{\sigma'}_{\ell'},{d^\sigma_\ell}^\dagger{d^{\sigma'}_{\ell'}}^\dagger$, etc. There is a local fermion gap $E_g^f\sim u$ or $\Delta$ below which all odd fermion excitations are forbidden.

Using the basis transformation $\phi^\sigma_\ell=M^{sj}_{\ell\sigma\sigma'}\Phi^{\sigma'}_{sj}$ from \eqref{eq:transformationUmklapp} or \eqref{eq:transformationSuper}, we see that Dirac fermion vertex operators, 
\begin{align}
d^\sigma_\ell \sim e^{i\phi^\sigma_\ell}=e^{i M^{sj}_{\ell\sigma\sigma'}\Phi^{\sigma'}_{sj}},
\label{eq:SODiracFermionOrigins}
\end{align}
are non-local.
This is because $M^{\uparrow1}_{\ell\sigma L},M^{\downarrow1}_{\ell\sigma R}=\pm 1/2$ are fractional (the rest of $M^{sj}_{\ell\sigma\sigma'}=\pm 1$). In addition, each Dirac fermion carries electric charge $Q=0$ and angular spin $S=\pm1/2$ in the Umklapp case, or $Q=\pm e$ (mod $2e$) and $S=0,\pm1$ in the superconducting case. Therefore, they must not be integral combinations of electrons, which has $Q=-e$ and $S=\pm1$. This reflects the fact that there are no longer any low-energy single-fermion excitations below the fermion gap $E_g^f$. On the other hand, all fermion bilinears, such as $d^\sigma_\ell d^{\sigma'}_{\ell'}$,
$d^{\sigma\dagger}_\ell d^{\sigma'\dagger}_{\ell'}$,
$d^{\sigma\dagger}_\ell d^{\sigma'}_{\ell'}$,
$d^{\sigma}_\ell d^{\sigma'\dagger}_{\ell'}$, are local. 
These Dirac fermions can be decomposed into real and imaginary Majorana components 
\begin{align}
    d^\sigma_\ell =\left(\psi^\sigma_{2\ell-1}+i\psi^\sigma_{2\ell}\right)/\sqrt{2},
\label{Majoranadecomposition1}\end{align}
where ${\psi^\sigma_j}^\dagger=\psi^\sigma_j$. All Majorana fermion bilinears, $\psi^\sigma_{j}\psi^{\sigma'}_k$, for $1\leq j\leq k\leq 4\mathcal{N}-2$, are also local. 
The momentum of these Majorana fermions are 0 (or $\pi/a$, where $a$ is the microscopic lattice constant, for the Umklapp case) if $k_{SO}=k_f$ in the Umklapp case or $k_{SO}=0$ in the superconducting case. 

Apart from fermion bilinears, conjugate pairs of spinor fields are local operators as well up to ground state expectation values of $\theta_\rho$ or $\varphi_\rho$. Spinor primary fields $[s_\pm]$ in $SO(4\mathcal{N}-2)_1$ are generated by vertex operators $s^\sigma_{\boldsymbol\varepsilon}=e^{i\sum_\ell\varepsilon^\ell\phi^\sigma_\ell/2}$, where $\varepsilon_\ell=\pm$. The sign $\prod_\ell\varepsilon_l=\pm$ sets the parity of the spinor field. A pair of spinor fields $s^\sigma_{\boldsymbol\varepsilon_1}s^\sigma_{\boldsymbol\varepsilon_2}$ in the same chiral sector $\sigma$ but with opposite parity (i.e.~$\prod_\ell\varepsilon^\ell_1\varepsilon^\ell_2=-1$) is an even product of fermions and thus is local. This shows $[s_+^\sigma]\times[s_-^\sigma]=1$. A pair of spinor fields $s^R_{\boldsymbol\varepsilon_1}s^L_{\boldsymbol\varepsilon_2}$ from opposite chiral sectors but with the same parity (i.e.~$\prod_\ell\varepsilon^\ell_1\varepsilon^\ell_2=1$) is effectively local. It differs from $e^{i\theta_\rho/2}$ ($e^{i\varphi_\rho/2}$) by an integral product of electron operators, and $\langle\theta_\rho\rangle$ (resp.~$\langle\varphi_\rho\rangle$) is pinned at its ground state expectation value by the Umklapp (resp.~pairing) potential \eqref{eq:SOfermionicgaps}. This shows $[s_\pm^R]\times[s_\pm^L]=1$.

The number of helical pairs of Majorana fermions can be reduced from $4\mathcal{N}-2$ to any integer $N$ below by a momentum-preserving local backscattering potential.
\begin{align}
    \mathcal{U}_{\rm mass}=im\sum_{j=N+1}^{4\mathcal{N}-2}\psi^L_j\psi^R_j.\label{eq:ExtraMajoranaDOF}
\end{align}
This gaps out $\psi^{L, R}_j$ for $j = N+1, \ldots 4 {\cal N} - 2$, where $4\mathcal{N}-2>N$.
The low-energy effective Hamiltonian density can be written in terms of remaining $N$ gapless non-chiral Majoranas as 
\begin{align}
\mathcal{H}_{\rm eff}
=\frac{iv}{2} \sum_y\sum_\sigma\sum_{j=1}^N\sigma\psi^\sigma_{y j}\partial_\mathsf{x}\psi^\sigma_{y j}.
\label{eq:effectiveKineticHamiltoniansSO}
\end{align}
We stress that, despite appearances, \eqref{eq:effectiveKineticHamiltoniansSO} does {\it not} describe a collection of free Majorana fermions, since single-particle Majorana fermion operators are non-local.

The gapless modes in \eqref{eq:effectiveKineticHamiltoniansSO} in each bundle of wires are effectively described by the non-chiral $SO(N)_1$ WZW CFT, which can now be used to construct the ${\rm SO}(N)_1$ coupled-wire model for states with spin liquid or topological superconducting order.
Within each bundle, the ${\rm SO}(N)_1$ WZW KM algebra for each chiral sector $\sigma$ is spanned by $N(N-1)/2$ current operators 
$J^\sigma_{y, jk}=i\psi^\sigma_{y a}(X^{ab})_{jk}\psi^\sigma_{y b}=i\psi^\sigma_{y j}\psi^\sigma_{y k}$. 
Here, $(X^{ab})_{jk}\equiv \delta^a_j\delta^b_k$ is a matrix representation of ${\rm SO}(N)_1$ with $1\leq j<k\leq N$ and $a,b=$$1,\ldots, N$.

Each bundle admits an internal $\mathbb{Z}_2$ symmetry:
\begin{align}
\begin{split}
\psi^\sigma_{yj}\rightarrow (-1)^{\delta_{y y'}}\psi^\sigma_{yj},\quad
\phi^\sigma_{ya}\rightarrow \phi^\sigma_{ya}+\sigma\pi \delta_{y y'},
\end{split}
\label{eq:SOinternalZ2}
\end{align}
for each fixed $y'$ that labels the bundle where this particular local $\mathbb{Z}_2$ symmetry applies.
An operator is only local if it is unchanged by the internal $\mathbb{Z}_2$ symmetries for all $y'$.
This internal symmetry is referred to as the local $\mathbb{Z}_2$ gauge symmetry.
For example, all single-fermion tunneling $\psi_{y_1}\psi_{y_2}$ between bundles is non-local and forbidden. On the other hand, single-fermion tunneling within a given bundle, like those in \eqref{eq:effectiveKineticHamiltoniansSO} are local. Single-fermion tunneling between bundles is not local, however two-fermion scattering between bundles is also local. In particular, the current backscattering ${\bf J}^R_{y}\cdot{\bf J}^L_{y+1} = \sum_{j<k} J^R_{y, jk} J^L_{y, jk}$ in \eqref{SON1currentbackscattering} below is an allowed process. In addition, counter-propagating pairs of spinor twist fields (with the same parity if $N\equiv0$ mod 4 or opposite parity if $N\equiv2$ mod 4) within the same bundle is effectively local at low energy because they differ from the local combination $s^R_{\boldsymbol\varepsilon}s^L_{\boldsymbol\varepsilon}$ of electron operators up to ground state expectation values of operators in the $\psi_{N+1,\ldots,4\mathcal{N}-2}$ sector, which is gapped by \eqref{eq:ExtraMajoranaDOF}.

We construct the chiral $SO(N)_1$ topological model,
\begin{align}
\mathcal{H}\left[{\rm SO}(N)_1\right]=\mathcal{H}_{\rm eff}+\mathcal{H}^{{\rm SO(N)_1}}_{\rm inter},
\label{SON1model}\end{align}
using the inter-bundle backscattering potential,
\begin{align}
\begin{split}
\mathcal{H}^{{\rm SO}(N)_1}_{\rm inter}&= u_{\rm inter}\sum_y{\bf J}^R_y
\cdot {\bf J}^L_{y+1}\\
&= -u_{\rm inter}\sum_y \sum_{1\leq j<k\leq N}\psi^R_{y,j}\psi^R_{y,k}\psi^L_{y+1,j}\psi^L_{y+1,k},
\end{split}\label{SON1currentbackscattering}
\end{align}
for $N\geq3$.
The potential between each neighboring pair of bundles is the $SO(N)_1$ Gross-Neveu interaction~\cite{GrossNeveu1974}, which is known to introduce a bulk excitation energy gap when $u_{\mathrm{inter}}>0$.
In terms of the bosonized variables defined by \eqref{eq:SODiracFermionOrigins} and \eqref{Majoranadecomposition1}, 
the two-body interactions are
\begin{align}
\begin{split}
&\mathcal{H}^{{\rm SO}(2r)_1}_{\rm inter}=u_{\rm inter}\sum_y\sum_{a=1}^r
\partial_\mathsf{x}\phi^R_{y,a}\partial_\mathsf{x}\phi^L_{y+1,a}\\
&\quad\quad\quad-u_{\rm inter}\sum_y\sum_{\epsilon=\pm}
\cos\left(\Theta_{y+1/2,a}+\epsilon\Theta_{y+1/2,b}\right),\\
&\mathcal{H}^{{\rm SO}(2r+1)_1}_{\rm inter}=\mathcal{H}^{SO(2r)_1}_{\rm inter}\\&\quad\quad-u_{\rm inter}\sum_y i \psi^R_{y,2r+1}\psi^L_{y+1,2r+1}
\sum_{a=1}^r\cos\Theta_{y+1/2,a},
\end{split}
\label{eq:SOinterforevenoddN}
\end{align}
where the angle variables are $\Theta_{y+1/2,a}=\phi^R_{y,a}-\phi^L_{y+1,a}$.
The sine-Gordon terms are relevant in the RG sense and do not mutually compete when $u_{\rm inter}>0$. 
The ground state expectation values of the sine-Gordon variables are simultaneously pinned at 
$\langle\Theta_{y+1/2,a}\rangle\sim m_{y+1/2,a}\pi$.
$m_{y+1/2,a}$ are either all odd or all even integers for all $a=1,\ldots, r$.
When $N=2r+1$, the last term in \eqref{eq:SOinterforevenoddN} introduces a Majorana mass to $\psi_{2r+1}$ and pins the expectation values of $\langle i\psi^R_{y,2r+1}(\mathsf{x})\psi^L_{y+1,2r+1}(\mathsf{x})\rangle\sim (-1)^{m_{y+1/2,a}}$. The model \eqref{SON1model} now has a finite excitation energy gap in the bulk and leaves behind chiral $SO(N)_1$ WZW CFT along the boundary edges. The bulk carries the corresponding $SO(N)_1$ topological order, which is explained in detail in ref.~\cite{kitaev2006anyons, PhysRevB.94.165142, Teo_2023, LimMulliganTeo2022A}.

We see that the current backscattering term \eqref{SON1currentbackscattering} is a gapping potential only when $N\geq3$. For $N=1$, there is no two-fermion backscattering process and \eqref{SON1currentbackscattering} is an empty sum. For $N=2$, the two-fermion backscattering process is a density-density interaction $\partial_{\mathsf{x}}\phi^R_y\partial_{\mathsf{x}}\phi^L_{y+1}$, which only renormalizes the kinetic Hamiltonian without introducing an energy gap. Before moving on to the next section, here we present the alternative coupled wire models for these two cases that represent the chiral Ising and $SO(2)_1$ topological phases.

First, we consider the case when there are $N=2$ Majorana fermions on each bundle in each chiral sector. Instead of the inter-bundle current backscattering potential \eqref{SON1currentbackscattering}, we consider the sine-Gordon potential, \begin{align}\mathcal{H}^{SO(2)_1}_{\mathrm{inter}}=-u_{\mathrm{inter}}\sum_y\cos\left(2\phi^R_y-2\phi^L_{y+1}\right),\label{SO21potential}\end{align} where $e^{i\phi^\sigma_y}=(\psi^\sigma_{y,1}+i\psi^\sigma_{y,2})/\sqrt{2}$. This term is symmetric under the internal $\mathbb{Z}_2$ symmetry \eqref{eq:SOinternalZ2} and is local. With a strong enough repulsive interaction $\partial_{\mathsf{x}}\phi^R_y\partial_{\mathsf{x}}\phi^L_{y+1}$, the sine-Gordon potential becomes RG relevant, gaps the bulk and leaves behind the chiral $SO(2)_1$ WZW CFT on its edges. 

Next, we consider the $N=1$ case. This case is special because there is no local inter-bundle gapping potential if there is only one (non-chiral) Majorana fermion in each bundle. Instead, we begin with $N=5$ gapless Majorana fermions $\psi^\sigma_{y,j=1,\ldots,5}$, and we now construct a coupled wire model with a bulk gap that leaves behind a single chiral Majorana fermion on each edge. The model consists of two sets of backscattering potentials, \begin{align}\begin{split}\mathcal{H}[SO(1)_1]&=\mathcal{H}_{\mathrm{eff}}^{N=5}-u_{\mathrm{inter}}\sum_y\cos\left(2\phi^R_y-2\phi^L_{y-1}\right)\\&\;\;\;-u_{\mathrm{inter}}\sum_y\sum_{3\leq j<k\leq5}\psi^R_{y,j}\psi^R_{y,k}\psi^L_{y+1,j}\psi^L_{y+1,k},\end{split}\end{align} where $e^{i\phi^\sigma_y}=(\psi^\sigma_{y,1}+i\psi^\sigma_{y,2})/\sqrt{2}$. Here, $SO(1)_1$ stands for the Ising model, which is a minimal CFT and not a WZW CFT. The first potential is similar to the $SO(2)_1$ sine-Gordon potential in \eqref{SO21potential} except with opposite chirality. The second potential is the $SO(3)_1$ current backscattering that gaps $\psi_{j=3,4,5}$. With these together, the bulk is gapped and, on each boundary edge, there are three forward propagating fermions $\psi^R_{j=3,4,5}$ and two backward propagating ones $\psi^L_{j=1,2}$. These two sectors are correlated so that the fermion pairs $\psi^L_j\psi^R_k$, for $j=1,2$ and $k=3,4,5$, are local and condensed in the anyon condensation sense. The counter-propagating fermions can be reduced by the edge potential, \begin{align}\mathcal{H}_{\mathrm{edge}}=u_{\mathrm{edge}}i(\psi^L_1\psi^R_3+\psi^L_2\psi^R_4).\end{align} This leaves behind a single chiral non-local Majorana fermion $\psi^R_5$ and the gapless theory is described by the chiral Ising CFT. Because of the anyon condensation of the fermion pairs $\psi^L_{j=1,2}\psi^R_{k=3,4,5}$, the bulk carries the Ising topological order ($\sigma\times\sigma=1+\psi$ and $h_\sigma=1/16$) instead of the tensor product $SO(3)_1\times\overline{SO(2)_1}$ topological order.

\subsection{Conformal Embedding and Descendant Topological Order}

\subsubsection{Symmetry embedding}
\label{sec:levelRankFromalismOrthogonalClass}

In the previous subsection, we constructed the bosonic $SO(N)_1$ WZW CFT from electrons, where the chiral current algebra is generated by bilinears of emergent Majorana fermions, $J^\sigma_{y,jk}=i\psi^\sigma_{y,j}\psi^\sigma_{y,k}=i\psi^\sigma_{y,a}(X^{jk})_{ab}\psi^\sigma_{y,b}$. The matrix representation of the generators of the ${\rm SO}(N)$ Lie algebra, for arbitrary integer $N\geq2$, are 
\begin{align}
    (X^{jk})_{rs}= \frac{1}{2}\left(
    \delta^j_r\delta^k_s-\delta^k_r\delta^j_s
    \right),
    \label{eq:SOmatrixrep}
\end{align}
where $r,s=1,\ldots, N$ and $1\leq j<k\leq N$.

Analogous to Sec.~\ref{sec:LRDAseries}, we consider the conformal embedding ${\rm SO}(m)_n \times {\rm SO}(n)_m \subseteq SO(mn)_1$, where we set $N = mn$ for $m,n\geq2$.
We will call ${\rm SO}(m)_n$ the $A$ sector and ${\rm SO}(n)_m$ the $B$ sector. 
The $m\times m$ matrix generators $X^{pq}$ of $SO(m)$ can be embedded into $SO(mn)$ by taking
$(X^A)^{pq}_{rs}\equiv \left(X^{pq}\otimes\mathbb{1}_n\right)_{rs}=(X^{pq})_{ab}\delta_{cd}\delta^{(a-1)n+c}_r\delta^{(b-1)n+d}_s$, where $r,s = 1, \ldots, mn$, $1 \leq p < q < m$, $a, b = 1, \ldots, m$, and $c, d = 1, \ldots, n$. 
The ${\rm SO}(m)_n$ current operators (for each chiral sector $\sigma$ on each bundle $y$) are 
\begin{align}
\begin{split}
J^A_{pq} &= i\sum_{r,s=1}^{mn}\psi_r\left(X^{pq}\otimes\mathbb{1}_n\right)_{rs}\psi_s\\
&=i\sum_{c=1}^n\psi_{(p-1)n+c}\psi_{(q-1)n+c},
\end{split}
\label{eq:AsectorSOCurrent}
\end{align}
which is a sum of $n$ decoupled copies of the $SO(m)_1$ currents.
Similarly for the $B$ sector, given any $n\times n$ anti-symmetric matrix $X^{pq}$
that generates $SO(n)$, we define
$(X^B)^{pq}_{rs}\equiv \left(\mathbb{1}_m\otimes X^{pq}\right)_{rs}=\delta_{ab}(X^{pq})_{cd}\delta^{(a-1)n+c}_r\delta^{(b-1)n+d}_s$, where $1 \leq p < q \leq n$ and $a,b,c,d,r,s$ each has the same range as before.
The ${\rm SO}(n)_m$ current operators are the fermion bilinears,
\begin{align}
\begin{split}
J^B_{pq} &= i\sum_{r,s=1}^{mn}\psi_r\left(\mathbb{1}_m\otimes X^{pq}\right)_{rs}\psi_s
\\
&=i\sum_{a=1}^m\psi_{(a-1)n+p}\psi_{(a-1)n+q}.
\end{split}
\label{eq:BsectorSOCurrent}
\end{align}

\subsubsection{Descendant states}
We now construct the coupled wire models for the $SO(m)_n$ topological phases. These are spin liquids or superconducting phases that can be viewed as partially occupying the $SO(mn)_1$ topological phases constructed before, and are descendant states of the $SO(mn)_1$ states. This is because $SO(m)_n$ is a WZW sub-algebra of $SO(mn)_1$. They are the analog of the fractional quantum Hall $U(1)_{mn}\times SU(m)_n$ phases, studied in Sec.~ \ref{SUdescendants}, that are descendants of the $U(1)_{mn}\times SU(mn)_1$ FQH state. Except, now there is no $U(1)$ sector that supports electric charge response. 

The coupled wire model begins with the effective Hamiltonian $\mathcal{H}_{\mathrm{eff}}$ that describes $N=mn$ non-chiral emergent Majorana fermions on each bundle. The bosonic $SO(N)_1$ WZW CFT on each bundle originates from a many-body Umklapp or superconducting pairing potential introduced in the previous subsection \ref{sec:BosoicWiresOfSOFamily}. Unlike the parent $SO(N)_1$ state where the entire current algebra is back-scattered to the neighboring bundle, here $SO(N)_1$ is split into the decoupled $SO(m)_n$ and $SO(n)_m$ sectors by the level-rank duality, and the two are separately back-scattered between neighboring bundles or within a bundle. Following Sec.~ \ref{SUdescendants}, we consider the Hamiltonian 
\begin{align}
\mathcal{H}\left[{\rm SO}(m)_n\right]=\mathcal{H}_{\rm eff}
+\mathcal{H}^{{\rm SO(m)_n}}_{\rm inter}
+\mathcal{H}^{{\rm SO(n)_m}}_{\rm intra},
\end{align}
where the current backscattering potentials are
\begin{widetext}
\begin{align}
\begin{split}
\mathcal{H}^{{\rm SO}(m)_n}_{\rm inter}&= u_{\rm inter}
\sum_y({\bf J}_{{\rm SO}(m)_n})^R_y\cdot ({\bf J}_{{\rm SO}(m)_n})^L_{y+1}\\
&=-u_{\rm inter}
\sum_y\sum_{c,c'=1}^n
\sum_{1\leq p<q\leq m}
\psi^R_{y,(p-1)n+c}\psi^R_{y,(q-1)n+c}
\psi^L_{y+1,(p-1)n+c'}\psi^L_{y+1,(q-1)n+c'},\\
\mathcal{H}^{{\rm SO}(n)_m}_{\rm intra}&= u_{\rm intra}
\sum_y({\bf J}_{{\rm SO}(n)_m})^R_y\cdot ({\bf J}_{{\rm SO}(n)_m})^L_{y+1}\\
&=-u_{\rm intra}
\sum_y\sum_{a,a'=1}^m
\sum_{1\leq p<q\leq n}
\psi^R_{y,(a-1)n+p}\psi^R_{y,(a-1)n+q}
\psi^L_{y+1,(a'-1)n+p}\psi^L_{y+1,(a'-1)n+q}.
\end{split}
\label{eq:SOmnbackscatterings}
\end{align}
\end{widetext}
Eq.~\eqref{eq:SOmnbackscatterings} applies for any $m,n\geq3$. Situations when $m=2$ or $n=2$ requires separate attention because the $SO(2)_l$ current back-scattering interaction is not a gapping potential. We will present the modified model for this specific case in the following subsection. We conjecture that for $m,n\geq3$, the potentials in \eqref{eq:SOmnbackscatterings} together create a finite excitation energy gap in the bulk and leave behind the gapless chiral $SO(m)_n$ WZW CFT on each boundary edge. The presence of the bulk energy gap can be proven with the help of previous results in parafermions for small $m,n\leq4$. 

The $SO(3)_3$ case was proven in Ref.~\cite{PhysRevB.94.165142} in the context of gapped topologically ordered surface states of topological superconductors. The $SO(3)_3$ KM current decomposes into $J^\pm=e^{\pm i\phi}\Psi$, where $e^{i\phi}$ is the vertex operator of the $SO(2)_3$ sub-algebra, and $\Psi$ is the $\mathbb{Z}_6$ parafermion~\cite{ZamolodchikovFateev85} in the coset CFT $SO(3)_3/SO(2)_3$. Subsequently, the current backscattering also decomposes, $J^\pm_R J^\mp_L=\cos(\phi_R-\phi_L)\Psi^R\Psi^L$. The sine-Gordon part pins the angle variable $\phi_R-\phi_L$ and gaps the $SO(2)_3=U(1)_{12}$ sector. The parafermion backscattering part gaps the $SO(3)_3/SO(2)_3$ coset CFT~\cite{Fateev91}.
A similar parafermion decomposition applies for $SO(3)_l=SO(2)_\times\mathbb{Z}_{2l}$ for any level $l\geq2$, and the $SO(3)_l$ current backscattering potential in general gaps all degrees of freedom. (The level $l=1$ case is a particular case of $SO(2r+1)_l$ and it was addressed previously below \eqref{eq:SOinterforevenoddN}.) The energy gap of the $SO(4)_l$ current backscattering potential can be shown to be finite by identifying $SO(4)_l=SU(2)_l\times SU(2)_l$ and the parafermion decomposition $SU(2)_l=U(1)_{2l}\times\mathbb{Z}_l$ for each of the two $SU(2)_l$ components. The proof of the energy gap for $SO(m\geq5)_n$ is beyond the scope of this paper and will be omitted.

\subsection{Topological Order Examples}
\subsubsection{\texorpdfstring{${\rm SO}(2)_n\times{\rm SO}(n)_2\subseteq{\rm SO}(2n)_1$}{SO(2)n x SO(n)2 = SO(2n)1}: anyon condensation, algebra extension, and \texorpdfstring{$\mathbb{Z}_2$}{Z2} gauge confinement}\label{sec:so2LnsonL2}

We present the coupled wire model for the $SO(n)_2$ and $SO(2)_n$ topological phases, for $n\geq2$. The current back-scattering interactions considered previously in \eqref{eq:SOmnbackscatterings} are not gapping potentials for $SO(2)$. Here we present a replacement for $\mathcal{H}_{\mathrm{inter}/\mathrm{intra}}^{SO(2)_n}$ using a sine-Gordon potential that, together with the current back-scattering $\mathcal{H}_{\mathrm{intra}/\mathrm{inter}}^{SO(n)_2}$ in \eqref{eq:SOmnbackscatterings}, opens a bulk excitation energy gap. Despite the suggestion of the conformal embedding ${\rm SO}(2)_n\times{\rm SO}(n)_2\subseteq{\rm SO}(2n)_1$, the coupled wire model $\mathcal{H}_{\mathrm{intra}}^{SO(2)_n}+\mathcal{H}_{\mathrm{inter}}^{SO(n)_2}$ does {\em not} represent the  non-Abelian $SO(n)_2$ topological order. Instead, as a result of a WZW algebra extension and the anyon condensation of a $\mathbb{Z}_2$ gauge charge, the topological order is of the Abelian $SU(n)_1$ type. The dual model  $\mathcal{H}_{\mathrm{inter}}^{SO(2)_n}+\mathcal{H}_{\mathrm{intra}}^{SO(n)_2}$ only carries the $SO(2)_n=U(1)_{4n}$ topological order if $n$ is odd. It reduces to $U(1)_n$ when $n$ is even. The reduction of topological orders in these special cases was previously reported in Ref.~\cite{Teo_2023}.

We begin with the bosonic $SO(2n)_1$ bundles.
Each decomposes according to the conformal embedding ${\rm SO}(2)_n\times {\rm SO}(n)_2\subseteq{\rm SO}(2n)_1$.
The ${\rm SO}(2)_n$ WZW algebra in the $A$ sector has a single current operator, which is the Majorana fermion bilinear combination 
\begin{align}
\begin{split}
\left[J_{SO(2)_n}\right]=i\sum_{a=1}^n\psi_{a}\psi_{n+a} = \sum_{a=1}^n\partial_\mathsf{x}\tilde\phi_{a},
\end{split}\label{SO2ncurrent}
\end{align} for each chiral sector $\sigma=R,L$ on each bundle $y$, where $e^{i\tilde\phi_a}=(\psi_{a}+i\psi_{n+a})/\sqrt{2}$. 
The current backscattering from \eqref{eq:SOmnbackscatterings} for $SO(2)_n$ consists solely of the density interactions $\partial_\mathsf{x}\tilde\phi^R_a\partial_\mathsf{x}\tilde\phi^L_{a'}$, which only renormalizes the boson velocities and does not generate a mass gap. Instead we consider the sine-Gordon potential (c.f.~\eqref{SO21potential} for ${\rm SO}(2)_1$ and similar construction in ref.~\cite{LimMulliganTeo2022A, PhysRevB.94.165142, Teo_2023}),
\begin{align}
\begin{split}
\mathcal{H}^{{\rm SO}(2)_n}_{y+\epsilon/2}&=
u_{\epsilon} \left[J_{SO(2)_n}\right]^{R\dagger}_{y}\left[J_{SO(2)_n}\right]^L_{y+\epsilon}\\
&\;\;\;\;- u'_{\epsilon} \cos (\vartheta_{y+\epsilon/2}),
\end{split}
\label{eq:SO2n}
\end{align}
where $\epsilon=0$ or 1 represents an intra-bundle or inter-bundle back-scattering interaction, respectively. The sine-Gordon variable is 
\begin{align}\vartheta_{y+\epsilon/2}=q\sum_{a=1}^n\left(\tilde\phi^R_{y,a}-\tilde\phi^L_{y+\epsilon,a}\right),
\end{align}
where $q=1$ if $\epsilon=0$ or $n$ is even, or $q=2$ if $\epsilon=1$ and $n$ is odd. $q$ is the smallest positive number so that $\cos\vartheta_{y+\epsilon/2}$ is an integral combination of electrons. The sine-Gordon potential is RG relevant when $u_{\epsilon}<-v(q^2n^2-4)/[2\pi n(q^2n^2+4)]$. In this case, all $SO(2)_n$ degrees of freedom are gapped. Together with the $SO(n)_2$ current back-scattering potential presented in \eqref{eq:SOmnbackscatterings}, for $n\geq3$, we arrive at the fully gapped topological models \begin{align}\begin{split}
&\mathcal{H}[U(1)_l]=\mathcal{H}_{\mathrm{eff}}+\mathcal{H}^{SO(n)_2}_{\mathrm{intra}}+\sum_y\mathcal{H}_{y+1/2}^{SO(2)_n},\\
&\mathcal{H}[SU(n)_1]=\mathcal{H}_{\mathrm{eff}}+\mathcal{H}^{SO(n)_2}_{\mathrm{inter}}+\sum_y\mathcal{H}_{y}^{SO(2)_n},
\end{split}\label{SO2nSOn2models}\end{align} where $l=n$ when $n$ is even, or $l=4n$ when $n$ is odd.

The topological orders and the WZW CFTs at a boundary are $U(1)_l$ and $SU(n)_1$, respectively. The former is distinct from $SO(2)_n$ when $n$ is even. The latter is different from $SO(n)_2$ for all $n$. We here deduce the topological orders and prove that the result is a consequence of electron locality. We begin with the first model in \eqref{SO2nSOn2models}. When $n$ is even, the primitive local boson on the edge CFT is $\exp(i\sum_{a=1}^n\tilde\phi_a)$. It is an integral product of electrons because it is an even product of the Dirac fermions $\tilde{d}_a=(\psi_a+i\psi_{n+a})/\sqrt{2}=e^{i\tilde\phi_a}$. The local boson has spin $h=n/2$, which sets the level of the $U(1)$ CFT to be $l=n$. When $n$ is odd, the vertex operator $\exp(i\sum_{a=1}^n\tilde\phi_a)$ is now an odd product of fermions and is therefore fractional. The primitive local boson is $\exp(2 i \sum_{a=1}^n\tilde\phi_a)$ instead. The CFT is $U(1)_{4n}$, which is identical to $SO(2)_n$. The distinction between the even and odd $n$ cases depends on the locality of the vertex operator $\exp(i\sum_{a=1}^n\tilde\phi_a)$. Its anyon condensation~\cite{PhysRevB.79.045316}, when $n$ is even, reduces the topological order from $SO(2)_n$ to $U(1)_n$.

Next, we move on to the second model in \eqref{SO2nSOn2models}. The KM current algebra of the edge chiral WZW CFT consists of the $SO(n)_2$ currents $J_{pq}=i\psi_p\psi_q+i\psi_{n+p}\psi_{n+q}$, for $1\leq p<q\leq n$. They are local operators because they are even products of Majorana fermions. In addition, the following fermion bilinear are also local: \begin{align}S_M=i\sum_{a,b=1}^n\psi_aM_{ab}\psi_{n+b},\label{SOnsymm}\end{align} where $M=(M_{ab})_{n\times n}$ is a real symmetric matrix. The simplest examples of such combinations are $S_{ab}=i\psi_a\psi_{n+b}+i\psi_b\psi_{n+a}$, where $1\leq a\leq b\leq n$. The $J$ and $S$ fields obey the operator product expansions (including only singular terms)
\begin{widetext}
\begin{align}\begin{split}
J_{SO(2)_n}(z)S_M(w)&=\frac{\mathrm{Tr}(M)}{(z-w)^2}+\ldots,\quad
J_{pq}(z)S_M(w)=\frac{i}{z-w}S_{[A_{pq},M]}(w)+\ldots,\\
S_M(z)S_{M'}(w)&=\frac{\mathrm{Tr}(MM')}{(z-w)^2}-\frac{i}{z-w}\sum_{1\leq p<q\leq n}[M,M']_{pq}J_{pq}(w)+\ldots,
\end{split}\label{JSOPE}\end{align}\end{widetext}
where $A_{pq}={\bf e}_p{\bf e}_q^T-{\bf e}_q{\bf e}_p^T$ and ${\bf e}_p^T=(0,\ldots,1,\ldots,0)$ is the $n$-dimensional unit vector whose $p^{\mathrm{th}}$ entry is non-zero.

There are two differences in symmetries between $J_{pq}$ and $S_M$. First, $J_{pq}=-J_{qp}$ is anti-symmetric, whereas $S_{ab}=S_{ba}$ is symmetric. Second, while both $J$ and $S$ are even under the internal $\mathbb{Z}_2$ symmetry \eqref{eq:SOinternalZ2}, they have opposite parities under the following global $\mathbb{Z}_2$ symmetry:  
\begin{align}\mathbb{Z}_2:\quad\psi_a\to\psi_a,\quad\psi_{n+a}\to-\psi_{n+a},\label{globalZ2}\end{align} for $a=1,\ldots,n$. $J_{pq}$ is even, while $S_M$ is odd under \eqref{globalZ2}. The diagonal symmetric combination $S_{\mathbb{1}}=\sum_{a=1}^nS_{aa}/2$ is identical to the $SO(2)_n$ current \eqref{SO2ncurrent}, where $\mathbb{1}$ is the $n\times n$ identity matrix. Subtracting this diagonal component, traceless symmetric combinations
span the primary field super-selection sector, 
\begin{align}[S]=\mathrm{span}\left\{S_M:M\in\mathbb{R}^{n\times n},M^T=M,\mathrm{Tr}(M)=0\right\},
\label{eq:localprimariesSOn2}
\end{align} that irreducibly represents $SO(n)_2$ and has non-singular operator product expansions with $J_{SO(2)_n}$. Since $[S]$ decouples from the $SO(2)_n$, it is unaffected by the sine-Gordon potential \eqref{eq:SO2n} and remains gapless on the edge. The primary fields in $[S]$ are local and, according to \eqref{JSOPE}, they extend the $SO(n)_2$ KM algebra on the edge to $SU(n)_1=SO(n)_2\oplus[S]$.

In the bulk, the anyon class corresponding to $[S]$ anyon condenses~\cite{PhysRevB.79.045316} because fields in $[S]$ are local and should belong in the trivial vacuum class. Consequently, the anyon classes in $SO(n)_2$ that carry a $\mathbb{Z}_2$ flux according to the $\mathbb{Z}_2$ symmetry \eqref{globalZ2} and have a $\pi$ monodromy braiding phase with $[S]$ are confined. This reduces the non-Abelian $SO(n)_2$ topological order to the Abelian $SU(n)_1$ order. We notice in passing that a coupled-wire model with the $SO(n)_2$ topological order can be constructed by starting with emergent Majorana fermion bundles with the $\mathbb{Z}_2$ symmetry \eqref{globalZ2} being an internal symmetry rather than a global one. In this case, each bundle carries the $[SO(n)_1]^2$ WZW CFT instead of $SO(2n)_1$, and the non-Abelian topological model is constructed based on the coset decomposition $[SO(n)_1]^2=SO(n)_2\times[SO(n)_1]^2/SO(n)_2$ instead of the level-rank duality $SO(2n)_1\supseteq SO(2)_n\times SO(n)_2$. Such a construction has already been done in Ref.~\cite{Teo_2023} and will not be repeated here.

\subsubsection{Common primary fields and anyon excitations}
\label{ss:somnchain}
We present some common primary field super-selection sectors that generically appear in the $SO(m)_n$ WZW CFT. We discuss how they arise from branching rules of primary field sectors in $SO(mn)_1\supseteq SO(m)_n\times SO(n)_m$, and subsequently demonstrate how their corresponding anyon excitations can be created in the topological bulk. We begin with the fermion sector $[\psi]$ in $SO(m)_1$. It is spanned by the $m$ Majorana fermions $\psi_{1,\ldots,m}$ and it forms the vector representation of $SO(m)$. For $SO(m)_n$ with general level $n\geq1$, we label the primary field sector associating with the vector representation by $[V]=\mathrm{span}\{V_a:a=1,\ldots,m\}$. The quadratic Casimir operator of the vector representation of $SO(m)$ is $C_V=m-1$. Therefore, the conformal scaling dimension of $[V]$ of $SO(m)_n$ is $h_V=\frac{C_V/2}{m+n-2}=\frac{m-1}{2(m+n-2)}$. For $m=2$, the vector representation is reducible and it decomposes into $[V]=[e^2]\oplus[e^{-2}]$, where $e^{\pm2}=e^{\pm i\sum_{a=1}^n\tilde\phi_a/n}$ is the spin $h=\frac{1}{2n}$ vertex field in $SO(2)_n=U(1)_{4n}$. For $m\geq3$, the vector representation is irreducible. For unit level $n=1$, the vector $[\psi]=[V]$ is a spin-$1/2$ fermion and obeys the Abelian fusion rule $[\psi]\times[\psi]=1$ because all fermion bilinear combinations are local $SO(m)_1$ currents. 

For higher levels $n\geq2$, $[V]$ in $SO(m)_n$ obeys the fusion rule $[V]\times[V]=1+[S]+[A]$, where $[S]$ is the traceless symmetric 2-tensor representation and $[A]$ is the anti-symmetric 2-tensor representation of $SO(m)$. $[S]$ has dimension $\dim[S]=m(m+1)/2-1$ and quadratic Casimir $C_S=2m$. In $SO(m)_n$, it has conformal scaling dimension $h_S=m/(m+n-2)$. $[A]$ is identical to the adjoint representation. It has dimension $\dim[A]=m(m-1)/2$, quadratic Casimir $C_A=2(m-2)$, and scaling dimension $h_A=(m-2)/(m+n-2)$ in $SO(m)_n$. For $m=2$, $[S]$ is reducible and decomposes as $[S]=[e^4]\oplus[e^{-4}]$, where $e^{\pm4}$ is the vertex operator $=e^{\pm i2\sum_{a=1}^n\tilde\phi_a/n}$ of $SO(2)_n$. $[A]$ rotates trivially under $SO(2)$ and is identical to the vacuum sector 1. 
Except when $m=4$, for which $[A]$ decomposes into the two spin-$j=1$ isoclinic rotations of $SO(4)=SU(2)\times SU(2)$, $[S]$ and $[A]$ are irreducible representations of $SO(m)$ when $m\geq3$. 

In the conformal embedding $SO(mn)_1\supseteq SO(m)_n\times SO(n)_m$, we have the branching rules (as vector spaces of current operators) \begin{align}\begin{split}SO(mn)_1&=SO(m)_n\oplus SO(n)_m\\&\quad\oplus\left([S]_{SO(m)_n}\times[A]_{SO(n)_m}\right)\\&\quad\oplus\left([A]_{SO(m)_n}\times[S]_{SO(n)_m}\right).\end{split}\label{SOmnbranching}\end{align} This means the current operators in $SO(mn)_1$ that are outside of $SO(m)_n$ and $SO(n)_m$ are combinations of tensor products between the traceless symmetric and anti-symmetric primary field sectors of $SO(m)_n$ and $SO(n)_m$. For instance, there are $mn(mn-1)/2$ fields on both sides of \eqref{SOmnbranching}. Moreover, the conformal scaling dimensions of products in the second and third line of \eqref{SOmnbranching} add up to \begin{align}\frac{m}{m+n-2}+\frac{n-2}{m+n-2}=\frac{m-2}{m+n-2}+\frac{n}{m+n-2}=1.\end{align} In the particular case when $n=2$, $[A]_{SO(2)_m}=1$ is trivial. Therefore, $[S]_{SO(m)_2}$ belongs in the $SO(2m)_1$ WZW algebra. It is local and extends $SO(n)_2$ to $SU(n)_1$. Level $n=2$ is the only situation where WZW algebra extension occurs. When $n\geq3$, the branching rule \eqref{SOmnbranching} shows there cannot be local spin-1 primary fields in $SO(m)_n$ that can extend the WZW algebra, and therefore the $SO(m)_n$ CFT on the edge as well as the bulk $SO(m)_n$ topological order persists.

Similar to the Fibonacci anyon creation by \eqref{FibonacciString} (see figure~\ref{FibonacciWire}) in $SU(2)_3$, an anyon pair belonging to the vector class $[V]$ can be created by acting on the ground state of the $SO(m)_n$ model with the string operator, \begin{align}\mathcal{S}=\prod_{y=y_1}^{y_2}J^L_y(\mathsf{x})J^R_y(\mathsf{x})^\dagger,\label{SOmnstring}\end{align} where every $J^\sigma_y$ is chosen to belong in either the second or the third component of the branching decomposition \eqref{SOmnbranching}. If the $J$'s belong in the second component in \eqref{SOmnbranching}, the  products $\left\langle\left(A_{SO(n)_m}\right)^L_y\left(A_{SO(n)_m}\right)^R_y\right\rangle$ and $\left\langle\left(S_{SO(m)_n}\right)^R_y\left(S_{SO(m)_n}\right)^L_{y+1}\right\rangle$ are pinned to their ground state expectation values at low energy. Dangling modes from $\left(S_{SO(m)_n}\right)^L_{y_1}$ and $\left(S_{SO(m)_n}\right)^R_{y_2}$ are left behind on the two ends of the string, creating the anyon pair of the $[S]$ class. If the currents $J$ in the operator string \eqref{SOmnstring} belong in the third component in \eqref{SOmnbranching}, the string creates an anyon pair of class $[A]$ on both ends.

The vector representations in the conformal embedding $SO(mn)_1\supseteq SO(m)_n\times SO(n)_m$ follow the single-channel branching rule \begin{align}[\psi]_{SO(mn)_1}=[V]_{SO(m)_n}\times[V]_{SO(n)_m}.\label{SOmnvectorbranching}\end{align} There are $mn$ fermions on both sides, and the scaling dimensions of the tensor product add up to \begin{align}\frac{m-1}{2(m+n-2)}+\frac{n-1}{2(m+n-2)}=\frac{1}{2}.\end{align} Following \eqref{Emstring}, the string of conjugate Majorana pairs, \begin{align}\mathcal{S}_\psi=\prod_{y=y_1}^{y_2}\psi_y^L(\mathsf{x})\psi^R_y(\mathsf{x}),
\label{eq:mjstringsomn}
\end{align} is an integral combination of electron operators because $\psi^L_y\psi^R_y$ on each bundle is local. When applied to the ground state of $SO(m)_n$, the bulk of the string takes ground state expectation values according to the branching rule \eqref{SOmnvectorbranching}, but the dangling modes $\left(V_{SO(m)_n}\right)^L_{y_1}$ and $\left(V_{SO(m)_n}\right)^R_{y_2}$ are left behind, creating the anyon pair belonging to class $[V]$.

The WZW CFT also carries spinor primary fields, which are representations of the double cover $\mathrm{Spin}(m)$. However, the properties of these spinor fields are out of the scope of this paper and will be omitted.

\section{\texorpdfstring{$C_r$}{C} Series and Symplectic Fermions}\label{Sec:CseriesTheSymplecticFermions}

In this section, we construct topological models with symplectic $Sp(2n)$ edge-state symmetry. Ref.~\cite{PhysRevLett.120.036801} studied surface topological order using the  
splitting of ${\rm SO}(4n^2)_1$ into two copies of ${\rm Sp}(2n)_n$, where the level and the rank of the algebra in each copy are identical.
We will construct $Sp(2m)_n$ spin liquids and topological superconductor models with arbitrary level and rank. 
This is achieved with the conformal embedding ${\rm Sp}(2m)_n\times {\rm Sp}(2n)_m \subseteq {\rm SO}(4mn)_1$ using the $SO(4mn)_1$ wires discussed previously in Sec.~\ref{sec:BosoicWiresOfSOFamily}. 

We begin with the matrix representation of the symplectic Lie algebra ${\rm Sp}(2n)$. It 
consists of $n\times n$ matrices $X$ with quaternion entries that satisfy $X+X^\dagger=0$.
The matrices can be written as $X=\sum_\mu X_\mu\mathfrak{q}^\mu$, where $\mu=0,\ldots,3$. $X_\mu$ are $n\times n $ real matrices and $\mathfrak{q}^{i=1,2,3}$ are quaternions, which can be represented in terms of Pauli matrices as
\begin{align}
\mathfrak{q}^0=\mathbb{1}_{2\times 2},\quad  \mathfrak{q}^1=-i\sigma_x,\quad \mathfrak{q}^2=-i\sigma_y,\quad \mathfrak{q}^3=-i\sigma_z.
\end{align}
The quaternions have the following properties:
\begin{align}
\begin{split}
(\mathfrak{q}^j)^\dagger = -\mathfrak{q}^j,
\quad&
\mathfrak{q}^i\mathfrak{q}^j=
-\delta^{ij}\mathbb{1}_{2\times 2}+
\epsilon^{ijk}\mathfrak{q}^k, 
\\
\mathfrak{q}^1\mathfrak{q}^2\mathfrak{q}^3&=(\mathfrak{q}^j)^2=-\mathbb{1}_{2\times2},
\end{split}
\end{align}
where $i,j,k=1,2,3$, and $\epsilon^{ijk}$ is the anti-symmetric Levi-Civita symbol.

The ${\rm Sp}(2n)$ Lie algebra is spanned by the following matrix generators $X_\mu(pq)$ (see also \cite{Baker2003MatrixGA}):
\begin{align}
[X_{\mu}(pq)]_{rs} = \frac{1}{4(2)^{\delta_{p,q}/2}}\left(\delta^p_r\delta^q_s-\eta_{\mu\mu}\delta^q_r\delta^p_s\right),
\label{eq:SymplecticMetric}
\end{align}
where $r,s=1,\ldots, n$, and $\eta_{\mu\nu}=\mathrm{diag}(1,-1,-1,-1)$ is the Minkowski metric.
For $\mu=0$, there are $n(n-1)/2$ real anti-symmetric matrices $X_0(pq)$, where $1\leq p<q\leq n$.
For $\mu\neq0$, $X_{j=1,2,3}(pq)$ are real symmetric 
matrices, where in this case $1\leq p\leq q\leq n$. There are $3\times n(n+1)/2$ such real symmetric matrices. Together, they generate the $n(2n+1)$ dimensional ${\rm Sp}(2n)$ Lie algebra.

The chiral $Sp(2n)_1$ WZW algebra can be represented by fermion bilinears.
We group $4n$ Majorana fermions $\psi^{r=1,\ldots, n}_\mu$ into $n$ symplectic fermions $\chi^r= \sum_\mu\psi^r_\mu\mathfrak{q}^\mu$. Their complex conjugations are $(\chi^r)^\dagger=\sum_\mu\eta_{\mu\mu}\psi^r_\mu\mathfrak{q}^\mu$.
Conversely, the Majoranas can be obtained from the symplectic fermions using:
\begin{align}
\psi^r_\mu = \frac{1}{2}{\eta_{\mu\nu}}\Tr[\chi^r \mathfrak{q}^\nu].
\label{eq:symplectifermionssp}
\end{align}
We are now ready to describe the current algebra of ${\rm Sp}(2n)_1$. The Hermitian current operators are defined by products of symplectic fermions:
\begin{align}
\begin{split}
J^\mu_{pq}&=\frac{i}{2}\Tr[\chi^\dagger X_\mu(pq) \mathfrak{q}^\mu \chi]
\\
&=\frac{i}{2}\sum_{\lambda,\nu, r,s}\eta_{\lambda\lambda}T^{\lambda\mu\nu}\psi^r_{\lambda} [X_\mu(pq)]_{rs} \psi^s_\nu.
\end{split}
\label{eq:SpCurrentOperator}
\end{align} 
We refer to
$T^{\lambda\mu\nu}\equiv\Tr\left[\mathfrak{q}^\lambda
\mathfrak{q}^\mu\mathfrak{q}^\nu\right]$ 
as the trace of the product of three quaternions, or simply the quaternion trace. It has the explicit form: 
\begin{align}
\begin{split}
T^{\lambda\mu\nu} &= 
2\eta^{\lambda\mu}\delta^{\nu 0}+
2\eta^{\mu\nu}\delta^{\lambda 0}+
2\eta^{\nu\lambda}\delta^{\mu 0}\\&\;\;\;\
-4\delta^{\lambda 0}\delta^{\mu 0}\delta^{\nu 0}
-2\epsilon^{0\lambda\mu\nu}
\end{split}
\end{align}
with Greek letters $\lambda,\mu,\nu = 0,1,2,3$ and lowercase letters $i,j,k = 1,2,3$. $\epsilon^{0\lambda\mu\nu}$ is the four dimensional the Levi-Civita symbol.

\subsection{Conformal Embedding and Descendant Topological Order}

\subsubsection{Symmetry embedding}
We now consider the conformal embedding ${\rm Sp}(2m)_n\times {\rm Sp}(2n)_m \subseteq {\rm SO}(4mn)_1$.
We refer to ${\rm Sp}(2m)_n$ as the $A$ sector and ${\rm Sp}(2n)_m$ as the $B$ sector.
The $m\times m$ matrix generator 
$X_\mu(pq)$, defined in \eqref{eq:SymplecticMetric}, of ${\rm Sp}(2m)_n$ can be embedded into ${\rm SO}(4mn)$ by taking
\begin{align}
\begin{split}
\left[X^A_\mu(pq)\right]_{rs}
&\equiv\left[X_\mu(pq)\otimes \mathbb{1}_n\right]_{rs}
\\
&=\sum_{a,b,c,d}
\left[X_\mu(pq)\right]_{ab}\delta_{cd}\delta^{(a-1)n+c}_r\delta^{(b-1)n+d}_s\\
&=
\frac{1}{4(2)^{\delta_{pq}/2}}\sum_{c=1}^n
\Big(\delta^{p'}_r\delta^{q'}_s-\eta^{\mu\mu}\delta^{q'}_r\delta^{p'}_s\Big).
\end{split}
\label{eq:SymplecticMatrixRepInEmbedding}
\end{align}
We denote $p'\equiv(p-1)n+c$ and $q'\equiv(q-1)n+c$,
where the indices are subject to the following conditions:
$r,s=1,\ldots, mn$, $a, b= 1,\ldots, m$, and $c, d=1,\ldots, n$.
Additionally, the indices $p$ and $q$ are constrained as follows:
$1\leq p<q\leq m$ when $\mu=0$; and $1\leq p\leq q\leq m$ when $\mu=1,2,3$. 
For each chiral sector $\sigma$, the current operators of the $A$ sector are a sum of $n$ $Sp(2m)_1$ currents:
\begin{align}
\begin{split}
J^{A,\mu}_{pq}&=
\frac{i}{2}\Tr\left[\chi^\dagger X^A_\mu(pq)\mathfrak{q}^\mu\chi\right]\\
&=\frac{i}{2}\sum_{\lambda,\nu,r,s}\eta_{\lambda\lambda}
T^{\lambda\mu\nu}
\psi^r_{\lambda}\left[X^A_\mu(pq)\right]_{rs}
\psi^s_{\nu}.
\end{split}
\label{eq:SpAcurrent}
\end{align}
Similarly for the $B$ sector, 
given any $n\times n$ anti-Hermitian matrix $X_\mu(pq)$ that generates ${\rm Sp}(2n)$, we define
\begin{align}
\begin{split}
\left[X^B_\mu(pq)\right]_{rs}
&\equiv\left[
\mathbb{1}_m\otimes X_\mu(pq)\right]_{rs}
\\
&= \sum_{a,b,c,d}
\delta_{ab}\left[X_\mu(pq)\right]_{cd}\delta^{(a-1)n+c}_r\delta^{(b-1)n+d}_s\\
&=
\frac{1}{4(2)^{\delta_{pq}/2}}
\sum_{a=1}^m
\Big(\delta^{p'}_r\delta^{q'}_s-\eta^{\mu\mu}\delta^{q'}_r\delta^{p'}_s\Big),
\end{split}
\end{align} with
$p'=(a-1)n+p$ and $q'=(a-1)n+q$.
The indices:  
$r,s = 1,\ldots, mn$, $a, b=1,\ldots, m$, and $c,d=1\ldots, n$. Similar to the $A$ sector, the indices $p$ and $q$ obey the constraints:
$1\leq p<q\leq n$ when $\mu=0$, and $1\leq p\leq q\leq n$ when $\mu=1,2,3$.
The current operators of the $B$ sector are 
\begin{align}
\begin{split}
J^{B,\mu}_{pq}&=
\frac{i}{2}\Tr\left[\chi^\dagger X^B_\mu(pq)\mathfrak{q}^\mu\chi\right]\\
&=\frac{i}{2}\sum_{\lambda,\nu,r,s}\eta_{\lambda\lambda}
T^{\lambda\mu\nu}
\psi^r_{\lambda}\left[X^B_\mu(pq)\right]_{rs}
\psi^s_{\nu}.
\end{split} 
\label{eq:SpBcurrent}
\end{align}

\subsubsection{Descendant states}\label{Sec:SymplecticWires}
We begin with the $SO(4 m n)_1$ wires discussed in Sec.\ref{sec:BosoicWiresOfSOFamily}. 
The $4mn$ chiral Majorana fermions $\psi^\sigma_{y,j=1,\ldots, 4mn}$ in each
${\rm SO}(4mn)_1$ wire are relabeled as
$\psi^{\sigma,r}_{y,\mu} = \psi^\sigma_{y, j=mn\mu+r}$,
with $\mu=0,1,2,3$ and $r=1,\ldots, mn$.
For a given bundle $y$, 
according to the discussion in Sec.~\ref{sec:BosoicWiresOfSOFamily}, an operator is local
invariant under the internal $\mathbb{Z}_2$ symmetry in \eqref{eq:SOinternalZ2}.
Consequently, the current operators, given in equations \eqref{eq:SpCurrentOperator}, \eqref{eq:SpAcurrent}, and \eqref{eq:SpBcurrent}, are local.

We now use these current operators to build the
topological states with ${\rm Sp}(2m)_n$ or ${\rm Sp}(2n)_m$ edge-state symmetry. 
We recall that the charge sector has previously been gapped out through Umklapp backscattering or superconducting pairing described by \eqref{eq:SOfermionicgaps}. Extra Majorana degrees of freedom were removed using fermions backscattering in \eqref{eq:ExtraMajoranaDOF}. Leaving us with the neutral ${\rm SO}(N=4mn)_1$ theory and $4mn$ Majoranas.
The microscopic $Sp(2m)_n$ Hamiltonian is
\begin{align}
\mathcal{H}\left[{\rm Sp}(2m)_n\right]=
\mathcal{H}_0+\mathcal{H}^{{\rm Sp}(2m)_n}_{\rm inter}+\mathcal{H}^{{\rm Sp}(2n)_m}_{\rm intra},
\label{Hspmodel}\end{align}
where the current-current backscattering terms are 
\begin{align}
\begin{split}
&\mathcal{H}^{{\rm Sp}(2m)_n}_{\rm inter}\\
&=u_{\rm inter}
\sum_{y}\sum_{\mu,p,q}
\left(J^{{\rm Sp}(2m)_n}\right)^{R,\mu}_{y,pq}
\left(J^{{\rm Sp}(2m)_n}\right)^{L,\mu}_{y+1,pq},
\end{split}
\label{eq:InterwireSpmnHamiltonian}
\end{align}
and 
\begin{align}
\begin{split}
&\mathcal{H}^{{\rm Sp}(2n)_m}_{\rm intra}\\
&=u_{\rm intra}\sum_{y}\sum_{\mu,p,q}
\left(J^{{\rm Sp}(2n)_m}\right)^{R,\mu}_{y,pq}
\left(J^{{\rm Sp}(2n)_m}\right)^{L,\mu}_{y,pq}.
\end{split}
\label{eq:IntrawireSpnmHamiltonian}
\end{align}
We postulate that the KM current backscatterings in \eqref{eq:InterwireSpmnHamiltonian} and \eqref{eq:IntrawireSpnmHamiltonian} generate finite bulk excitation energy gap.

\subsection{Topological Order Examples}

In this subsection, we discuss examples of the resulting $Sp(2m)_n$ topological order.
The simplest conformal embedding in the ${\rm SO}(4mn)_1\supseteq{\rm Sp}(2m)_n\times{\rm SO}(2n)_m$ family is ${\rm SO}(4)_1={\rm Sp}(2)_1\times {\rm Sp}(2)_1$, where ${\rm Sp}(2)_1={\rm SU}(2)_1$. Using this embedding, the corresponding topological phase described by \eqref{Hspmodel} has the Abelian topological order ${\rm SU}(2)_1$ that supports semionic quasiparticle excitations. In the remainder of this section, we will examine two sets of examples of non-Abelian topological phases using the symplectic algebra decompositions, ${\rm SO}(8)_{1} \supseteq {\rm Sp}(2)_{2} \times {\rm Sp}(4)_{1}$ and ${\rm SO}(16)_1 \supseteq {\rm Sp}(2)_4 \times {\rm Sp}(8)_1$.

\subsubsection{\texorpdfstring{SO(8)$_1 \supseteq$ Sp(2)$_2 \times$ Sp(4)$_1$}{SO(8)_{1} \supseteq Sp(2)_{2} \times Sp(4)_{1}}}

We begin by considering the conformal embedding of the $3$-dimensional ${\rm Sp}(2)_2$ algebra and the $10$-dimensional ${\rm Sp}(4)_1$ algebra into the $28$-dimensional ${\rm SO}(8)_1$. The symplectic algebras are isomorphic to orthogonal ones, ${\rm Sp}(2)_2={\rm SU}(2)_2={\rm SO}(3)_1$ and ${\rm Sp}(4)_1={\rm SO}(5)_1$. Both carry an Ising-like topological order with anyon classes $1,[f],[\sigma]$. The Ising twist field in ${\rm SO}(N)_1$, for odd $N$, has conformal scaling dimension $N/16$. It obeys the fusion rule $[\sigma]\times[\sigma]=1+[f]$, where $[f]$ is the spin-1/2 fermion of $SO(N)_1$.

Although the $[f]$ super-selection sector forms a vector representation of the orthogonal algebra ${\rm SO}(N)_1$, in the context of the symplectic algebra ${\rm Sp}(2)_2$, it is the adjoint representation, $[f]_{SO(3)_1}=[A]_{Sp(2)_2}$, which is also the anti-symmetric representation consisting of skew-Hermitian quaternion $2$-tensors. In the symplectic algebra ${\rm Sp}(4)_1$, the fermion is the traceless symmetric representation $[f]_{SO(5)_1}=[S]_{Sp(4)_1}$ consisting of traceless-Hermitian quaternion $2$-tensors. The Ising twist fields for $N=3,5$ are vector representations of the symplectic algebras, $[\sigma]_{SO(3)_1}=[V]_{Sp(2)_2}$ and $[\sigma]_{SO(5)_1}=[V]_{Sp(4)_1}$.

Fermion pair products in $[f]_{{\rm SO(3)_1}}\times[f]_{{\rm SO}(5)_1}=[A]_{{\rm Sp}(2)_2}\times [S]_{{\rm Sp}(4)_1}$ are local bosons in ${\rm SO}(8)_1$ with conformal scaling dimension $h=1/2+1/2=1$. They account for the remaining ${\rm SO}(8)_1$ currents outside of $SO(3)_1$ and $SO(5)_1$, and complete the branching rule of ${\rm SO}(8)_1$ currents:
\begin{align}
\begin{split}
{\rm SO}(8)_1&={\rm SO}(3)_1\oplus {\rm SO}(5)_1 
\oplus \left([f]_{{\rm SO(3)_1}}\times[f]_{{\rm SO}(5)_1}\right)\\
&=Sp(2)_2\oplus Sp(4)_1 \oplus \left(
[A]_{Sp(2)_2}\times[S]_{Sp(4)_1}\right).
\end{split}
\label{eq:SOdimensionalityIdentity}
\end{align}
In Sec. \ref{ss:IVB3}, we will explore how certain anyon excitations can emerge from the branching rules and how they can be generated within the topological bulk.

In the symplectic decomposition of ${\rm SO}(8)_1$, it may be more natural to associate the fermion fields in $[A]_{Sp(2)_2}\oplus[S]_{Sp(4)_1}$ as one of the two $SO(8)$ spinors, say $[s_+]_{SO(8)_1}$. So that fields in the $SO(8)$ vector representation $[V]_{SO(8)_1}$ (as well as those in the opposite spinor sector $[s_-]_{SO(8)_1}$) branch into pair products of the Ising twist fields of the two sectors in $[\sigma]_{SO(3)_1}\times[\sigma]_{SO(5)_1}=[V]_{Sp(2)_2}\times[V]_{Sp(4)_1}$.

\subsubsection{\texorpdfstring{${\rm SO}(16)_1 \supseteq {\rm Sp}(2)_4 \times {\rm Sp}(8)_1$}
{{\rm SO}(16)_{1} \supseteq {\rm Sp}(2)_{4} \times {\rm Sp}(8)_{1}}}
\label{sec:sp2L4sp8L1}

The two symplectic subalgebras of $SO(16)_1$ are identical to the orbifold CFTs: \begin{align}{\rm Sp}(2)_4=\frac{{\rm SU}(3)_1}{\mathbb{Z}_2},\quad{\rm Sp}(8)_1=\frac{(E_6)_1}{\mathbb{Z}_2}.\end{align}
The orbifold structure arises because there are non-local spin-1 boson fields that extend ${\rm Sp}(2)_2$ to ${\rm SU}(3)_1$, and ${\rm Sp}(8)_1$ to $(E_6)_1$. The bosons are individually fractional, but boson pairs are local. Each boson carries a $\mathbb{Z}_2$ gauge charge and have non-trivial $\pi$ monodromy when braiding around certain non-Abelian anyons in ${\rm Sp}(2)_2$ and ${\rm Sp}(8)_1$ that carry a $\mathbb{Z}_2$ gauge flux.

We demonstrate using ${\rm Sp}(2)_4$.
Given the isomorphism between ${\rm Sp}(2)_4$ and ${\rm SO}(3)_2$, we can rewrite the ${\rm Sp}(2)_4$ current operators in a ${\rm SO}(3)_2$ form similar to those presented in Sec.~\ref{Sec.BDseriesAndTopOrder}. Subsequently, we express the spin-1 boson fields that carry non-trivial $\mathbb{Z}_2$ gauge charge using \eqref{SOnsymm}. We will see that they are fractional. They obey the current OPE \eqref{JSOPE} and extend the ${\rm Sp}(2)_4$ WZW algebra into ${\rm SU}(3)_1$.

First we express the ${\rm Sp}(2)_4$ currents $\left[J_{{\rm Sp}(2)_4}\right]^{\mu=1,2,3}_{p=q=1}$ as a bilinear combination of a new set of Majorana fermions $\tilde\psi_{j=1,\ldots,6}=\cos\tilde\varphi_j$:
\begin{align}
\begin{split}
\left[J_{{\rm Sp}(2)_4}\right]^{\mu=1}
&\sim\cos(\tilde{\varphi}_1)\cos(\tilde{\varphi}_2)+\cos(\tilde{\varphi}_4)\cos(\tilde{\varphi}_5)\\
&\simeq i\tilde{\psi}_1\tilde{\psi}_2+
i\tilde{\psi}_4\tilde{\psi}_5,
\\
\left[J_{{\rm Sp}(2)_4}\right]^{\mu=2}
&\sim\cos(\tilde{\varphi}_1)\cos(\tilde{\varphi}_3)+\cos(\tilde{\varphi}_4)\cos(\tilde{\varphi}_6)
\\
&\simeq i\tilde{\psi}_1\tilde{\psi}_3+
i\tilde{\psi}_4\tilde{\psi}_6,
\\
\left[J_{{\rm Sp}(2)_4}\right]^{\mu=3}
&\sim\cos(\tilde{\varphi}_2)\cos(\tilde{\varphi}_3)+\cos(\tilde{\varphi}_5)\cos(\tilde{\varphi}_6)
\\
&\simeq i\tilde{\psi}_2\tilde{\psi}_3+
i\tilde{\psi}_5\tilde{\psi}_6,
\end{split}
\label{eq:rotatingSp2L4toSO3L2}
\end{align} 
with angle variables $\tilde{\varphi}$ defined by
\begin{align}
\begin{split}
\Tilde{\varphi}_1&=\frac{\phi_1-\phi_3-\phi_5+\phi_7}{2}, \quad 
\Tilde{\varphi}_4=\frac{\phi_2-\phi_4-\phi_6+\phi_8}{2},
\\
\Tilde{\varphi}_2&=\frac{\phi_1-\phi_3+\phi_5-\phi_7}{2}, \quad 
\Tilde{\varphi}_5=\frac{\phi_2-\phi_4+\phi_6-\phi_8}{2},
\\
\Tilde{\varphi}_3&=\frac{\phi_1-\phi_5+\phi_3-\phi_7}{2}, \quad 
\Tilde{\varphi}_6=\frac{\phi_2-\phi_6+\phi_4-\phi_8}{2}.
\label{eq:redefinitionofbosonizedvaraibles}
\end{split}
\end{align}
Here for simplicity, we have temporarily suppressed the $y$ and $\sigma$ indexes.
The notation ``$\simeq$" indicates equivalence up to a normalization factor after refermionization. The current operators now matches that of the ${\rm SO}(3)_2$ in Sec.~\ref{Sec.BDseriesAndTopOrder}.

Following \eqref{SOnsymm} and \eqref{eq:localprimariesSOn2}, the spin-1 boson fields that carry non-trivial $\mathbb{Z}_2$ gauge charge are $\mathcal{Z}_M=i\tilde\psi_aM_{ab}\tilde\psi_{3+b}$, where $M=(M_{ab})_{3\times3}$ is a real traceless symmetric matrix. They span a super-selection sector $[\mathcal{Z}]$ that irreducibly represents the ${\rm SO}(3)_2$ algebra, and they extend the algebra to ${\rm SU}(3)_1={\rm SO}(3)_2\oplus[\mathcal{Z}]$. (See the OPEs in \eqref{JSOPE}.)

The boson fields in the $\mathbb{Z}_2$ gauge charge sector $[\mathcal{Z}]$ are non-local. For example, the fermion pair 
\begin{align}
\begin{split}
i\tilde{\psi}_a\tilde{\psi}_{3+b}&\simeq
\cos(\tilde{\varphi}_a+\tilde{\varphi}_{3+b})+
\cos(\tilde{\varphi}_a-\tilde{\varphi}_{3+b}).
\end{split}
\end{align}
consists of fractional vertex operators with half-integral angle variable combinations $\tilde{\varphi}_a\pm\tilde{\varphi}_{3+b}$ from \eqref{eq:redefinitionofbosonizedvaraibles}. Therefore
field operators in $[\mathcal{Z}]$ are non-local and do not condense in the anyon condensation sense~\cite{PhysRevB.79.045316}. Consequently, the $\mathbb{Z}_2$ orbifold structure of ${\rm Sp}(2)_4$ remains. 

A similar $\mathbb{Z}_2$ orbifold structure applies to ${\rm Sp}(8)_1=(E_6)_1/\mathbb{Z}_2$. There is a non-local super-selection sector $[\mathcal{Z}]$ of spin-1 boson fields that irreducibly represent the ${\rm Sp}(8)_1$ WZW algebra and extend it to $(E_6)_1={\rm Sp}(8)_1\oplus[\mathcal{Z}]$. However, only the currents in ${\rm Sp}(8)_1$ are local. Since the $\mathbb{Z}_2$ charge sector $[\mathcal{Z}]$ is fractional, it does not anyon condense. The non-Abelian ${\rm Sp}(8)_1$ anyons that carry a non-trivial $\mathbb{Z}_2$ flux component remain deconfined.

\subsubsection{Common primary fields and anyon excitations}\label{ss:IVB3}

In closing, we present some common features in the ${\rm Sp}(m)_n$ topological order and some branching rules in the conformal embedding ${\rm SO}(4mn)_1 \supseteq {\rm Sp}(2m)_n \times {\rm Sp}(2n)_m$.

In the simplest case, ${\rm SO}(4)_1={\rm Sp}(2)_1\times {\rm Sp}(2)_1$, because ${\rm Sp}(2)_1$ is identical to ${\rm SU}(2)_1$, its
fusion group of Abelian anyons is the vacuum 1 and the semion $[V]$, which forms the 
vector representation of ${\rm Sp}(2)_1$. It obeys the fusion rule $[V]\times[V]=1$. The vector representation of ${\rm Sp}(2)_n$, at a general level $n\geq1$, is identical to the $j=1/2$ representation of ${\rm SU}(2)_n$. For $n\geq2$, it obeys the fusion rule $[V]\times[V]=1+[A]$, where $[A]$ is the adjoint representation of ${\rm Sp}(2)_n$ that is also the $j=1$ representation of ${\rm SU}(2)_n$.

We presented two level cases ${\rm Sp}(4)_1={\rm SO}(5)_1$ and ${\rm Sp}(8)_1=(E_6)_1/\mathbb{Z}_2$ above. We have seen that the vector representation $[V]$ of ${\rm Sp}(4)_1$ is identical to the Ising spinor representation of ${\rm SO}(5)_1$ that carries spin $h_V=5/16$. The vector representation $[V]$ of ${\rm Sp}(8)_1$ is one of the two anyon classes in the $(E_6)_1/\mathbb{Z}_2$ orbifold that carries a non-trivial $\mathbb{Z}_2$ flux. It carries spin $h_V=3/8$. For general ${\rm Sp}(2m)_1$, the vector obeys the fusion rule $[V]\times[V]=1+[S]$, where $[S]$ is the traceless symmetric representation. For example, we have seem that $[S]_{{\rm Sp}(4)_1}$ is identical to the spin $1/2$ fermion $[f]_{{\rm SO}(5)_1}$. $[S]_{{\rm Sp}(8)_1}$ has spin $2/3$ and is associated with the conjugate pair of Abelian anyons $\mathcal{E}$ and $\mathcal{E}^\dagger$ with the same spin in $(E_6)_1$~\cite{LimMulliganTeo2022A} that are related by the $\mathbb{Z}_2$ symmetry. 

For general $Sp(2m)$, we label the primary field sector corresponding to the vector representation $[V]={\rm span}\left\{V_a: a=1,\ldots, 2m\right\}$. The quadratic Casimir of $[V]$ is $C_V=2(2m+1)/4$. The dual Coxeter of $Sp(2m)$ is $g=m+1$. Therefore, the scaling dimension of $[V]$ in $Sp(2m)_n$ is $h_V=\frac{C_V/2}{m+n+1}=\frac{2m+1}{4(m+n+1)}$.
The vector representation of $Sp(2m)$ is irreducible for $m\geq 1$ \cite{francesco2012conformal}. For higher level $n\geq 2$ and rank $m\geq2$, the vector $[V]$ in $Sp(2m)_n$ obeys the fusion rule 
$[V]\times[V] = 1 + [A] +[S]$. 
The adjoint representation $[A]$ of $Sp(2m)$ is the skew-Hermitian quaternion $2$-tensor with dimension $\dim[A]=2m^2+m$ and quadratic Casimir $C_A=2(m+1)$. It has conformal scaling dimension $h_{A}=(m+1)/(m+n+1)$ in $Sp(2m)_n$.
The traceless Hermitian quaternion $2$-tensor $[S]$ has dimension $\dim[S]=m(2m-1)-1$ and quadratic Casimir $C_S=2m$.
It has scaling dimension $h_{S}=m/(m+n+1)$ in $Sp(2m)_n$.

Generically, when the ranks $m,n\geq2$, the ${\rm SO}(4mn)_1$ currents in the conformal embedding ${\rm SO}(4mn)_1\supseteq{\rm Sp}(2m)_n\times{\rm Sp}(2n)_m$ splits according to the branching rule
\begin{align}
\begin{split}
SO(4mn)_1 &= Sp(2m)_n\oplus Sp(2n)_m\\
&\oplus \left(
[S]_{Sp(2m)_n}\times[A]_{Sp(2n)_m}
\right)\\
&\oplus \left(
[A]_{Sp(2m)_n}\times[S]_{Sp(2n)_m}
\right).
\end{split}
\label{eq:branchingruleSpmn}
\end{align}
There are $2mn(4mn-1)$ fields on both sides of \eqref{eq:branchingruleSpmn}.
The $(2m+1)(2n+1)(2mn-m-n)$ 
current operators of ${\rm SO}(4mn)_1$
that lie outside ${\rm Sp}(2m)_n$ and ${\rm Sp}(n)_m$ are linear 
combinations of tensor products of the the traceless symmetric and adjoint representations $[S]$ and $[A]$ of ${\rm Sp}(2m)_n$ and ${\rm Sp}(2n)_m$. The conformal scaling dimensions of the tensor products in \eqref{eq:branchingruleSpmn} sum up to
\begin{align}
\begin{split}
\frac{m+1}{m+n+1}+\frac{n}{m+n+1}
=\frac{n+1}{m+n+1}+\frac{m}{m+n+1}=1.
\end{split}
\end{align}

In the lower rank cases where $m=1$ and $n\geq 2$, $Sp(2)_n$ does not admit the the traceless symmetric representation $[S]$.
The branching rule in \eqref{eq:branchingruleSpmn} becomes 
\begin{align}
SO(4n)_1 = Sp(2)_n\oplus Sp(2n)_1 \oplus
\left([A]_{Sp(2)_n}\times[S]_{Sp(2n)_1}\right),
\end{align} 
as demonstrated in the example of \eqref{eq:SOdimensionalityIdentity}. 
The dimension difference between 
${\rm SO}(4n)_1$ and ${\rm Sp}(2)_n\times {\rm Sp}(2m)_1$ equals the number of independent tensor product fields in $[A]_{Sp(2)_n}\times[S]_{Sp(2n)_1}$, each has the
conformal scaling dimension
\begin{align}
h_{A}^{Sp(2)_n} +h_{S}^{Sp(2n)_1} =
\frac{2}{n+2} + \frac{n}{n+2} = 1.
\end{align}

Using a similar Wilson string construction in Sec.~\ref{ss:somnchain}, a pair of anyons in class $[A]$ or in class $[S]$ in the $Sp(m)_n$ topological model can be generated by applying to the ground state a string of electron operator, similar to that in \eqref{SOmnstring}. Each current operator $J^\sigma_y$ in the string must be selected from either the second or third component of the branching rules in \eqref{eq:branchingruleSpmn}. For example, if the $J$ operators are in the second component of \eqref{eq:branchingruleSpmn}, the adjoint pair $\langle A_{Sp(2n)_m}^{R,y}A_{Sp(2n)_m}^{L,y}\rangle$ within a wire and the symmetric pair $\langle S_{Sp(2m)_n}^{R,y}S_{Sp(2m)_n}^{L,y+1}\rangle$ between adjacent wire are pinned to their ground state expectation values. This leaves behind unpaired non-local fields from $\left(S_{Sp(2m)_n}\right)^L_{y_1}$ and $\left(S_{Sp(2m)_n}\right)^R_{y_2}$ at the two ends of the string, and leads to the creation of an anyon pair of the $[S]$ class. Conversely, if the currents $J$ belong to the third component in \eqref{eq:branchingruleSpmn}, the string operator generates a pair of anyon excitation of the $[A]$ class.

Furthermore, in the conformal embedding of $SO(4mn)_1\supseteq Sp(2m)_n\times Sp(2n)_m$, the vector representations obey a single-channel branching rule:
\begin{align}
    [\psi]_{SO(4mn)_1}=[V]_{Sp(2m)_n}\times[V]_{Sp(2n)_m},
\label{eq:spmjfbranchingrule}
\end{align}
with $4mn$ fermions on both sides. The conformal scaling dimension of the tensor product sums to
\begin{align}
    \frac{2m+1}{4(m+n+1)}+\frac{2n+1}{4(m+n+1)}=\frac{1}{2}.
\end{align}
The non-chiral fermion pair $\psi^L_y\psi^R_y$ on each wire is a local operator in the wire construction of $SO(4mn)_1$ that is equivalent to a combination of electron operators in low energy. 
Consequently, we can construct a string of electron operator equivalent to conjugate Majorana pairs similar to the one in \eqref{eq:mjstringsomn}. 
When applied to the ground state of the $Sp(2m)_n$ model,
operator pairs $\langle V_{Sp(2n)_m}^{R,y}V_{Sp(2n)_m}^{L,y}\rangle$ and $\langle V_{Sp(2m)_n}^{R,y}V_{Sp(2m)_n}^{L,y+1}\rangle$ in bulk of the string are pinned to their ground state expectation values.
This leaves behind the dangling modes $\left(V_{Sp(2m)_n}\right)^L_{y_1}$ and 
$\left(V_{Sp(2n)_m}\right)^R_{y_2}$
at both ends of the string, creating an anyon pair of the $[V]$ class.

\section{Discussion and Conclusion}\label{Sec:DiscussionAndConclusion}


In this paper, we used exactly solvable coupled-wire Hamiltonians to construct 2D topological phases
with gapless chiral edge states described by WZW CFT of affine classical Lie algebras $A_r={\rm SU}(r+1)$, $B_r={\rm SO}(2r+1)$, $C_r={\rm Sp}(2r)$, $D_r={\rm SO}(2r)$ at arbitrary ranks and levels (except for $SO(n)_2$ which was previously constructed in ref.~\cite{Teo_2023}).
These topological phases include chiral states of quantum Hall liquids in the unitary $A$ class, or topological superconductors and spin liquids in the orthogonal and symplectic classes $B$, $C$ and $D$. The model Hamiltonians consist of many-electron interactions that backscatter the WZW Lie algebra currents. It has been known and proven -- in some cases in this paper and other cases in previous works~\cite{PhysRevB.94.165142, Teo_2023, LimMulliganTeo2022A} -- that the current backscattering interactions introduce a finite energy gap for ${\rm SU}(2)_n={\rm Sp}(2)_n$, ${\rm SU}(m)_1$, ${\rm SO}(m)_1$, ${\rm SO}(m)_2$, ${\rm SO}(3)_n$, ${\rm SO}(4)_n$, ${\rm Sp}(4)_1$, and ${\rm Sp}(8)_1$. 
We postulate that the current backscattering interactions of ${\rm SU}(m)_n$, ${\rm SO}(m)_n$ and ${\rm Sp}(2m)_n$ create a finite bulk excitation energy gap, for general level and rank (except for ${\rm SO}(2)_m=U(1)_{4m}$ which can be gapped by a sine-Gordon interaction instead). We defer the proof to a later work.

In addition to the classical Lie algebras, there are the exceptional simple Lie algebras $E_6$, $E_7$, $E_8$, $F_4$ and $G_2$. Our previous work in Ref.~\cite{LimMulliganTeo2022A} addressed the fractional quantum Hall states whose edge states are described by WZW CFTs of these affine exceptional Lie algebras at level 1. It would be interesting to construct coupled-wire models at arbitrary levels and investigate what topological phases they may represent.

It would be natural to expect the method of coupled wire construction to encompass a wider set of 2D topological phases, whose boundary states are beyond WZW CFTs. For example, in ref.~\cite{LimMulliganTeo2022A}, we constructed a topological model with Ising topological order and Ising CFT on its edge. The Ising theory is the smallest unitary minimal CFT. We postulate that all chiral unitary minimal CFT models $\mathcal{M}(m,m-1)$, with central charge $c=1-6/m(m+1)$ for $m\geq3$, can be realized as the boundary of a topological phase constructed as a coupled wire model. In particular, minimal CFTs are cosets $\mathcal{M}(m,m-1)=SU(2)_1\times SU(2)_{m-2}/SU(2)_{m-1}$. We anticipate all chiral coset CFTs $\mathcal{G}_k/\mathcal{H}$ can be realized using the coupled wire construction as well.

In \cite{PhysRevB.104.115155},
we studied the multipartite entanglement of the Laughlin and Moore-Read states using  coupled-wire models.  
We found that a general Moore-Read ground state cannot be disentangled even when the disentangling condition \cite{PhysRevA.91.012339} holds.
Future efforts could focus on
extending our work \cite{PhysRevB.104.115155} to the phases we studied in this paper.

It is interesting to explore how some of these constructions could be realized in more realistic systems. Recent studies \cite{PhysRevB.110.L201106,paul2024designinghigherhallcrystals} demonstrated that coupled-wire models can emerge in graphene-like materials. However, it remains unclear how one might experimentally induce the specific combination of attractive and repulsive interactions between different electron channels required to produce the phases considered in this paper within such materials.

\section*{Acknowledgments}
We thank Bo Han and Yichen Hu for insightful discussions. This research was partly conducted while JCYT was visiting the Okinawa Institute of Science and Technology (OIST) through the Theoretical Sciences Visiting Program (TSVP). 
This work was supported by the U.S.~Department of Energy, Office of Science, Office of Basic Energy Sciences, under Award No.~DE-SC0020007 (M.M. and P.K.L.). 
M.M.~acknowledges the kind hospitality of the Kavli Institute for Theoretical Physics, which is supported in part by the National Science Foundation under Grants No.~NSF PHY-2309135, during the completion of this work.

\newpage
\appendix

\section{$U$-matrix Construction}

We detail the construction of the $U$-matrix that maps $\tilde \Phi^\sigma_{y \mu} = \sum_{\sigma' j} U^{\sigma \sigma'}_{\mu j} \Phi^{\sigma'}_{y j}$, cf.\eqref{bigphibigtildephi}. It is an
$(2N+4)\times (2N+4)$ integral matrix taking the form of
\begin{align}
U =\begin{pmatrix}
U^{++} & U^{+-}\\
U^{-+} & U^{--}
\end{pmatrix},
\label{eq:partonUmatrix}
\end{align} 
with entries depending on the parity of $N$.

When $N$ is odd, its block matrices are given by \eqref{eq:genSUNblocks}. Its diagonal block matrices, and off diagonal block matrices are related by inversion symmetry, i.e., $U^{--}=U^{++}\Lambda$ and $U^{-+}=U^{+-}\Lambda$.
$\Lambda$ is the inversion symmetry operator, 
a square matrix with all anti-diagonal elements being 1 and elsewhere zeros.
The choice of $U$ satisfies $U\eta U^T=K\oplus (-K)$ where $K\equiv(\oplus_{i=1}^3 N)\oplus K_{{\rm SU}(N)_1}$ characterizes the affine Lie algebra ${\rm U}(1)^3_N\times {\rm SU}(N)_1$ of the integrated channels.
The metric
$\eta=\mathbb{1}_{N+1}\oplus(-\mathbb{1}_{N+1})$ separates the left and right chiral sectors of the theory.
The row vectors of $U$ also generate appropriate charge and spin assignments for the integrated electronic operators. So in the unit of electric charge $e$,
$q^\sigma_\rho=Q\big(e^{i\tilde{\Phi}_{y\rho}^\sigma}\big)=1$, 
$q^\sigma_{c_{1}}=Q\big(e^{i\tilde{\Phi}_{yc_{1}}^\sigma}\big)=1$,
$q^\sigma_{c_{2}}=Q\big(e^{i\tilde{\Phi}_{yc_{2}}^\sigma}\big)=N$,
and
$q^\sigma_{n_J}=Q\big(e^{i\tilde{\Phi}_{yn_{J=1,\ldots, N-1}}^\sigma}\big)=0$. The spin of an integrated electronic operator can be read off from
$h\big(e^{i\Tilde{\Phi}^\sigma_{y\mu}}\big)=\sum_{J}\eta_{\sigma'\sigma'}(U^{\sigma\sigma'}_{\mu J})^2/2$. The corresponding spins for each integrated electronic species are $h(e^{\Tilde{\Phi}^\sigma_{y\rho}})= h(e^{\Tilde{\Phi}^\sigma_{y c_{1,2}}})=\sigma N/2$, and $h\big(e^{i\Tilde{\Phi}^\sigma_{yn_j}}\big)=\sigma $. 

When $N$ is even and $N=mn\geq 4$ with $m,n>1$, 
the integrated channel of $\rho$ need to be rearranged such that it carries charge $2e$ and spin $\sigma N/2$. Analogous to the Luther-Emery liquid \cite{PhysRevLett.33.589,PhysRevB.71.045113}, let the electronic channel $c^\sigma_0$ symbolizes the spin-up electron annihilation operator $c^\sigma_{\uparrow}$, and the channel $c^\sigma_{N+1}$ symbolizes the spin-down annihilation operator $c^\sigma_{\downarrow}$. The corresponding bosonized variables can be rearranged into a local field $\varphi$ and its dual $\vartheta$. Such that they obey the commutation relation 
\begin{align}
\left[\varphi(\mathsf{x}),\partial_{\mathsf{x}'}\vartheta(\mathsf{x}')\right] = i\pi\delta(\mathsf{x}-\mathsf{x}').
\end{align}
Because of the polarization of spins under the strong transverse magnetic field, the integrated electronic channel $c_1$ can be expressed in terms of bosonized variables of the $c^\sigma_0$ and $c^\sigma_{N+1}$ electronic operators by
\begin{align}
\begin{split}
\begin{pmatrix}
    \varphi \\ 2\vartheta\\ \Tilde{\Phi}^R_{c_1}\\ \Tilde{\Phi}^L_{c_1}
\end{pmatrix}=
\begin{pmatrix}
    0&1&1&0\\-1&0&1&0\\-1&1&1&0\\0&0&0&1
\end{pmatrix}
\begin{pmatrix}
    \Phi^R_\uparrow \\ \Phi^R_\downarrow\\ \Phi^L_\uparrow\\ \Phi^L_\downarrow
\end{pmatrix}.
\end{split}
\label{eq:LutherEmeryTransformation}
\end{align}
We can then define $\Tilde{\Phi}^\sigma_{y\rho}=\varphi-\sigma N\vartheta$ based on \eqref{eq:LutherEmeryTransformation}. So that the $\rho$ mode carries the desired charge and spin. The $c_1$ mode, on the other hand, has charge $e$ and spin $\sigma/2$. These two integrated modes remain local as integral combinations of electrons.
The rest of the integrated channels, namely the $c_2, n_{J=1,\ldots,N-1}$ modes, follow the same definitions as those in the odd $N$ case. The integral matrix $U$ now takes the form of \eqref{eq:genSUNevenblockSet1}.
It satisfies $U\eta U^T=K\oplus (-K)$ where $K\equiv N\oplus 1\oplus N\oplus K_{{\rm SU}(N)_1}$ characterizes the affine Lie algebra ${\rm U}(1)_N\times$ ${\rm U}(1)\times {\rm U}(1)_N\times {\rm SU}(N)_1$ 
for the integrated  channels. 
Based on the spin-charge assignments for either $N$ odd or even cases, the vertex operators of species $\rho,c_1,$ and $c_2$ are charged fermions while $n_1,\ldots, n_{N-1}$ are neutral bosons.
This completes the construction of $U$-matrix for all parities of $N$.

\section{Momentum Conservation }
\label{sec:AseriesMomentumConserv}

In this appendix, we address the momentum conservation of the coupled-wire models for the quantum Hall states in section~\ref{Sec:AseriesPartonWireConstruction}. The models are constructed on an array of bundles of electron wires. The vertical position of the $j^{\mathrm{th}}$ wire in the $y^{\mathrm{th}}$ bundle is $\mathsf{y}=yd+\delta_j$. The Fermi momentum is $p_{F,j}=\hbar k_{F,j}$. With the perpendicular magnetic field turned on and under the vector potential ${\bf A}=-B\mathsf{y}\hat{\bf x}$, the momentum of the $R/L$-moving electron on each wire is shifted to \eqref{eq:generalShiftFermiMomenta}, which we repeat here \begin{align}
k^\sigma_{yj}=\frac{eB}{\hbar c} \left(y d+\delta_j\right) +\sigma k_{F,j},
\label{app:generalShiftFermiMomenta}
\end{align} where $\sigma=R/L=+/-$. Using the basis transformation $\tilde\Phi=U\Phi$ in \eqref{bigphibigtildephi} defined by the $U$ matrix in \eqref{eq:genSUNblocks} and \eqref{eq:genSUNevenblockSet1}, the momentum of the $\mu=\rho,c_1,c_2,I=1,\ldots,N-1$ modes are $\tilde{k}^\sigma_{y\mu}=\sum_{j\sigma'}U^{\sigma\sigma'}_{\mu j}k^\sigma_{yj}$. With the electric charges $q_\mu=Q(e^{i\tilde\Phi^\sigma_{y\mu}})=\sum_{j\sigma'}U^{\sigma\sigma'}_{\mu j}$ (in units of $e$), \begin{align}\tilde{k}^\sigma_{y\mu}=q_\mu\frac{eBd}{\hbar c}y+\sum_{j\sigma'}U^{\sigma\sigma'}_{\mu j}\left(\kappa_j+\sigma' k_{F,j}\right),\end{align} where $\kappa_j=\frac{eB}{\hbar c}\delta_j$. The only modes that carry electric charges are $\rho,c_1,c_2$, and the rest have $q_{I=1,\ldots,N-1}=0$. Therefore, for the neutral modes $I=1,\ldots,N-1$, their momentum $\tilde{k}^\sigma_{yI}=\tilde{k}^\sigma_I$ are $y$-independent as they do not couple with the magnetic field. 

The momentum conservation conditions for all quantum Hall models in section~\ref{Sec:AseriesPartonWireConstruction} are the same. They are \begin{align}\tilde{k}^R_{y,\rho}=\tilde{k}^L_{y+1,\rho},\quad\tilde{k}^R_{yc_{1,2}}=\tilde{k}^L_{yc_{1,2}},\quad\tilde{k}^R_I=\tilde{k}^L_I.\label{kconscond}\end{align} The first condition ensures the $U(1)_N$ inter-bundle sine-Gordon interaction $\mathcal{H}^{U(1)_N}_{\mathrm{inter}}$ in \eqref{eq:U1Ninter} preserves momentum. When expressing the interaction back in terms of the electrons, the oscillating factors $e^{ik\mathsf{x}}$ in \eqref{eq:N2electronicchannels} cancel, and the potential density $\mathcal{H}^{U(1)_N}_{\mathrm{inter}}$ survives the $\mathsf{x}$-integration in the thermodynamic limit. Similarly, the second condition in \eqref{kconscond} guarantees momentum conservation for the intra-bundle potentials in \eqref{cgapping} that gap the $c_{1,2}$ modes. The third condition in \eqref{kconscond} ensures momentum conservation for the $SU(N)_1$ inter-bundle current backscattering potential in \eqref{eq:SUNinterlevel1}. In fact, we will find a solution for the more restrictive condition \begin{align}\tilde{k}^R_{I}=\tilde{k}^L_{I}=0\label{kconscondn}\end{align} for the neutral modes so that the $SU(n)_m$ and $SU(m)_n$ current backscattering potentials in \eqref{eq:SUnmintra} and \eqref{eq:SUmninter} also preserve momentum conservation.

The momentum conservation conditions \eqref{kconscond} and \eqref{kconscondn} can be met by tuning the electron density, which sets the bare Fermi momentum $k_{F,j}$, and the relative vertical position $\delta_j$ of each wire. Small deviations of $k_{F,j}$ and $\delta_j$ away from the fine-tuned solution to \eqref{kconscond} and \eqref{kconscondn} are perturbations that do not change the topological phase as long as they do not close the bulk excitation energy gap of the models. We present a particular solution of $k_{F,j}$ and $\delta_j$ starting with the ansatz \begin{align}\kappa_j=\frac{eB}{\hbar c}\delta_j=k_{F,j}.\end{align} The conditions in \eqref{kconscond} under \eqref{app:generalShiftFermiMomenta} becomes \begin{align}\left(U^{++}-U^{-+}\right)2{\bf k}_F=q_\rho\frac{eBd}{\hbar c}{\bf e}_1\end{align} where ${\bf k}_F=(k_{F,0},\ldots,k_{F,N+1})^T$, ${\bf e}_1=(1,0,\ldots,0)^T$, and $q_\rho=1$ if $N$ is odd or $q_\rho=2$ if $N$ is even. The matrix difference $U^{++}-U^{-+}$ is invertible for all $N\geq2$ and the linear system has the unique solution \begin{align}\begin{split}&k_{F,0}=\left(\frac{eBd}{\hbar c}\right)\frac{1-N}{4N},\quad k_{F,j=1,\ldots,N}=0\\&k_{F,N+1}=\left(\frac{eBd}{\hbar c}\right)\frac{1+N}{4N}\end{split}\end{align} for odd $N$, or \begin{align}k_{F,0}=k_{F,N+1}=\left(\frac{eBd}{\hbar c}\right)\frac{1}{N},\quad k_{F,j=1,\ldots,N}=0\end{align} for even $N$. From the $U$ matrices in \eqref{eq:genSUNblocks} and \eqref{eq:genSUNevenblockSet1}, the patterns of zeros $k_{F,j=1,\ldots,N}=0$ implies the vanishing of the momentum of the neutral modes \eqref{kconscondn} because $\sum_{j=0}^{N+1}U^{\sigma\sigma'}_{Ij}k_{F,j}=0$, for $I=1,\ldots,N-1$ and any $\sigma,\sigma'=\pm$.

The Fermi momentum solution gives $\sum_j2k_{F,j}=\frac{eBd}{\hbar c}$. The electron number density (per unit length) on each wire is $n^0_j=2k_{F,j}/(2\pi)$, and the total number density of each bundle $n_e=\sum_jn^0_j$. The number of magnetic flux quanta per unit length passing between consecutive bundles is $n_B=Bd/\phi_0$, where $\phi_0=2\pi\hbar c/e$ is the magnetic flux quantum. The filling fraction is the ratio \begin{align}\nu=\frac{n_e}{n_B}=\frac{\hbar c}{eBd}\sum_{j=0}^{N+1}2k_{F,j}\end{align} which is $\nu=1/N$ for odd $N$ or $\nu=4/N$ for even $N$.

\section{Parafermion Decomposition of \texorpdfstring{${\rm SU}(2)_n$}{SU(2)n}}
\label{Appendix:ParafermionDecomp}
Starting with $n=2$, the current in \eqref{eq:SU2LevelnCurrents}
becomes 
\begin{align}
\begin{split}
\mathtt{E}^+_{{\rm SU}(2)_2} &=
e^{i(\phi_1-\phi_3)}+e^{i(\phi_2-\phi_4)}=
\xi\times \Psi,
\end{split}
\label{eq:CurrentSU2level2plus}
\end{align}
with corresponding components given by  
\begin{align}
\begin{split}
\xi &= \sqrt{2}e^{i\frac{(\theta_1+\theta_2)}{2}}\\
\Psi &=
\frac{1}{\sqrt{2}}\left(
e^{i\frac{(\theta_1-\theta_2)}{2}}+e^{-i\frac{(\theta_1-\theta_2)}{2}}
\right)\\
\theta_1&\equiv \phi_1-\phi_3, \quad 
\theta_2\equiv \phi_2-\phi_4.
\end{split}
\label{eq:parafdecompSU2L2}
\end{align}
The parafermion primary fields obey the OPE
\begin{align}
\begin{split}
\Psi(z)\Psi^\dagger(w) &=
\frac{1}{(z-w)}\left[1+(z-w)^2T_{\mathbb{Z}_2}\right]+\ldots,
\end{split}
\end{align}
where the energy-momentum tensor of the $\mathbb{Z}_2$ parafermion CFT is $T_{\mathbb{Z}_2}=\cos(\theta_1-\theta_2)$.
Therefore the parafermions satisfy fusion rules $\Psi\times\Psi = \Psi\times \Psi^\dagger =1$ with $\Psi=\Psi^\dagger$.

For $n=3$, the current operator is 
\begin{align}
\begin{split}
\mathtt{E}^+_{{\rm SU}(2)_3} &=
e^{i\theta_1}+e^{i\theta_2}+e^{i\theta_3}.
\end{split}
\label{eq:SU2level4CurrentOp}
\end{align}
The angle variables are now defined by $\theta_1\equiv \phi_1-\phi_4$, 
$\theta_2\equiv \phi_2-\phi_5$, and 
$\theta_3\equiv \phi_3-\phi_6$.
\eqref{eq:SU2level4CurrentOp} can be factorized in terms of
\begin{align}
\begin{split}
\xi &= \sqrt{3}e^{i(\theta_1+\theta_2+\theta_3)/3},\\
\Psi&=\frac{1}{\sqrt{3}}
\left(
e^{i\frac{2\theta_1-\theta_2-\theta_3}{3}}
+
e^{i\frac{-\theta_1+2\theta_2-\theta_3}{3}}
+
e^{i\frac{-\theta_1-\theta_2+2\theta_3}{3}}
\right).
\end{split}
\end{align} 
The $\mathbb{Z}_3$ parafermion primary fields obey OPEs
\begin{align}
\begin{split}
\Psi(z)\Psi^\dagger(w)&=
\frac{1}{(z-w)^{4/3}}\left[
1+\frac{2}{3}(z-w)^2T_{\mathbb{Z}_3}
\right]+\ldots,\\
\Psi(z)\Psi(w)&=\frac{2/\sqrt{3}}{(z-w)^{2/3}}\Psi^\dagger(w)+\ldots,
\end{split}
\end{align}
with fusion rules
$\Psi\times \Psi=\Psi^\dagger$, $\Psi\times \Psi^\dagger = \Psi^{3}=(\Psi^\dagger)^3=1$.
The energy-momentum tensor of the $\mathbb{Z}_3$ parafermion CFT takes the form of $T_{\mathbb{Z}_3}=\sum_{1\leq a<b\leq 3}\cos(\theta_a-\theta_b)$.

Finally, for $n=4$, the current operator is
\begin{align}
\mathtt{E}^+_{{\rm SU}(2)_4}
&=\sum_{c=1}^4 e^{i\theta_c},
\label{eq:SU2L4parafermiondecomposition1}
\end{align} 
where $\theta_c\equiv \phi_c-\phi_{4+c}$.
The corresponding parafermion decomposition has
\begin{align}
\begin{split}
\xi &= \sqrt{4}e^{i\sum_{j=1}^4\theta_j/4},\\
\Psi &=
\frac{1}{\sqrt{4}}\Big(
e^{i(3\theta_1-\theta_2-\theta_3-\theta_4)/4}
+e^{i(-\theta_1+3\theta_2-\theta_3-\theta_4)/4}
\\&\;\;\;\;
+e^{i(-\theta_1-\theta_2+3\theta_3-\theta_4)/4}
+e^{i(-\theta_1-\theta_2-\theta_3+3\theta_4)/4}
\Big),
\end{split}
\label{eq:SU2L4parafermiondecomposition2}
\end{align}
following by the OPEs
\begin{align}
\begin{split}
\Psi(z)\Psi^\dagger(w) &=
\frac{1}{(z-w)^{3/2}}\Big(1+\frac{2}{4}(z-w)^2T_{\mathbb{Z}_4}
\Big)+...,
\\
\Psi(z)\Psi(w) &=
\frac{\sqrt{3/2}}{(z-w)^{1/2}}\Psi_2+\ldots,\\
\Psi(z)\Psi_2(w) &=
\frac{\sqrt{3/2}}{(z-w)}\Psi^\dagger+\ldots,
\end{split}
\end{align}
where the primary field
\begin{align}
\begin{split}
\Psi_2 &=
\frac{1}{\sqrt{6}}
\Big(
e^{i(\theta_1+\theta_2-\theta_3-\theta_4)/2}
+e^{i(\theta_1-\theta_2+\theta_3-\theta_4)/2}
\\&\;\;\;\;
+e^{i(\theta_1-\theta_2-\theta_3+\theta_4)/2}
+e^{i(-\theta_1-\theta_2+\theta_3+\theta_4)/2}
\\&\;\;\;\;
+e^{i(-\theta_1+\theta_2-\theta_3+\theta_4)/2}
+e^{i(-\theta_1+\theta_2+\theta_3-\theta_4)/2}
\Big).
\end{split}
\end{align}
This yields the fusion rules of $\mathbb{Z}_4$ parafermion CFT:
$\Psi\times\Psi=\Psi_2$, $\Psi\times \Psi_2 =\Psi_3= \Psi^\dagger$, and $\Psi^4=(\Psi^\dagger)^4=1$.

In general, the ${\rm SU}(2)$ current at any given level $n$ can be decomposed into 
\begin{align}
\begin{split}
\xi &=\sqrt{n} e^{i\theta_\perp/n}, 
\quad 
\Psi = \frac{1}{\sqrt{n}}
\sum_{ c=1}^n
e^{i(\theta_{c}-\theta_\perp/n)},
\end{split}
\end{align} 
where $\theta_\perp = \theta_1+\dots+\theta_{n}$, and $\theta_c=\phi_c-\phi_{n+c}$.
A primary field $\Psi_k$, upon fusing $\Psi$ $k$ times, where 
$k=0\;{\rm mod}\;n$,
has the normalized constant $\sqrt{C^n_k}$ where
$C^n_k = n!/[k!(n-k)!]$. 
Its bosonization is the sum of vertex operators,  
\begin{align}
\begin{split}
\Psi_k = \frac{1}{\sqrt{C^n_k}} \sum_{1\leq {j_1} <\ldots < {j_k}\leq n}
e^{i(\theta_{j_1} +\ldots + \theta_{j_k}-k\theta_\perp/n)}.
\end{split}
\end{align}
The primary fields of $\mathbb{Z}_n$ parafermion CFT have OPEs 
\begin{align}
\begin{split}
\Psi(z)\Psi^\dagger(w) &= \frac{1}{(z-w)^{\frac{2(n-1)}{n}}}
\left(1+\frac{2}{n}(z-w)^2T_{\mathbb{Z}_n}\right)+\ldots,\\
\Psi_k(z)\Psi_\ell(w)&=\frac{c_{k,\ell}}{(z-w)^{\frac{2k\ell}{n}}}\Psi_{k+\ell}(w)+\ldots,
\end{split}
\end{align}
and carry spins $h(\Psi_k) = k(n-k)/n$.
The energy-momentum tensor can be generalized to $T_{\mathbb{Z}_n} = \sum_{1\leq a<b\leq n}\cos(\theta_a-\theta_b)$.
The structure constants of the parafermion current algebra,
\begin{align}
\begin{split}
c_{k,\ell} &= 
C^{k+\ell}_\ell\sqrt{\frac{C^n_{k+\ell}}{C^n_k C^n_\ell}}=
\sqrt{\frac{(k+\ell)!(n-k)!(n-\ell)!}{k!\ell!(n-k-\ell)!n!}},
\end{split}
\end{align}
rely upon the normalization factors of fields $\Psi_k$, $\Psi_\ell$ and $\Psi_{k+\ell}$, with $C^{k+\ell}_\ell$ duplicate terms as a result of operator product expansion between $\Psi_k$ and $\Psi_\ell$. 
The conformal embedding, in the form of the parton construction, allows us to write down the closed form of $\mathbb{Z}_n$ parafermion primary fields.
This is demonstrated through explicit calculations of parafermion current algebra. Although Zamolodchikov and Fateev had proved the general result
\cite{ZamolodchikovFateev85,FATEEV199191}, we wrote down the explicit parafermion decomposition here, specifically and pedagogically for our study of topological orders in ${\rm SU}(2)_{n=2,3,4}$.

\bibliographystyle{unsrtnat}

\end{document}